\documentclass[11pt,a4paper]{article}
\usepackage{jheppub}
\usepackage{amsmath}
\usepackage[table]{xcolor}
\usepackage{bbm}
\usepackage[utf8]{inputenc}
\usepackage{graphicx}
\usepackage{tabularx}
\usepackage{multirow}
\usepackage{hyperref}
\usepackage{comment}
\usepackage{amssymb,amsfonts}
\usepackage{mathrsfs}
\usepackage{subcaption}

\usepackage{booktabs, comment}

\def \sltz {\text{SL}_2(\mathbb Z)}
\usepackage{array}   
\newcolumntype{L}{>{$}l<{$}} 
\usepackage{float} 
\usepackage{chngpage} 
\newcommand\scalemath[2]{\scalebox{#1}{\mbox{\ensuremath{\displaystyle #2}}}}

\title{Bootstrapping Fermionic Rational CFTs with Three Characters}
\author[1]{Jin-Beom Bae,}
\author[2]{Zhihao Duan,}
\author[2]{Kimyeong Lee,}
\author[2]{Sungjay Lee,}
\author[2]{and Matthieu Sarkis}

\affiliation[1]{Mathematical Institute, University of Oxford \\
$~~ $Andrew Wiles Building, Radcliffe Observatory Quarter \\
$~~ $Woodstock Road, Oxford, OX2 6GG, U.K.}
\affiliation[2]{Korea Institute for Advanced Study \\
$~~ $85 Hoegiro, Dongdaemun-Gu, Seoul 02455, Korea}

\preprint{KIAS-P21028}

\abstract{
Recently, the modular linear differential equation (MLDE) for level-two congruence subgroups $\Gamma_\theta, \Gamma^{0}(2)$ and $\Gamma_0(2)$ of $\text{SL}_2(\mathbb{Z})$ was developed and used to classify the fermionic rational conformal field theories (RCFT). Two character solutions of the second-order fermionic MLDE without poles were found and their corresponding CFTs are identified. Here we extend this analysis to explore the landscape of three character fermionic RCFTs obtained from the third-order fermionic MLDE without poles. Especially, we focus on a class of the fermionic RCFTs whose Neveu-Schwarz sector vacuum character has no free-fermion currents and Ramond sector saturates the bound $h^{\text{R}} \ge \frac{c}{24}$, which is the unitarity bound for the supersymmetric case. Most of the solutions can be mapped to characters of the fermionized WZW models. We find the pairs of fermionic CFTs whose characters can be combined to produce $K(\tau)$, the character of the $c=12$ fermionic CFT for $\text{Co}_0$ sporadic group. 
}

\begin{document}

\maketitle

\section{Introduction and Summary}

The classification of two-dimensional rational conformal field theory(RCFT) has been a longstanding problem. Often, two-dimensional RCFTs are known to possess an extended chiral algebra, beyond the Virasoro algebra, which has been widely used to explored various observables such as the torus partition function and correlation functions. Altogether with the modular invariance of the partition function or the crossing symmetry of the correlation function, the unitary RCFT with $c<1$ has been completely classified \cite{Belavin:1984vu}. However, a full classification of RCFT with $c>1$ is still far-fetched. 

A promising approach to classify RCFT is to utilize the modular property of the characters. Due to the fact that the characters, whose number is assumed to be $N$, can be regarded as components of a vector-valued modular form of weight-zero, the characters are identified with the independent solutions of an $N$-th order differential equation which is invariant under $\sltz$ \cite{Eguchi:1987qd,Mathur:1988na, Mathur:1988gt}. Based on this observation, significant progress has been made for the classification of two-character and three-character RCFTs \cite{Naculich:1988xv, hoehn2007selbstduale, Tuite:2008pt, Hampapura:2015cea, Hampapura:2016mmz, Chandra:2018pjq, Mukhi:2019xjy, Franc_2020}.

It is natural to extend the classification problem to the fermionic RCFT. This problem has been recently initiated by present authors with help of the fermionic modular linear differential equation(MLDE) \cite{Bae:2020xzl}. To introduce a fermionic RCFT, we start by putting the theory on a torus with a spin structure. The torus entails four distinct spin structures that are characterized by the anti-periodic(A) and periodic(P) boundary conditions along the two cycles of torus. In this paper, we denote four spin structures associated with (A,A), (P,A), (A,P) and (P,P) as NS, $\widetilde{\text{NS}}$, R and $\widetilde{\text{R}}$-sectors. Specifically, the characters of the NS, $\widetilde{\text{NS}}$ and R sectors are known to form a vector-valued modular form under the congruence subgroups $\Gamma_\theta$, $\Gamma^{0}(2)$ and $\Gamma_{0}(2)$, respectively. Again, $N$-characters are expected to be identified with the solutions of an $N$-th order MLDE associated with the congruence subgroups $\Gamma_\theta$, $\Gamma^{0}(2)$ and $\Gamma_{0}(2)$.

The main purpose of this paper is to explore the theory space of three-character fermionic RCFT by analyzing the fermionic third-order MLDE. The classification program is closely related to the modular tensor categories (MTC) which encode the algebraic structures of both RCFT and three-dimensional topological quantum field theory(TQFT)\cite{Moore:1991ks, Wen:2016lrc, Schoutens:2015uia}. Apart from the low rank MTCs \cite{Gepner:1994bb, Rowell:2007dge}, the classification of the low rank fermionic MTCs have been recently studied in \cite{Bruillard:2016yio}. By analyzing the fermionic RCFT with MLDE, we aim to see if our classification arrives at the same consequences as the fermionic MTCs. As the first step, we restrict to the cases where the modular forms involved in the MLDE do not have poles inside the fundamental domain of congruence subgroups. In addition, we focus on the two special types of solutions. The first type solutions can be identified with the fermionic characters constructed from the characters of three-character bosonic RCFTs\cite{Hampapura:2016mmz,Gaberdiel:2016zke,Franc_2020}. More precisely, we take the tensor product of $\mathcal{N}$-copies of the Majorana-Weyl free fermion and arbitrary three-character bosonic RCFT. Regardless of $\mathcal{N}$, we show that characters of the above fermionic RCFT can be regarded as the solutions of a third-order fermionic MLDE. 

\begin{table}[]\label{tab:3rd}
    \centering
     \arraycolsep=5pt \def\arraystretch{1.5}
\begin{tabular}{c|c|c|c}
Class & $(c,h_1,h_2)$ & $\mathcal{B}$ & Comments \\
\hline \hline
I & $(\frac{3m}{2}, \frac{1}{2}, \frac{m}{8})$ & $(\text{SO}(m)_1)^{3}$ & $m$ runs for $2,3,\cdots, 16$\\ \hline 
\multirow{8}{*}{II} & $(2,\frac{1}{6},\frac{1}{3})$ & $(\text{U}(1)_{12}/{\mathbb{Z}_2})^2$ & $(\text{The first unitary} \ \mathcal{N}=2 \ \text{minimal model})^2$ \\[0.3em]
& $(\frac{18}{5},\frac{3}{10},\frac{2}{5})$ & $(\text{SU}(2)_3)^2$ & \\[0.3em]
& $(\frac{9}{2},\frac{1}{4},\frac{1}{2})$ & $(\text{SU}(2)_6)^2/{\mathbb{Z}_2}$ & \\[0.3em]
& $(5,\frac{1}{3},\frac{1}{2})$  & $\text{Sp}(4)_3$ & \\[0.3em]
& $(7,\frac{1}{2},\frac{2}{3})$  & $\text{Sp}(6)_2$ & Dual pair of $c=5$ solution \\[0.3em]
& $(\frac{15}{2},\frac{1}{2},\frac{3}{4})$ & $\text{SU}(4)_4/{\mathbb{Z}_2}$ & Dual pair of $c=\frac{9}{2}$ solution \\[0.3em]
& $(\frac{42}{5},\frac{3}{5},\frac{7}{10})$ & $(\text{Sp}(6)_1)^2$ & Dual pair of $c=\frac{18}{5}$ solution \\[0.3em]
& $(10,\frac{2}{3},\frac{5}{6})$ & $(\text{SU}(6)_1)^2$ & Dual pair of $c=2$ solution \\[0.3em] \hline
\multirow{2}{*}{III} & $(\frac{39}{2},\frac{3}{4},\frac{3}{2})$ & $(\text{Sp}(12)_1)^2/{\mathbb{Z}_2}$ & \\[0.3em]
& $(22,\frac{5}{6},\frac{5}{3})$ & $(\text{SU}(12)_1)^2$ & \\[0.3em]
& $(\frac{66}{5},\frac{4}{5},\frac{11}{10})$ & Unknown &  \\[0.3em]
\end{tabular}    
    \caption{\label{Summary of Result} Summary of the BPS solutions of the third-order MLDE. Most of the solutions listed in this table are related to the bosonic RCFTs with chiral algebras, except the solution with $c=66/5$.}
\end{table} 

On the other hand, the feature of the second type solution is that it saturates the Ramond sector supersymmetric unitarity bound, i.e., $h^{R} \ge \frac{c}{24}$. For this reason, we refer them as to the BPS solutions and they are listed in table \ref{Summary of Result}. We claim that most of the BPS solutions have relations with the WZW model or its $\mathbb{Z}_2$ orbifold theory. More concretely, the BPS solutions turn out to realize the characters of the fermionized WZW models. Those fermionic RCFTs arose as a consequence of the generalized Jordan-Wigner transformation, which was employed in recent studies\cite{Runkel:2020zgg, Hsieh:2020uwb, Kulp:2020iet}. One exception is the BPS solution with $c=\frac{66}{5}$. It is not clear if this solution is related to any WZW model or its orbifold theory via fermionization.

The characters of RCFT are the main ingredients of the partition function. Nevertheless, the method of MLDE does not explicitly tell us how to combine the solutions to construct a modular invariant partition function. In addition to solving the MLDE, we need to know their $S$-matrix to construct a partition function. To this end, we rewrite the BPS third-order MLDE in terms of the modular $\lambda$ function and show that the BPS solutions can be expressed in terms of the hypergeometric function in $\lambda$. With the closed-form solutions, we determine the $S$-matrix using the monodromy matrix. Such $S$-matrix enables us to study the fusion rule algebra via the Verlinde formula.

Sometimes, it has been known that the characters satisfy a novel bilinear relation. For instance, the characters of the Ising model and babymonster CFT are combined to produce $J(\tau)$\cite{Hampapura:2016mmz}. The bilinear relation has been utilized to analyze the monstralizing commutant pair in \cite{Bae:2020pvv}. It turns out that the solutions of the fermionic MLDE also exhibit bilinear relations for some cases. Especially, we find that the fermionized WZW models with SO$(m)_3$ for $m=2,3,4,5$ and their bilinear pairs span the BPS-type solutions of the fermionic second and third-order MLDEs. Based on these observations, we conjecture that the fermionized WZW models with SO$(6)_3$ and SO$(7)_3$ can be considered as the BPS-type solutions of the fermionic fourth-order MLDE. 

The fermionic RCFT involves the superconformal field theories. Thus, the fermionic MLDE provides a way of classifying the supersymmetric RCFT. The classification of the supersymmetric RCFT based on the WZW algebra has been discussed in \cite{Johnson-Freyd:2019wgb, Bae:2021lvk}. It would be plausible to explore the supersymmetric RCFT that cannot be mapped to the WZW models via bosonization. To promote fermionic RCFT to the superconformal field theories, their NS-sector vacuum character should possess chiral primary of weight $h=3/2$ which can be interpreted as a supersymmetric partner of the stress-energy tensor. Furthermore, the primaries of the R-sector ought to satisfy the unitarity bound $h^\textsc{r}\ge c/24$ and the $\widetilde{\text{R}}$-sector partition function would be constant due to the boson-fermion cancellation. The solutions listed in table \ref{Summary of Result} turn out to fulfill the supersymmetry conditions\footnote{Except the solution with $c=66/5$. Since its identification is unknown, it is not easy to see if the $\widetilde{\text{R}}$ sector partition function becomes constant or not.}. Some of them are not listed in \cite{Bae:2021lvk} since they are related to the WZW models with product groups.

This paper is organized as follows. In section \ref{Preliminaries} we review the construction of MLDE, the bosonization and fermionization of two-dimensional field theory, and the Rademacher expansion applied to the vector-valued modular form. The classification of solutions of the BPS third-order MLDE is provided in section \ref{Classification}. We further discuss the closed-form expression of the BPS solutions and their $S$-matrix. We focus on the bilinear relation of the BPS solutions in section \ref{deconstrution}, adopting a view of the deconstruction. The technical details are presented in appendices.

{\bf Note Added:} While this work is completed, \cite{Kaidi:2021ent,Das:2021uvd} appeared which contain some overlap in the classification of bosonic RCFT with three characters. In this paper, we use the analytic expressions of the solutions to establish the classification.

\section{Preliminaries} \label{Preliminaries}

\subsection{Review on the Modular Linear Differential Equations}\label{section:MLDEintro}

In this subsection, we review an approach of classifying bosonic RCFTs via MLDE \cite{Mathur:1988na}. We also discuss an extension of MLDE to the fermionic RCFTs \cite{Bae:2020xzl} using the congruence subgroup of level two. 
 
To illustrate the main idea of the classification scheme, we start by putting a RCFT on a torus. The torus partition function can be regarded as a trace over the Hilbert space of states on $S^1$, which can be further decomposed in terms of conformal characters $f_i(\tau)$ and $\bar{f}_j(\bar{\tau})$,
\begin{equation}\label{eq:torus_pf}
\begin{aligned}
Z(\tau,\bar{\tau}) &= \text{Tr}_{\mathcal{H}_{S^1}}[q^{L_0 - c/24}\bar{q}^{\bar{L}_0 - c/24}]\\
& = \sum_{i, j} M_{i j}\text{Tr}_{V_{h_i}}[q^{L_0 - \frac{c}{24}}]\text{Tr}_{V_{\bar{h}_{j}}}[\bar{q}^{\bar{L}_0 - \frac{c}{24}}] = \sum_{i, j} M_{i j} f_{i}(\tau) \bar{f}_{j}(\bar{\tau})\,.
\end{aligned}
\end{equation}
In the second line of \eqref{eq:torus_pf}, we decompose ${\mathcal{H}_{S^1}}$ into irreducible representations $V_{h_i}$, $V_{\bar{h}_j}$ of the left and right copies of an extended chiral algebra with multiplicity $M_{h, \bar{h}}$. The characteristic feature of RCFT is that its partition function consists of finitely many characters $f_i(\tau)$. Each character represents an irreducible representations $V_{h_i}$($V_{\bar{h}_j}$) and is characterized by the conformal weight $h_i(\bar{h}_j)$.

It is well-established that the torus has a global diffeomorphism group known as the modular group $\sltz$. Thus the torus partition function ought to be modular invariant and the characters should transform linearly under the $\sltz$. More concretely, the $S$ and $T$ transformation rules of the characters are given by
\begin{align}\label{eq:modularmatrices}
    f_a(-1/\tau)  = \sum_{b=0}^{N-1}S_{ab}f_b(\tau)\,,  \ \ 
    f_a(\tau+1)  = \sum_{b=0}^{N-1}T_{ab}f_b(\tau) \,, 
\end{align}
where $N$ denotes the number of characters. The modular matrices $S_{ab}$ and $T_{ab}$ satisfy the relations,
\begin{align}
\label{eq:modularRelations}
    S^2 = \big(ST\big)^3 = C,
\end{align}
where $C$ is the charge conjugation matrix, as well as
\begin{align}
  \big( S^\dagger \mathcal{M}   S\big)_{ab}=\mathcal{M}_{ab}, \quad 
  T_{ab}=q^{2\pi i h_a} \delta_{ab}\,, 
\end{align}
where $h_a$ means the conformal weight of the primary associated to $f_a(\tau)$. In other words, the set of characters $f_a(\tau)$ forms a vector-valued modular form of weight zero. It has been known that the characters $f_a(\tau)$ can be identified with the solutions of a modular linear differential equation (MLDE) of order $N$ \cite{Mathur:1988gt}.

An explicit form of the MLDE is given by \begin{align}\label{MLDE:wronsk}
    \left[ {\cal D}^N  + \sum_{s=0}^{N-1} \phi_s(\tau) {\cal D}^s \right] f(\tau) =0\, ,
\end{align}
where the Ramanujan-Serre derivative is defined as
\begin{equation}
    {\cal D} = \frac{1}{2\pi i } \frac{\text{d}}{\text{d}\tau} -\frac{k}{12}\, E_2(\tau)\,, 
\end{equation}
and $\phi_s(\tau)= (-1)^{N-s} W_s/W_N$ are modular forms of weight $2N - 2s$. 
Here we define the Wronskian $W_s$ as
\begin{align}
\label{eq:wronskian0}
    W_s = {\rm det} \left( \begin{array}{ccc}
    f_0 & \cdots &f_{N-1}  \\
    {\cal D} f_0 & \cdots & {\cal D} f_{N-1} \\
    \vdots & & \vdots \\
    {\cal D}^{s-1}f_0 & \cdots &D^{s-1}f_{N-1}  \\
    {\cal D}^{s+1}f_0 & \cdots & D^{s+1}f_{N-1} \\
    \vdots & & \vdots \\
    {\cal D}^N f_0 & \cdots & {\cal D}^n f_{N-1}  \\
\end{array}\right)\,,
\end{align}
therefore the coefficient function $\phi_s(\tau)$ can have the pole when $W_N$ becomes zero.

A useful way to understand the order of poles in $\phi_s$ is to utilize the \textit{valence} formula. To phrase it more precisely, let us suppose $g$ is a modular form of weight $k$. The valence formula relates the order of $g$ at the various points $p$ in the fundamental domain, which is denoted as $\nu_p(g)$. An explicit relation among $\nu_p(g)$ is given by \cite{Serre:1993,ranestad20081}
\begin{align}
\label{eq:valencebosonic}
   \nu_{i\infty}(g)+ \frac12 \nu_i(g)+\frac13\nu_{\omega}(g) +
    \sum_{p\in{\rm SL}_2(\mathbb{Z})\backslash \mathbb{H}
\atop p\neq i, \omega, i\infty} 
 \nu_{ p}(g)  = \frac{k}{12},
\end{align}
where $\omega = \exp(\tfrac{2\pi i}{3})$. Let us  take a modular form $g$ as the Wronskian $W_N$ with modular weight $k=N(N-1)$\footnote{Strictly speaking $W_N$ is modular only up to some constant factor. Fortunately, the proof of \eqref{eq:valencebosonic} still works in this case.}. Since the leading behavior of the Wronskian in $q$-expansion is given by
\begin{align}
    W_N(\tau)\sim q^{-\frac{Nc}{24}+\sum_a h_a} \Big(1+ {\mathcal O}(q) \Big)\, ,
\end{align}
it is straightforward to see $\nu_{i\infty}(g) = -\frac{Nc}{24}+\sum_a h_a$. Therefore, the valence formula reads
\begin{align}\label{eq:sl2zvalence}
     \frac{\ell}{6} = \frac{N(N-1)}{12} + \frac{Nc}{24} - \sum_a h_a,
\end{align}
where
\begin{align}
\label{eq:ellsl2z}
    \frac{\ell}{6}= \frac12 \nu_i+\frac13 \nu_\omega + \sum_{p\in{\rm SL}_2(\mathbb{Z})\backslash \mathbb{H}
\atop p\neq i, \omega, i\infty}\nu_{p}\, . 
 \end{align}
Note that $\ell$ runs for $2,3,4,\cdots$. Since the ring of holomorphic modular forms is generated by Eisenstein series $E_4$ and $E_6$, that vanish at $\tau = i$ and $\tau = \omega$ respectively, one can often deduce the form of $W_N$ in terms of $E_4$ and $E_6$ once $\ell$ is specified.

Next, we turn to constructing the MLDE whose solution can be identified with the characters of fermionic RCFT. We start from the fact that  defining fermionic RCFT needs the presence of spin structure. The spin structure arose since one can assign periodic (R) or anti-periodic (NS) boundary conditions along the non-trivial one-cycles. On the torus, there are four possible spin structures (NS,NS), (R,NS), (NS,R) and (R,R) sectors. For the notational convenience, we denote them as NS, $\widetilde{\text{NS}}$, R and $\widetilde{\text{R}}$, respectively. The first three sectors are transformed into each other by the modular transformation, therefore the partition function of each sector is not invariant under the $\sltz$. In fact, the partition functions of NS, $\widetilde{\text{NS}}$ and R sector are invariant under the level-two congruence subgroups $\Gamma_\theta$, $\Gamma^0(2)$ and $\Gamma_0(2)$, respectively.

\begin{figure}[h!]
     \centering
     \begin{subfigure}[b]{0.3\textwidth}
         \centering
         \includegraphics[width=\textwidth]{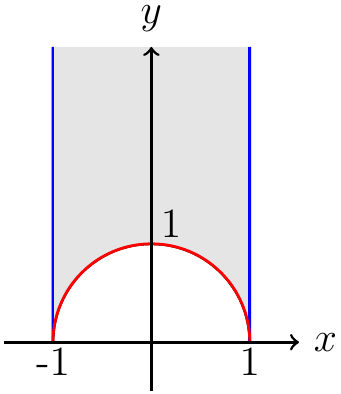}
         \caption{$\Gamma_\theta$}
     \end{subfigure}
     \hfill
     \begin{subfigure}[b]{0.3\textwidth}
         \centering
         \includegraphics[width=\textwidth]{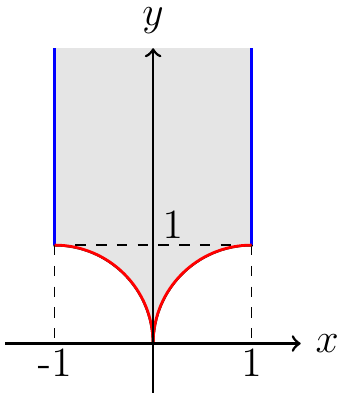}
          \vspace*{-5.3mm}
         \caption{$\Gamma^0(2)$}
     \end{subfigure}
     \hfill
     \begin{subfigure}[b]{0.3\textwidth}
         \centering
         \includegraphics[width=\textwidth]{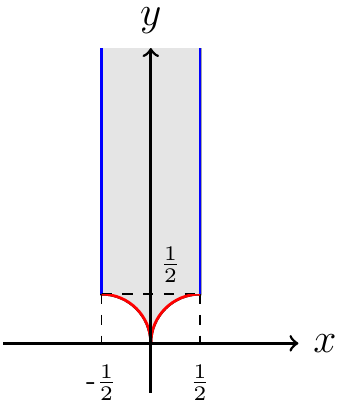}
         \vspace*{-6.7mm}
         \caption{$\Gamma_0(2)$}
     \end{subfigure}
        \caption{Fundamental domains of $\Gamma_\theta$, $\Gamma^0(2)$ and $\Gamma_0(2)$ respectively ($\tau = x + iy$).}
        \label{fig:three graphs}
\end{figure}

The level-two congruence subgroups are defined as follows,
\begin{equation}
\begin{aligned} 
    \Gamma_\theta=  &\left.\Big\{ \gamma\in {\rm SL}_2(\mathbb{Z})\right.| \gamma\equiv \big({_1 \ _0 \atop ^0 \ ^1}\big) \ {\rm or}\   \big({_0 \ _1  \atop ^1 \ ^0} \big) \ {\rm mod} \ 2 \Big\}, \\
    \Gamma^0(2)&=  \left.\Big\{ \gamma\in {\rm SL}_2(\mathbb{Z})\right.| \gamma\equiv \big({_\star \ _0 \atop ^\star \ ^\star }\big)  \   {\rm mod} \ 2\Big\},\\
    \Gamma_0(2)&=  \left.\Big\{ \gamma\in {\rm SL}_2(\mathbb{Z})\right.| \gamma\equiv \big({_\star \ _\star \atop ^0 \ ^\star}\big)   \   {\rm mod}\ 2 \Big\}.
\end{aligned}
\end{equation}
Their fundamental domains can be chosen as in figure \ref{fig:three graphs}. 

It is necessary to understand the space of modular forms of each congruence subgroup. The modular forms of each subgroup are known to be generated by the Jacobi theta functions as presented below.
\begin{align}
\begin{split}
\label{eq:modular form basis}
    \mathcal{M}_{2k}(\Gamma_\theta) &=\langle\   (-\vartheta_2^4)^r\vartheta_4^{4s} +  (-\vartheta_2^4)^s\vartheta_4^{4r}, \ r\leq s , \ r+s=k  \ \rangle , \\
    \mathcal{M}_{2k}(\Gamma^0(2)  ) &=\langle\   (-1)^{r+s}  ( \vartheta_3^{4r}\vartheta_4^{4s} +  \vartheta_3^{4s}\vartheta_4^{4r}), \ r\leq s, \ r+s=k \ \rangle , \\
  \mathcal{M}_{2k}(\Gamma_0(2)) &=\langle\  \vartheta_2^{4r}\vartheta_3^{4s} +   \vartheta_2^{4s}\vartheta_3^{4r} ,\ r\leq s,  \ r+s=k \ \rangle .
\end{split}
\end{align}

Having classified the basis of modular forms of each congruence subgroups, one can utilize the MLDE \eqref{MLDE:wronsk} to search the characters of fermionic RCFT. Let us first restrict our interest to the NS-sector partition function. The relevant MLDE again has a form of 
\begin{align}
    \left[ {\cal D}^N  + \sum_{s=0}^{N-1} \phi_s(\tau) {\cal D}^s \right] f_i^{\rm NS}(\tau) =0\,.
\end{align}
However, now the coefficient function $\phi_s(\tau)$ should be a meromorphic modular form of $\Gamma_\theta$ instead of $\sltz$. The fermionic MLDE can be written in terms of the basis of modular form, i.e., \eqref{eq:modular form basis}. For instance, the structure of the second-order NS-sector MLDE reads
\begin{align}
\label{eq:2MLDEThe}
\begin{split}
& \Big[\mathcal{D}^2+ \mu_1  \left(-\vartheta_2^4(\tau)+\vartheta_4^4(\tau)\right) \mathcal{D} + \mu_2 \theta_3^8 + \mu_3 E_4 \Big]f^{\text{NS}}_i(\tau)=0\,, 
\end{split}
\end{align}
where we assumed that the coefficient function does not involve any pole in the fundamental domain. The MLDE for the $\widetilde{\text{NS}}$ sector or R-sector is immediately obtained by applying $T$ or $ST$ transformation to \eqref{eq:2MLDEThe}. 

Let us briefly discuss the valence formula and its consequence. For the details, we refer readers to \cite{Bae:2020xzl}. The valence formula for $\Gamma_\theta$ is known to have a form of
 	\begin{equation}
 	\label{eq:valence_formula_gammatheta}
 	2\nu_{i\infty}(g)+\nu_1(g) + 	\frac{\nu_{i}(g)}{2}+\sum_{\substack{\tau\in\Gamma_\theta\backslash\mathbb H\\ \tau\neq i}}\nu_\tau(g)=\frac{k}{4}\,.
	\end{equation}
We now apply \eqref{eq:valence_formula_gammatheta} to the Wronskian $W_N$ constructed from $f_{i}^{\rm NS}(\tau)$. One can show that the order of Wronskian at $\tau = i \infty$ and $\tau = 1$ is given by
\begin{equation}\label{eq:W_Ninf}
		\nu_{i\infty}(W_N)=-\frac{Nc}{24} + \sum_{i}h_i^{\textsc{ns}}\,, \qquad \nu_{1}(W_N)=-\frac{Nc}{24} + \sum_{i}h_i^{\textsc{r}}\,.
\end{equation}
Therefore the valence formula leads to the following relation,
\begin{equation}
\label{eq:valencetheta}
     \frac{\tilde{\ell}}{2}= \frac{N(N-1)}{4} + \frac{Nc}{8} - 2\sum_i h_i^{\rm NS} - \sum_i h_i^{\rm R},
\end{equation}
where
\begin{equation}
\label{eq:valtheta}
    \frac{\tilde{\ell}}{2} = \frac12 \nu_i(W)+ \sum_{  \tau\in\Gamma_\theta\backslash \mathbb{H} 
    \atop \tau\neq i}\nu_\tau(W)\, .
\end{equation}

\subsection{Bosonization and Fermionization}\label{sec:fermionization}

We briefly review the modern perspective of  bosonization and fermionization in two-dimensional spacetime. The bosonization is known as the GSO projection and we will refer to fermionization as the generalized Jordan-Wigner transformation following \cite{Karch:2019lnn,Hsieh:2020uwb}.

Let us denote a bosonic theory with a non-anomalous $\mathbb{Z}_2$ symmetry as $\mathcal{B}$. To construct a fermionic theory $\mathcal{F}$ from $\mathcal{B}$, we first introduce a non-trivial two-dimensional invertible fermionic topological order on the Riemann surface $\Sigma_g$ with spin structure $\rho$. This theory is known as the Kitaev Majorana chain. Its partition function is described by the mod 2 index that is often referred to as the Arf invariant. The Arf invariant is 1 for even $\rho$ and 0 for odd $\rho$. An explicit form for the partition function of the Kitaev Majorana chain is,
\begin{align}
Z_{KM}[\rho] = e^{i \pi \text{Arf}[\rho]}.
\end{align}
When the Kitaev Majorana chain is coupled to the background gauge field $T$ of the fermionic parity $(-1)^F$, the partition function becomes $e^{i \pi \text{Arf}[T + \rho]}$.

The idea of fermionization is to couple the bosonic theory $\mathcal{B}$ with the Kitaev Majorana chain. Then the corresponding partition function has a form of
\begin{align}
Z_{\mathcal{B}}[S] e^{i \pi \text{Arf}[S + \rho]},
\end{align}
where $S$ is the background field for the non-anomalous $\mathbb{Z}_2$ symmetry of $\mathcal{B}$. The next step is gauging the diagonal $\mathbb{Z}_2$, which amounts to promoting the background field $S$ to a dynamical field $s$. Then one obtains the partition function of $\mathcal{F}$, 
\begin{align}\label{fermionization01}
    Z_{\cal F }\big[T + \rho \big] = \frac{1}{2^g}\sum_{s \in H^1(\Sigma_g, \mathbb{Z}_2)} 
    Z_{\cal B} \big[s\big] ~ \text{exp} \Big[i\pi \left( \text{Arf}[s+ \rho] + \text{Arf} [\rho]  +\int  s \cup T \right) \Big],
\end{align}
where the sum $s$ counts all distinct gauge fields for the $\mathbb{Z}_2$ and the product $\cup$ denotes the cup product on cohomology classes. The map \eqref{fermionization01} will be called as the generalized Jordan-Wigner transformation.

To bosonize a given fermionic theory, we gauge the fermion parity $(-1)^F$ so the resulting theory does not depend on the spin structure. After this procedure, the partition function takes the form of
\begin{align}
Z_{\cal B} = \frac{1}{2^g} \sum_{t \in H^1(\Sigma_g, \mathbb{Z}_2)} Z_{\cal F}[t + \rho],
\end{align}
where $t$ is a dynamical gauge field for the fermion parity. On the other hand, it is known that gauging the fermion parity yields a $\mathbb{Z}_2$ symmetry in the gauged theory. Thus the partition function $Z_{\cal B}$ should be associated with the background field $S$. A full expression of the partition function with the background field $S$ is given by
\begin{align}
Z_{\cal B}[S] = \frac{1}{2^g} \sum_{t \in H^1(\Sigma_g, \mathbb{Z}_2)} Z_{\cal F}[t + \rho] \ \text{exp} \left[ i \pi \bigg( \text{Arf}[S + \rho] +  \text{Arf}[\rho]  + \int t  \cup S \bigg)\right].
\end{align}
One can show that $Z_{\cal B}[S]$ is independent of the spin structure $\rho$, as desired.

Let us comment on the $\mathbb{Z}_2$ orbifold theory $\widetilde{\mathcal{B}} = \mathcal{B}/{\mathbb{Z}_2}$. We use $s$ for the dynamical gauge field corresponding to the non-anomalous $\mathbb{Z}_2$ symmetry of $\mathcal{B}$. We also use the known fact that the orbifold theory $\widetilde{\mathcal{B}}$ possesses a quantum $\mathbb{Z}_2^Q$ symmetry and denote $T$ as a background field for $\mathbb{Z}_2^Q$. Then, the partition function of $\widetilde{\mathcal{B}}$ reads
\begin{align}\label{orbifold}
    Z_{\tilde {\cal B} }\big[T\big] = \frac{1}{2^g}\sum_{s \in H^1(\Sigma_g, \mathbb{Z}_2)} Z_{\cal B} \big[s\big] ~ \text{exp} \Big[i\pi \int  s \cup T\Big].
\end{align}
The above relation is often called the Kramers-Wannier duality.

We also note that the new fermionic theory $\tilde{\cal F }$ can be obtained from the fermionic theory $\mathcal{F}$, by attaching $Z_{KM}$ to the partition function of
$\mathcal{F}$. More precisely, the partition function of $\tilde{\cal F }$ is given by
\begin{align}\label{arf transformation}
     Z_{\tilde{\cal F }}\big[ \rho \big] = Z_{\cal F }\big[\rho \big] Z_{KM}\big[\rho \big] = Z_{\cal F }\big[\rho \big] 
     \text{exp} \Big[i\pi \text{Arf}[ \rho]  \Big].
\end{align} 

Let us illustrate an explicit application of the generalized Jordan-Wigner transformation on $\Sigma_g = T^2$. We start with the bosonic theory $\mathcal{B}$ having a non-anomalous $\mathbb{Z}_2$ symmetry. Its Hilbert space can be decomposed into the untwisted sector($\mathcal{H}_u$) and twisted sector($\mathcal{H}_o$). Since we choose $\Sigma_g = T^2$, there are four distinct gauge fields for the discrete symmetry. Let us define the partition functions for four distinct configurations as follows, 
\begin{align}
\label{line defects}
\begin{split}
  Z_{(1,1)}(\tau,\bar \tau) & \equiv Z_{\cal B}(\tau,\bar \tau) = \text{Tr}_{{\cal H}_u}\Big[ q^{L_0 -c/24} {\bar q}^{\bar L_0 -c/24} \Big] ,
  \\ 
  Z_{(g,1)}(\tau,\bar \tau) & \equiv  \text{Tr}_{{\cal H}_u}\Big[ {\mathbb L}_g q^{L_0 -c/24} {\bar q}^{\bar L_0 -c/24} \Big],
  \\ 
  Z_{(1,g)}(\tau,\bar \tau) & \equiv  \text{Tr}_{{\cal H}_t}\Big[  q^{L_0 -c/24} {\bar q}^{\bar L_0 -c/24} \Big],
  \\
  Z_{(g,g)}(\tau,\bar \tau) & \equiv  \text{Tr}_{{\cal H}_t}\Big[ {\mathbb L}_g q^{L_0 -c/24} {\bar q}^{\bar L_0 -c/24} \Big].
\end{split}
\end{align} 
Here $g$ is a group element of the discrete group of our interest. $Z_{(g,h)}$ denote the partition function with a topological line defect $\mathbb{L}_g$ and $\mathbb{L}_{g'}$ inserted along the spatial direction and time direction, respectively. 

From \eqref{orbifold}, it is straightforward to check that the partition function of an orbifold theory $\widetilde{\mathcal{B}}$ can be expressed as a sum of the partition functions listed in \eqref{line defects}. Explicitly, we have
\begin{align}
Z_{\widetilde{\mathcal{B}}} = \frac{1}{2} \left( Z_{(1,1)} + Z_{(g,1)} + Z_{(1,g)} + Z_{(g,g)}\right).
\end{align}
After applying the Jordan-Wigner transformation to the partition function of $\mathcal{B}$, the transformation formula \eqref{fermionization01} suggests that the partition function of each spin structure can be expressed as,
\begin{align}
\begin{split}
Z_{{\mathcal{F}}}^{\text{NS}} &= \frac{1}{2} \left( Z_{(1,1)} + Z_{(g,1)} + Z_{(1,g)} - Z_{(g,g)}\right), \\
Z_{{\mathcal{F}}}^{\widetilde{\text{NS}}} &= \frac{1}{2} \left( Z_{(1,1)} + Z_{(g,1)} - Z_{(1,g)} + Z_{(g,g)}\right), \\
Z_{{\mathcal{F}}}^{\text{R}} &= \frac{1}{2} \left( Z_{(1,1)} - Z_{(g,1)} + Z_{(1,g)} + Z_{(g,g)}\right), \\
Z_{{\mathcal{F}}}^{\widetilde{\text{R}}} &= \frac{1}{2} \left( Z_{(1,1)} - Z_{(g,1)} - Z_{(1,g)} - Z_{(g,g)}\right).
\end{split}
\end{align}
Once we compute $Z_{(g,1)}$, the other partition functions $Z_{(1,g)}$ and $Z_{(g,g)}$ are followed by applying modular transformations to $Z_{(g,1)}$. Therefore, the main challenge here is to compute the $Z_{(g,1)}$ with insertion of a certain line defect $\mathbb{L}_g$.

In a rational CFT, a topological line defect $\mathbb{L}_g$
can be realized as a Verlinde line operator. 
The Verlinde line operators are in one-to-one correspondence with chiral primaries,
and preserve the left and right chiral algebra separately
We denote by ${\cal L}_{h_i}$ a Verlinde line associated with a primary $|\phi_i\rangle$ of conformal weight $h_i$.  
Its action on a primary state $\big| \phi_k \big\rangle$ is then given by
\begin{align}
\label{Velinde line}
{\cal L}_{h_i} \big| \phi_k \big\rangle = \frac{S_{i,k}}{S_{0,k}} \big| \phi_k \big\rangle,
\end{align}
where $S_{i,k}$ denotes the $(i,k)$ entry of the $S$-matrix. 
When the eigenvalues of ${\cal L}_{h}$ are either $1$ or $-1$, 
it can be regarded as a $\mathbb{Z}_2$ symmetry  generator. 

In the case of solvable models such as the WZW models, it is easy to compute $Z_{(g,1)}$ since the $S$-matrix is well-known.

\subsection{Rademacher Expansion}
Let us introduce an effective technical tool to compute the characters of a given bosonic RCFT. The main idea is originated from the work of the Rademacher \cite{10.2307/2371313}, who showed that the exact Fourier coefficients of any modular form of non-positive weight can be determined by the singular terms and modular covariance. Here we present the Fourier coefficients of the vector-valued modular form by exploiting Rademacher's idea. An application of the Rademacher expansion to the vector-valued modular form has been discussed in the literature, e.g., \cite{Cheng:2011ay, Bae:2020pvv}.

Let us consider a weight-zero vector-valued modular form $f_{\mu}(q)$, whose series expansion is given by
\begin{align}
f_{\mu}(q) = q^{p_\mu} \sum_{n \ge 0} F_{\mu}(n) q^n,
\end{align}
where $q = e^{2\pi i \tau} = e^{-\beta + 2\pi i \frac{r}{s}}$ and $p_\mu$ denote the leading exponent of $f_{\mu}(q)$\footnote{Here, our notation for an action of the modular group on $\tau$ is
$
\tau \rightarrow \frac{a\tau + b}{s\tau - r}, \quad a r + b s=-1,
$
where $a,b,s,r$ are integers and $(r,s)=1$.
}. The characters of bosonic RCFT  $f_{\mu}(q)$ transform under the $\sltz$ with a unitary $n$-dimensional representation $M(\gamma) : \sltz \rightarrow GL(n,\mathbb{C})$. More explicitly, 
\begin{align}
f_\mu(\gamma \tau) = M(\gamma)_{\mu \nu} f_{\nu}(\tau),
\end{align}
where $\gamma$ denotes a group element of $\sltz$. We also use the notation $q' = e^{2\pi i \tau'} = e^{-\frac{4\pi^2}{\beta s^2} + 2\pi i \frac{a}{s}}$.

The Fourier coefficient $F_{\mu}(n)$ can be read off from the contour integral. An explicit formula for the Fourier coefficient $F_{\mu}(n)$ is given by
\begin{align}
\label{eq:redemacherexp}
\begin{split}
F_{\mu}(n) &= \sum_{s,\nu} \sum_{m+p_\mu <0} M(\gamma)^{-1}_{\mu \nu} F_\nu(m) K\ell(s) \frac{2\pi}{s} \sqrt{\frac{|p_\mu + m|}{p_\mu + n}} I_1\left( \frac{4\pi\sqrt{|p_\mu + n|}\sqrt{p_\mu + m}}{s}\right).
\end{split}
\end{align}
In the above formula, $K\ell(s)$ denotes the Kloosterman sum, which is defined as
\begin{align}
K\ell(s) =\sum_{\substack{(r,s)=1 \\ 0 \le r < s}} e^{-2\pi i \frac{r}{s} (p_\mu +m)} e^{2\pi i \frac{a}{s}(n+p_\mu)},
\end{align}
and $I_{\alpha}(z)$ is the modified Bessel function of the first kind,
\begin{equation}
I_{\alpha}(z) = \sum_{n \geq 0} \frac{1}{n!\, \Gamma(n+ \alpha + 1)} \left(\frac{z}{2}\right)^{\alpha + 2n}\,.
\end{equation}
Clearly, the input data are the singular part of $f_\mu(q)$ satisfying $m+p_\mu <0$ and the representation $M(\gamma)$.

\section{Classification} \label{Classification}

The main goal of this section is to explore the solution space of the fermionic third-order MLDEs and  identify the solutions with the  characters of a certain fermionic RCFT. The general  fermionic 3rd order MLDE  for each spin structure is given as following:
\begin{align}
\begin{split}
\label{eq:order3MLDE}
      \bigg[\mathcal D^3+\mu_1 e_2 \mathcal D^2+\left(\mu_2 e_2^{\ 2}+\mu_3 E_4\right)\mathcal D+\mu_4 e_2 E_4 + \mu_5 e_2^3\bigg]f^{\textsc{ns}}_i(q)&=0, \\
      \bigg[\mathcal D^3+\mu_1 e_2' \mathcal D^2+\left(\mu_2 e_2'^{2}+\mu_3 E_4\right)\mathcal D+\mu_4 e_2' E_4 + \mu_5 e_2'^3\bigg]f^{\widetilde{\textsc{ns}}}_i(q)&=0, \\
      \bigg[\mathcal D^3+\mu_1 e_2'' \mathcal D^2+\left(\mu_2 e_2''^{2}+\mu_3 E_4\right)\mathcal D+\mu_4 e_2'' E_4 + \mu_5 e_2''^3\bigg]f^{\textsc{r}}_i(q)&=0, \\
\end{split}
\end{align}
where $e_2,e_2'$ and $ e_2''$ are weight two modular forms of $\Gamma_\theta,\Gamma^0(2),$ and $\Gamma_0(2)$, respectively and given below:
\begin{equation}
    e_2(\tau)=\vartheta_4(\tau)^4-\vartheta_2(\tau)^4, \  e_2'(\tau)=\vartheta_3(\tau)^4+\vartheta_2(\tau)^4,   e_2''(\tau)=-\vartheta_3(\tau)^4-\vartheta_4(\tau)^4.
\end{equation} 

The solutions of the above fermionic MLDE can be expressed in q-series. For example, the  characters on NS sector of fermionic CFT with central charge $c$ and conformal weights $h_1^\textsc{ns},h_2^\textsc{ns}$ of NS-sector primaries other than the vacuum would have the following expansion;
\begin{align}
\begin{split}
    \label{series1}
    f_0^\textsc{ns}&=q^{-\frac{c}{24}} (1+a_1q^\frac12+a_2 q+ a_3q^\frac32 +a_4q^2+\cdots), \\
    f_1^\textsc{ns}&=q^{-\frac{c}{24}+h^\textsc{ns}_1} (1+b_1q^\frac12+b_2 q+ b_3q^\frac32+b_4q^2+\cdots), \\
    f_2^\textsc{ns}&=q^{-\frac{c}{24}+h^\textsc{ns}_2} (1+c_1q^\frac12+c_2 q+ c_3q^\frac32+c_4q^2+\cdots). 
    \end{split}
\end{align}

We will focus on the two special cases. First, we find that the infinitely many solutions of \eqref{eq:order3MLDE} can be constructed from the characters of the three-character bosonic RCFTs. More precisely, we consider the tensor product between the three characters of bosonic RCFT and $\mathcal{N}$-copies of the Majorana-Weyl fermion. Based on the known classification of the bosonic RCFT with three-characters \cite{Gaberdiel:2016zke,Hampapura:2016mmz,Franc_2020}, we claim that there are  infinitely many solutions of \eqref{eq:order3MLDE}.

Second, we consider the special case of $\mu_5 = -\frac{1}{4} \mu_4$. We refer to this case as the BPS equation since there is an R-sector primary saturating the unitarity bound $h^{R} = \frac{c}{24}$. In addition, we require that there is no free fermion contribition and  so the coefficient $a_1$ of $q^\frac12$ in the vacuum character \eqref{series1} of the NS sector vanishes. We provide a full classification of this  "BPS fermionic MLDE" and identify them with the WZW models via the generalized Jordan-Wigner transformation.

\subsection{Fermionic solutions from the Bosonic MLDE}

\paragraph{Bosonic third-order MLDE}

Let us make a few comments on the classification of bosonic RCFT with three characters. It has been known that there are finitely many bosonic RCFTs with $\ell = 0$ having no Kac-Moody algebra \cite{Tuite:2008pt,Hampapura:2016mmz}. More recently, the classification has been extended to the case where the theory involves Kac-Moody currents \cite{Gaberdiel:2016zke,Franc_2020}. In addition to the solutions listed in \cite{Gaberdiel:2016zke,Franc_2020}, we further find a few more unitary solutions that have positive integer coefficients in the $q$-expansion.


Let us start with the third-order MLDE with $\ell=0$ whose general form is given by
\begin{align}
\label{3rd MLDE}
\begin{split}
\left[\mathcal{D}^3 + \mu_1 E_4 \mathcal{D} + \mu_2 E_6 \right] f_i(q) = 0.
\end{split}
\end{align}
By introducing a parameter $K= \frac{12^3}{j}$ and $\theta_K = K \frac{d}{dK}$, the third-order differential equation \eqref{3rd MLDE} can be recast as \cite{franc2015hypergeometric, Franc_2020},
\begin{align}
\label{eqtninK}
\begin{split}
\bigg[ (1-K) \theta_K^3 - \left(K+\frac{1}{2}\right) \theta_K^2 + \left(\mu_1 + \frac{1-4K}{18}\right) \theta_K + \mu_2 \bigg] f_i(q) = 0,
\end{split}
\end{align}
with help of the identity $q \frac{dK}{dq} = \frac{E_6}{E_4} K$. Utilizing the initial equation of \eqref{3rd MLDE}, one can express the coefficients $\mu_1$ and $\mu_2$ by $c,h_1,h_2$ as follows,\footnote{With help of the valence formula, the conformal weight $h_2$ can be fixed in terms of $c$ and $h_1$ by $h_2 = \frac{c+4-8 h_1 }{8}$.}
\begin{align}
\label{mu1mu2}
\begin{split}
\mu_1 &= \frac{1}{576} \left(3 c^2-48 c h_1-48 c h_2+576 h_1 h_2-32\right), \\
\mu_2 &= \frac{c^3}{13824}-\frac{c^2 h_1}{576}-\frac{c^2 h_2}{576}+\frac{c h_1 h_2}{24}.
\end{split}
\end{align}
After plugging \eqref{mu1mu2} in \eqref{eqtninK}, the differential equation becomes
\begin{align}
\label{hypergeometric eqtn}
\begin{split}
\left( \theta_K + \frac{c}{24}\right) \left( \theta_K + \frac{c}{24} - h_1\right) \left( \theta_K + \frac{c}{24} - h_2\right) f_i(q) - K \theta_K \left(\theta_K + \frac{1}{3} \right) \left(\theta_K + \frac{2}{3} \right) f_i(q)= 0.
\end{split}
\end{align}
which takes the form of a hypergeometric equation given by \cite{Beukers1989}. As far as the difference between any two of conformal weights is not integer, three independent solutions of \eqref{hypergeometric eqtn} are given by 
\begin{align}
\label{vacuum character of 3rd}
\begin{split}
f_0(q) &= K^{-\frac{c}{24}} {_3F_2}\left(-\frac{c}{24}, \frac{8-c}{24}, \frac{16-c}{24};1-h_1,1-h_2;K\right) \\
       &= q^{-\frac{c}{24}}\bigg( 1 + \left(31 c-\frac{(c-16) (c-8) c}{8 (h_1-1) (h_2-1)}\right) q + \cdots \bigg), \\
f_1(q) &= N_1 K^{h_1-\frac{c}{24}} {_3F_2}\left(h_1-\frac{c}{24}, h_1 + \frac{8-c}{24}, h_1 + \frac{16-c}{24};1+h_1,1+h_1-h_2;K\right), \\
f_2(q) &= N_2 K^{h_2-\frac{c}{24}} {_3F_2}\left(h_2-\frac{c}{24}, h_2 + \frac{8-c}{24}, h_2 + \frac{16-c}{24};1+h_2,1+h_2-h_1;K\right). 
\end{split}
\end{align}
The overall normalization is tuned to provide integer coefficients in $q$-expansion.

Now the classification can be done by exploiting the fact that any characters of unitary RCFT should possess non-negative integer coefficients in $q$-expansion\footnote{With appropriate choice of the normalization, two characters of primaries also should exhibit non-negative integer coefficients in $q$-expansion.}.
Let us note that the coefficient of the linear term of the vacuum character, which will be denoted as $a_1$, can be expressed in terms of $c$ and $h_1$. The idea is to consider rational variable $c = m_1/m_2$ and non-negative integer $a_1$ as two free parameters. By running the non-negative integers $m_1, m_2$ and $a_1$ in the following range,
\begin{align}
\label{parameter range}
0 < a_1 < 1500, \quad 0 < m_1 < 850, \quad 0 < m_2 < 200,
\end{align}
we search for the case of the Fourier coefficients of the characters exhibit non-negative integers. The list of solutions are presented in table \ref{bosonic third order classification} of appendix \ref{app:bosonic_3MLDE_appendix}. 

In addition to the solutions reported in \cite{Hampapura:2016mmz,Gaberdiel:2016zke, Franc_2020}, we further find 11 unitary solutions of \eqref{3rd MLDE} that admit non-negative integer coefficients in $q$-expansion. The new solutions are can be found in table \ref{new bosonic solutions}. We remark that some of the new solutions can be identified with characters of the WZW models. For instance, the solution number 84 of table \ref{new bosonic solutions} corresponds to characters of the WZW model for $(\text{SU}(3)_1)^2$. One can show that the solutions number 84 and 85 satisfy a bilinear relation of the form
\begin{align}
f_0^{(84)}(\tau) f_0^{(85)}(\tau) + 8748 f_{h=\frac{2}{3}}^{(84)}(\tau) f_{h=\frac{4}{3}}^{(85)}(\tau) + 26244 f_{h=\frac{1}{3}}^{(84)}(\tau) f_{h=\frac{5}{3}}^{(85)}(\tau) = j(\tau) + 96,
\end{align}
and use it to guess an identification of the solution number 85. The right-hand side of the above bilinear relation can be interpreted as the partition function of $c=24$ self-dual CFT \cite{Schellekens:1992db}. More precisely, three self-dual theories number 24, 26, 27 of Schelleken's list that are associated with the algebra $A_{2,1}^{12}$, $A_{5,2}^2 C_{2,1} A_{2,1}^{2}$, $A_{8,3}^2 C_{2,1} A_{2,1}^{2}$ can admit the WZW model for $(SU(3)_1)^2$ in the bilinear relation. We propose that the seventh solution has a relation with the algebra $A_{2,1}^{10}$. As a consistent check, we find that the vacuum solution $f_0^{(85)}(\tau)$ is decomposed into the characters of $A_{2,1}$ as follows,
\begin{align}
f_0^{(85)}(\tau) = \left(f_0^{\mathfrak{a}_{2,1}}(\tau)\right)^{10} + 60 \left(f_0^{\mathfrak{a}_{2,1}}(\tau)\right)^{4} \left(f_{h=\frac{1}{3}}^{\mathfrak{a}_{2,1}}(\tau)\right)^{6} + 20 \left(f_0^{\mathfrak{a}_{2,1}}(\tau)\right)^{1} \left(f_{h=\frac{1}{3}}^{\mathfrak{a}_{2,1}}(\tau)\right)^{9}.
\end{align}
It would be interesting to check if the $\mathbb{Z}_{N}$ orbifold applied to the $(SU(3)_1)^{10}$ WZW model can leads to the three-character fermionic RCFT with $f_0^{(85)}(\tau)$, $f_{h=\frac{4}{3}}^{(85)}(\tau)$ and $f_{h=\frac{1}{3}}^{(85)}(\tau)$ via the generalized Jordan-Wigner transformation. 

We next focus on the solution number 87 with $(c,h_1,h_2) = (12,2/3,4/3)$. It is noteworthy that the solution number 87 forms a self-dual relation, namely
\begin{align}
\label{bilinear for IV}
\begin{split}
f_0^{(87)}(\tau) f_0^{(87)}(\tau) + 157464 f_{h=\frac{2}{3}}^{(87)}(\tau) f_{h=\frac{4}{3}}^{(87)}(\tau) = j(\tau) + 312.
\end{split}
\end{align}
We interpret $j(\tau) + 312$ as the partition function of self-dual theory number 58 of the Schelleken's list. Since an algebra of it is given by $(E_{6,1})^4$, the solution $f_i^{(87)}(\tau)$ is identified to characters of the WZW model for $(E_{6,1})^2$.

\paragraph{Fermionic solutions} Our next target is to analyze the solutions of \eqref{eq:order3MLDE} that are constructed from the characters of the above bosonic RCFTs. To this end, we take the tensor product of  $N$ copies of the Majorana-Weyl fermions to the three characters of bosonic RCFT. The characters of the individual spin structures take the form of 
\begin{equation}
\label{tensored character}
f_i^{\rm NS} = f_i^{\rm B}  (\psi^{\rm NS})^N,\quad f^{\widetilde{\rm NS}}_i = f_i^{\rm B}  (\psi^{\widetilde{\rm NS}})^N, \quad f_i^{\rm R} = f_i^{\rm B}  (\psi^{\rm R})^N\,.
\end{equation}
where the characters of the Majorana-Weyl fermions are given by
\begin{align}\label{aaa}
    \psi^\text{NS} = \sqrt{\frac{\vartheta_3}{\eta}}, \quad
    \psi^{\widetilde{\text{NS}}} = \sqrt{\frac{\vartheta_4}{\eta}}, \quad 
    \psi^\text{R} = \sqrt{\frac{\vartheta_2}{\eta}} .
\end{align}
As an illustrative example, let us consider the characters of the babyMonster CFT with $c=47/2$. The three bosonic characters are known to have the following $q$-expansion \cite{Hampapura:2016mmz}\footnote{This theory itself can be regarded a bosonization of $c=47/2$ fermionic single character CFT with supersymmetry~\cite{Lin:2019hks} and explicit expression of characters given in~\cite{Bae:2020xzl}.},
\begin{align}
\begin{split}
f_0^B(q) &= q^{-47/48} + 96256 q^{49/48} + 9646891 q^{97/48} + 
 366845011 q^{145/48} + \cdots, \\
f_1^B(q) &= 4371 q^{25/48} + 1143745 q^{73/48} + 64680601 q^{121/48} + 
 1829005611 q^{169/48} + \cdots , \\
f_2^B(q) &= 96256 q^{23/24} + 10602496 q^{47/24}+ 420831232 q^{71/24}  + \cdots, 
\end{split}
\end{align}
and tensoring them with the one Majorana-Weyl fermion yields the following characters,
\begin{align}
\label{BMtensor}
\begin{split}
f_0^{NS}(q) &= q^{-1} + q^{-1/2} + q^{1/2} +96257 q + 96257 q^{3/2}+9646892 q^2 + \cdots, \\
f_1^{NS}(q) &= 4371 q^{1/2} +4371 q + 1143745 q^{3/2}+1148116 q^2+64684972 q^{5/2}+ \cdots , \\
f_2^{NS}(q) &= 96256 q^{15/16}+96256 q^{23/16}+10602496 q^{31/16}+10698752 q^{39/16}  + \cdots.
\end{split}
\end{align}
It is easy to see that the characters \eqref{BMtensor} solve the fermionic MLDE \eqref{eq:order3MLDE} with 
\begin{align}
\mu_1 = \frac{1}{16}, \quad \mu_2 = \frac{3}{256}, \quad \mu_3 = -\frac{2363}{2304}, \quad \mu_4 = \frac{365}{512}, \quad \mu_5 = -\frac{125}{512}. 
\end{align}
We further remark that tensoring the bosonic theory with the arbitrary number $\mathcal{N}$ of the Majorana-Weyl fermion provides the solutions of \eqref{eq:order3MLDE}. To see this point, let us revisit the valence formula. From \eqref{tensored character}, one can see that the central charge $c_F$ and conformal weights of the tensored theory are given by
\begin{align}
c_F = c + \frac{\mathcal{N}}{2}, \quad h_i^{NS} = h_i^{B}, \quad h_i^{\widetilde{NS}} = h_i^{B} , \quad h_i^{R} = h_i^{B} + \frac{\mathcal{N}}{16},
\end{align}
and therefore the valence formula reads
\begin{align}
\label{Valence tensor product}
\frac{3}{2} + \frac{3c_F}{8} - 2\sum_j h_j^{\rm NS} - \sum_j h_j^{\rm R} = -\frac{1}{3} \bigg(\sum_{i} \left( h^B_i - \frac{c}{24}\right) - \frac{1}{2} \bigg) = 0.
\end{align}
In the last equation, we use the Valence formula of the bosonic RCFT with three characters and $\ell = 0$. Equation \eqref{Valence tensor product} shows that the tensored theory with an arbitrary number of Majorana-Weyl fermions arose as the solutions of a holomorphic fermionic MLDE with $\tilde{\ell}=0$.

\subsection{The third-order MLDE of BPS Type}

For this type of solutions with $\mu_5=-\frac14 \mu_4$,  \eqref{eq:order3MLDE} becomes
\begin{align}
\begin{split}
\label{eq:BPS_3order}
	&\left[\mathcal D^3+\mu_1 e_2 \mathcal D^2+\left(\mu_2  e_2^{2}+\mu_3 E_4\right)\mathcal D+\frac{3\mu_4}{4}  e_2\vartheta_3^8\right]f^\textsc{ns}_i(\tau)=0, \\
	&\left[\mathcal D^3+\mu_1 e_2' \mathcal D^2+\left(\mu_2  e_2'^{2}+\mu_3 E_4\right)\mathcal D+\frac{3\mu_4}{4}  e_2'\vartheta_4^8\right]f^{\widetilde{\textsc{ns}}}_i(\tau)=0, \\
	&\left[\mathcal D^3+\mu_1 e_2'' \mathcal D^2+\left(\mu_2  e_2''^{2}+\mu_3 E_4\right)\mathcal D+\frac{3\mu_4}{4}  e_2''\vartheta_2^8\right]f^\textsc{r}_i(\tau)=0.
\end{split}
\end{align}
The differential equations for $\widetilde{\text{NS}}$ and R-sector are obtained by taking $T$ and $TS$ transformation to the NS-sector MLDE. In other words, the solutions of both $\widetilde{\text{NS}}$ and R-sectors can be obtained from the NS-sector solutions. For this reason, we mainly focus on the NS-sector solutions with series expansion \eqref{series1}.

By using the MLDE in  the NS-sector,  we can fix the coefficients $\mu_i$ of \eqref{eq:BPS_3order}  in terms of $c$, $h_1^\textsc{ns}$ and $h_2^\textsc{ns}$  and the condition $a_1=0$ of the NS vacuum character  \eqref{series1} as follows;
\begin{equation}
\begin{split}
\mu_1&=\frac{1}{8} (c-8 (h_1^{\textsc{ns}}+ h_2^{\textsc{ns}} ) +4), \\
\mu_2&=\frac{5 c^2-12 c (4  h_1^{\textsc{ns}} +4  h_2^{\textsc{ns}}  -3)+72 (2  h_1^{\textsc{ns}} -1) (2  h_2^{\textsc{ns}}  -1)}{1728},\\
\mu_3&=\frac{c^2-24 c ( h_1^{\textsc{ns}} + h_2^{\textsc{ns}}  )+ 360h_1^\textsc{ns}h_2^\textsc{ns} -36 (   h_1^{\textsc{ns}}  +  h_2^{\textsc{ns}} ) -6}{432},\\
  \mu_4&=\frac{c (c-24  h_1^{\textsc{ns}} ) (c-24  h_2^{\textsc{ns}}  )}{10368}.
\end{split}
\end{equation}
Since the differential equation \eqref{eq:BPS_3order} is invariant under 
the action of the the congruence subgroup $\Gamma_\vartheta$, the NS-sector conformal characters $f_i^{\text{NS}}(\tau)$ form a vector-valued modular form under $\Gamma_\vartheta$.

Although the most generic third-order MLDE \eqref{eq:order3MLDE} for $\Gamma_\vartheta$ with $\tilde{\ell}=0$
needs two parameters $\mu_4,\mu_5$ for the weight-six coefficient function, we require the characters of the R-sector primaries  have conformal weights satisfy the unitarity bound. To see this, we note that the saturation of the unitarity bound requires the MLDE for the R-sector, which is associated with the principal Hecke subgroup of level two, $\Gamma_0(2)$, 
to have weight-six coefficient functions vanishing at the cusp at infinity which happens when $\mu_5=-\frac14 \mu_4$.  It is easy to show that the weight-six coefficient function of \eqref{eq:BPS_3order} maps to the $\Gamma_0(2)$ modular form that vanishes at $\tau = i \infty$. More concretely, we have
\begin{align}
    T S : 	 \left(e_2E_4-\frac{e_2^{\ 3}}{4} \right)=\frac{3}{4}\,  e_2 \vartheta_3^8  \longrightarrow -\frac{3}{4}  \left(\vartheta_3^4+\vartheta_4^4\right)\vartheta_2^8 = -384 q + {\cal O}(q^2),
\end{align}
and actually the image under $TS$ transformation is the only weight-six modular form  of $\Gamma_0(2)$ that vanishes at $\tau = i \infty$. 
As $h_0^\textsc{r}=\frac{c}{24}$ and $h^\textsc{r}_{1,2}>\frac{c}{24}$, we refer to \eqref{eq:BPS_3order} as the BPS third-order MLDE, in what follows.

With help of the R-sector MLDE \eqref{eq:BPS_3order}, we can find the conformal weights  $h^\textsc{r}_{1,2}$ of two other primaries in the R-sector in terms of $c,h^\textsc{ns}_{1,2}$. Explicitly, we find that $h^\textsc{R}_i$ have the form of
\begin{align}
\begin{split}
\label{eq:3rdOrderRamondWeights}
   &\ h^{\textsc{r}}_0 =\frac{c}{24}\,,\\
   &\  h^\textsc{r}_{1,2} =\frac{2c+9}{12}-(h^\textsc{ns}_1+h^\textsc{ns}_2) \\
    &\  \  \pm \frac{1}{24}\sqrt{c^2+36(c+3)-48(c+9)(h^\textsc{ns}_1+h^\textsc{ns}_2) +288(2(h^\textsc{ns}_1)^2+2(h^\textsc{ns}_2)^2+h^\textsc{ns}_1h^\textsc{ns}_2) }\, , 
\end{split}
\end{align}
and of course they are consistent with the valence formula. The rationality of $h^\textsc{r}_{1,2}$ restricts the allowed values of  $c, h^\text{ns}_{1,2}$. The condition $h^\textsc{r}_{1,2}>c/24$ constrains the possible $c,h^\textsc{ns}_{1,2}$ further. 

The main goal of this section is to classify the solutions of the BPS third-order MLDE. To this end, let us first find the closed-form expression for the solutions of \eqref{eq:BPS_3order}. We start with recasting the BPS third-order MLDE \eqref{eq:BPS_3order} in terms of the modular $\lambda(\tau)$ function as follows.
\begin{equation}\label{thirdMDE by lambda}
    \left[\frac{\text{d}^3}{\text{d}\lambda^3}+A_\alpha(\lambda)\frac{\text{d}^2}{\text{d}\lambda^2}+B_\alpha(\lambda)\frac{\text{d}}{\text{d}\lambda}+C_\alpha(\lambda)\right]f^\alpha_i(\lambda)=0.
\end{equation}
Here, the superscript $\alpha$ denote ths spin structure, namely $\alpha\in\{\text{NS},\widetilde{\text{NS}},\text{R}\}$. For $\alpha = \text{NS}$, we choose the coefficients as
\begin{align}
\begin{split}
\label{coeffNS}
    A_\textsc{ns}(\lambda)&=\frac{2(\mu_1+1) (1-2 \lambda) }{\lambda (1-\lambda)}\,,\\
    B_\textsc{ns}(\lambda)&=\frac{12 \mu_1+36 \mu_2+36 \mu_3+2-4(12 \mu_1+36 \mu_2+9 \mu_3+5)  \lambda(1-\lambda)  }{9\lambda ^2 (1-\lambda)^2 }\,,\\
    C_\textsc{ns}(\lambda)&=\frac{6 \mu_4(1-2\lambda) }{\lambda ^3(1-\lambda)^3 }\,,
\end{split}
\end{align}
to identify \eqref{thirdMDE by lambda} and NS-sector differential equation of \eqref{eq:BPS_3order}. For the other spin structures, the coefficients are given by
\begin{align}
\label{CoeffNSt}
    A_{\widetilde{\textsc{ns}}}(\lambda)&=\frac{2 (\lambda  (\mu_1-2)+\mu_1+1)}{ \lambda(1-\lambda) }, \qquad 
    C_{\widetilde{\textsc{ns}}}(\lambda)=-\frac{6  \mu_4(\lambda +1)}{ \lambda ^3(1-\lambda)}, \\
    B_{\widetilde{\textsc{ns}}}(\lambda)&=\frac{4 \lambda  (\lambda  (-6 \mu_1+9 \mu_2+9 \mu_3+5)-3 \mu_1+18 \mu_2-9 \mu_3-5)+12 \mu_1+36 \mu_2+36 \mu_3+2}{9  \lambda ^2(1-\lambda)^2}, \nonumber
\end{align}
and
\begin{align}
\label{CoeffR}
    A_\textsc{r}(\lambda)&=\frac{2 \lambda  \mu_1-4 \lambda -4 \mu_1+2}{\lambda(1-\lambda)}, \qquad C_\textsc{r}(\lambda)=-\frac{6  \mu_4(2-\lambda)}{(1-\lambda)^3 \lambda }, \\
    B_\textsc{r}(\lambda)&=\frac{4 \lambda  (\lambda  (-6 \mu_1+9 \mu_2+9 \mu_3+5)+15 \mu_1-36 \mu_2-9 \mu_3-5)-24 \mu_1+144 \mu_2+36 \mu_3+2}{9  \lambda ^2(1-\lambda)^2}. \nonumber
\end{align}
Note that the coefficients \eqref{CoeffNSt} and \eqref{CoeffR} are obtained from \eqref{coeffNS} by applying aforementioned $T$ and $TS$ transformations of the $\lambda$ variable 
\begin{equation}
\label{ST transformation}
    \lambda\xrightarrow{T}\frac{\lambda}{\lambda-1}\,,\ \ \ \ \ \ \ \  \lambda\xrightarrow{TS}1-\frac{1}{\lambda}\, .
\end{equation}
Therefore, it is sufficient to focus on the analytic solution of the NS-sector. 

For the technical convenience, let us write the characters as
\begin{equation}
	f^\textsc{ns}_i=\lambda^{-\frac{c}{12}}(1-\lambda)^{-\frac{c}{12}}\tilde{f}^\textsc{ns}_i,
\end{equation}
and introduce a new variable $z:=4\lambda(1-\lambda)$. In terms of the $z$ variable, one can rewrite \eqref{thirdMDE by lambda} as the following ordinary differential equation,
\begin{align}
\label{eq2:hypergeomEq}
	\left[\left(z \frac{d}{dz}\right)^3+\left(b_2+\frac{3}{2(z-1)}\right)\left(z \frac{d}{dz}\right)^2+\left(b_{11}+\frac{b_{12}}{z-1}\right)\left(z \frac{d}{dz}\right)+b_0\left(1+\frac{1}{z-1}\right)\right]\tilde{f}^\textsc{ns}_i=0\,,
\end{align}
with
\begin{equation}
\label{coeff of ODE}
	\begin{split}
		b_0&=-\frac{c(2h^\textsc{ns}_1-1)(2h^\textsc{ns}_2-1)}{32}\,, \qquad b_2=\frac{3-4(h^\textsc{ns}_1+h^\textsc{ns}_2)}{2},\\
		b_{11}&=\frac{54-18c-c^2-(108-24c)(h^\textsc{ns}_1+h^\textsc{ns}_2)+216h^\textsc{ns}_1h^\textsc{ns}_2}{144}\,,\\
		b_{12}&=\frac{54-18c-c^2-(108-24c)(h^\textsc{ns}_1+h^\textsc{ns}_2)-360h^\textsc{ns}_1h^\textsc{ns}_2}{144}\,,
	\end{split}
\end{equation}
The solution of \eqref{eq2:hypergeomEq} can be qualified as the hypergeometric function of order three, namely $_3F_2$ function. It turns out that the behavior of the ODE \eqref{eq2:hypergeomEq} depends on whether one of the NS-sector conformal weight is equal to $1/2$ or not. If none of the conformal weights of primaries equals $1/2$, the solutions of \eqref{eq2:hypergeomEq} have the form of
\begin{equation}
\label{three solutions}
\begin{split}
	&f^\textsc{ns}_i(\lambda)=2^{\frac{c}{3} - 12 h^\textsc{ns}_i}\lambda^{-\frac{c}{12}}(1-\lambda)^{-\frac{c}{12}}\left(4\lambda(1-\lambda)\right)^{1-\beta_i}\cdot \\
	&\left._3F_2\right.(1+\alpha_0-\beta_i,1+\alpha_1-\beta_i,1+\alpha_2-\beta_i;1+\beta_0-\beta_i,\hat{\dots},1+\beta_2-\beta_i;4\lambda(1-\lambda))\,,
\end{split}
\end{equation}
where the hat meaning omission of the $1 \equiv 1+\beta_i-\beta_i$. The coefficients of the analytic solutions \eqref{three solutions} are expressed in terms of the central charge and NS-sector conformal weights as follows.
\begin{equation}
\label{eq:HyperParametersGeneric}
\begin{split}
	&\alpha_0=-\frac{c}{12}\,, \quad  \alpha_1=\frac{c}{24}+\frac{3}{4}-h^\textsc{ns}_1-h^\textsc{ns}_2-\frac{h^\textsc{r}_1-h^\textsc{r}_2}{2}\,,  \quad \beta_0=1\, ,   \\
	&\alpha_2=\frac{c}{24}+\frac{3}{4}-h^\textsc{ns}_1-h^\textsc{ns}_2+\frac{h^\textsc{r}_1-h^\textsc{r}_2}{2}\,, \quad \beta_1=1-2h^\textsc{ns}_1\,, \quad \beta_2=1-2h^\textsc{ns}_2\,.
\end{split}
\end{equation}
The $S$-matrix of the characters can be obtained with help of the monodromy of the hypergeometric function. We find that an element of $S$-matrix is given by
\begin{equation}
\label{Smatrix three character}
\begin{split}
	    &S_{mn}=2^{12(h^\textsc{ns}_n-h^\textsc{ns}_m)}\times\\
	    &\times\left\{\delta_{mn}-2ie^{-i\pi\sum_{k}(\alpha_k-\beta_k)}\prod_{k}\frac{\Gamma(1+\alpha_k-\beta_n)\Gamma(1+\beta_k-\beta_m)}{\Gamma(1+\beta_k-\beta_n)\Gamma(1+\alpha_k-\beta_m)}\frac{\prod_{k}\sin\left(\pi(\alpha_k-\beta_n)\right)}{\prod_{\substack{k(\neq n)}}\sin\left(\pi(\beta_k-\beta_n)\right)}\right\}\,.
\end{split}
\end{equation}
where detailed derivation is presented in the appendix \ref{app:smatrix}.

Suppose the NS-sector involves the primary of weight $h^{NS}=1/2$. In that case, the characters are expressed as the regular hypergeometric function,
\begin{equation}
\label{three sol with hhalf}
	\begin{split}
    	f^\textsc{ns}_0&=2^\frac{c}{3}\lambda^{-\frac{c}{12}}(1-\lambda)^{-\frac{c}{12}}\left[1-\frac{9 (2 h-1) }{ 2(c-24 h+6)}\left[\, _2F_1\left(-\frac{c}{12},\frac{c-24 h+6}{12} ;1-2 h;4\lambda(1-\lambda)\right)-1\right]\right],\\
		f^\textsc{ns}_1&=2^{\frac{c}{3}-8h}\,\lambda^{-\frac{c}{12}+2h}(1-\lambda)^{-\frac{c}{12}+2h}\, _2{F}_1\left(\frac{c+6}{12},2 h-\frac{c}{12};2 h+1;4\lambda(1-\lambda)\right),\\
		f^\textsc{ns}_2&=\frac{9 (2 h-1)2^{\frac{c}{3}} }{4c (c-24 h+6)}\,\lambda^{-\frac{c}{12}}(1-\lambda)^{-\frac{c}{12}}\left[\, _2F_1\left(-\frac{c}{12},\frac{c-24 h+6}{12} ;1-2 h;4\lambda(1-\lambda)\right)-1\right],
	\end{split}
\end{equation}
and the $S$-matrix for the above characters is given by
\begin{equation}
\scalemath{0.8}{
S=
\begin{pmatrix}
 \frac{9 (2 h-1)  }{2 (c-24 h+6)}\left[\frac{\sin \left(\frac{1}{6} \pi  (c-12 h)\right)}{\sin (2 \pi  h)}+\frac{2 c-30 h+3}{9 (2 h-1)}\right] &  2^{8 h-2}\frac{ 3\Gamma (2-2 h) \Gamma (-2 h)}{\Gamma \left(-\frac{c}{6}\right) \Gamma \left(\frac{c}{6}-4 h+2\right)} & \frac{c (2 c-30 h+3) }{c-24 h+6}\left[\frac{\sin \left(\frac{1}{6} \pi  (c-12 h)\right)}{\sin (2 \pi  h)}+1\right] \\
 2^{-8 h}\frac{ \Gamma (2 h) \Gamma (2 h+1)}{\Gamma \left(\frac{c}{6}+1\right) \Gamma \left(4 h-\frac{c}{6}\right)} &  \frac{\sin \left(\frac{1}{6} \pi  (c-12 h)\right)}{\sin (2 \pi  h)} & 2^{-8 h+2}(2 c-30 h+3)\frac{  \Gamma (2 h-1) \Gamma (2 h+1)}{ 3\Gamma \left(\frac{c}{6}\right) \Gamma \left(4 h-\frac{c}{6}\right)} \\
 -\frac{9 (2 h-1) }{4 c (c-24 h+6)}\left[\frac{\sin \left(\frac{1}{6} \pi  (c-12 h)\right)}{\sin (2 \pi  h)}+1\right] & 2^{8 h-4}\frac{ \Gamma (2-2 h) \Gamma (-2 h)}{\Gamma \left(1-\frac{c}{6}\right) \Gamma \left(\frac{c}{6}-4 h+2\right)} & \frac{(-2 c+30 h-3) }{2 (c-24 h+6)}\left[\frac{ \sin \left(\frac{1}{6} \pi  (c-12 h)\right)}{\sin (2 \pi  h)}+\frac{18 h-9}{2 c-30 h+3}\right] \\
\end{pmatrix}
}.
\end{equation}

Having constructed the closed-form solutions \eqref{three solutions} and \eqref{three sol with hhalf}, we are now ready to make a classification on the BPS third-order MLDE. The strategy is simple : we find the rational parameters $c, h_1^{\text{NS}}$ and $h_2^{\text{NS}}$ that allow the characters \eqref{three solutions} and \eqref{three sol with hhalf} to have non-negative coefficients in $q$-expansion. To this end, the free parameters are set to be the rational numbers of non-negative integers $m_i, n_i, l_i$, 
\begin{align}
c=m_1/m_2, \quad h_1 = n_1/n_2, \quad h_2 = l_1/l_2,
\end{align}
and we search if there are integers $(m_i, n_i, l_i)$ in the range of 
\begin{align}
0 < m_1 \le 200, \quad 0 < m_2 \le 18, \quad 0< n_1, l_1 \le 25, \quad 0< n_2, l_2 \le 18
\end{align}
that allow characters to have non-negative integer coefficients in the $q$-expansion. The exhaustive list of solutions is given in table \ref{tab:thirdOrderClassification}, where we divide them into four classes for illustration purposes. We refer the reader to the appendix \ref{app:3MLDE_appendix} for the complete list of characters of the NS and R-sector.

\begin{table}[]\label{tab:3rd}
    \centering
     \arraycolsep=5pt \def\arraystretch{1.5}
\begin{tabular}{c|c}
Class & $(c, h_1^{\text{NS}}, h_2^{\text{NS}},h_0^{\text{R}},h_1^{\text{R}},h_2^{\text{R}})$ \\
\hline \hline
I & $\left(\frac{3m}{2}, \frac{1}{2}, \frac{m}{8}, \frac{m}{16}, \frac{1}{16} (2 m+4+|m-4|), \frac{1}{16} (2 m+4-|m-4|\right)$\\ \hline 
I$'$ & $\left(12, \frac{1}{2}, 1, \frac{1}{2}, \frac{3}{2}, 1\right)$, $\left(18,\frac{1}{2},\frac{3}{2},\frac{3}{4},\frac{9}{4},\frac{5}{4}\right)$, $\left( 24,\frac{1}{2},2,1,3,\frac{3}{2} \right)$\\ \hline 
\multirow{2}{*}{II} & $(2,\frac{1}{6},\frac{1}{3},\frac{1}{12},\frac{3}{4},\frac{5}{12})$ , $(\frac{18}{5},\frac{3}{10},\frac{2}{5},\frac{3}{20},\frac{3}{4},\frac{11}{20})$, $(\frac{9}{2},\frac{1}{4},\frac{1}{2},\frac{3}{16},\frac{15}{16},\frac{9}{16})$ , $(5,\frac{1}{3},\frac{1}{2},\frac{5}{24},\frac{7}{8},\frac{5}{8})$, \\
& $(7,\frac{1}{2},\frac{2}{3},\frac{7}{24},\frac{7}{8},\frac{5}{8})$ , $(\frac{15}{2},\frac{1}{2},\frac{3}{4},\frac{5}{16},\frac{15}{16},\frac{9}{16})$, $(\frac{42}{5},\frac{3}{5},\frac{7}{10},\frac{7}{20},\frac{19}{20},\frac{3}{4})$ , $(10,\frac{2}{3},\frac{5}{6},\frac{5}{12},\frac{13}{12},\frac{3}{4})$ \\ \hline
III & $(\frac{39}{2}, \frac{3}{4}, \frac{3}{2},\frac{13}{16},\frac{33}{16},\frac{23}{16})$, $(\frac{66}{5},\frac{4}{5},\frac{11}{10},\frac{11}{20},\frac{27}{20},\frac{3}{4})$, $(22,\frac{5}{6},\frac{5}{3},\frac{11}{12},\frac{9}{4},\frac{19}{12})$\\ \hline
\multirow{2}{*}{IV} & $(1,\frac{1}{6},\frac{1}{2},\frac{1}{24},\frac{3}{8},\frac{1}{8}), (\frac{3}{2},\frac{1}{2},\frac{1}{4},\frac{1}{16},\frac{5}{16},\frac{3}{16}), (3,\frac{1}{2},\frac{1}{3},\frac{1}{8},\frac{11}{24},\frac{3}{8})$,\\  &  $(9,\frac{1}{2}, \frac{2}{3},\frac{3}{8},\frac{9}{8},\frac{25}{24}), (\frac{21}{2},\frac{1}{2}, \frac{3}{4},\frac{7}{16},\frac{21}{16},\frac{19}{16}) ,(11,\frac{1}{2}, \frac{5}{6},\frac{11}{24},\frac{11}{8},\frac{9}{8})$
\end{tabular}    
    \caption{\label{tab:thirdOrderClassification} List of BPS solutions of the third-order MLDE. The solutions of the class I are identified with the characters of the fermionized $\left(\text{SO}(m)_1\right)^{3}$ WZW models for $m=2,3,\cdots, 16$. The class I$^\prime$ involves the one-parameter solutions. The solutions of  class II have relations with the various WZW models. No identification is known for the $(c=\frac{66}{5})$, while the solutions with $c=39/2$ and $c=22$ can be identified with help of $c=39/4$ and $c=11$ theory discussed in \cite{Bae:2020xzl,Bae:2021lvk}. The solutions in class IV cannot have a consistent fusion rule algebra.}
\end{table} 

We present below the solutions of classes I, II, and III with details. We will not make comments on class IV separately, since their $S$-matrix cannot provide a consistent fusion rule algebra.

\subsubsection{Identification of Class I solutions}
\label{sec:Identification I}

Having classified the solutions of the BPS third-order MLDE, our next goal is to verify the solutions listed in table \ref{tab:thirdOrderClassification}. Especially, it turns out that the class I solutions of table \ref{tab:thirdOrderClassification} are related to the $\left(\text{SO}(m)_1\right)^{3}$ WZW model, after the generalized Jordan-Wigner transformation is applied. 

\paragraph{$\left(\text{SO}(m)_1\right)^{3}$ theory}

Let us consider the level-one WZW models for ${\text{SO}(m)}$ which are the bosonic RCFTs with $c=\frac{m}{2}$. Regardless of $m$, the theories of interest involve the primaries of weight $0, \frac{1}{2}, \frac{m}{16}$. The degeneracy of weight $m/16$ primary is two for $m$ even and one for $m$ odd. The characters of the $\text{SO}(m)_1$ WZW models have the form of
\begin{align}
\label{spin(m) characters}
\begin{split}
    \chi^{(m)}_0 &= \frac12 \left( (\psi^{\rm{NS}})^m +(\psi^{\widetilde{\rm{NS}}})^m  \right)  =  \frac{1}{2} \left[ \left( \frac{16}{\lambda(1-\lambda)}\right)^{\frac{m}{24}} + \left(\frac{16(1-\lambda)^2}{\lambda} \right)^{\frac{m}{24}} \right], \\
   \chi^{(m)}_1 &=\frac12 \left( (\psi^{\rm{NS}})^m -(\psi^{\widetilde{\rm{NS}}})^m  \right) =  \frac{1}{2} \left[ \left( \frac{16}{\lambda(1-\lambda)}\right)^{\frac{m}{24}} - \left(\frac{16(1-\lambda)^2}{\lambda} \right)^{\frac{m}{24}} \right],  \\
   \chi^{(m)}_2 &=  \frac{1}{2} (\psi^{\rm{R}})^m = \frac{1}{2} \left( \frac{16\lambda^2}{1-\lambda}\right)^{\frac{m}{24}},
\end{split}
\end{align}
and their $S$-transformation rule is given by
\begin{equation}
\label{spin(m) Smatrix}
    \left( \begin{array}{c} \chi^{(m)}_0(-1/\tau)  \\ \chi^{(m)}_1(-1/\tau)  \\ \sqrt{2}\chi^{(m)}_2(-1/\tau)
    \end{array}\right) 
    = \frac12\left( \begin{array}{ccc} 1 & 1 &  \sqrt{2} \\ 1 & 1 & -\sqrt{2} \\
    \sqrt{2} & -  \sqrt{2}   & 0 
    \end{array}\right)     
    \left( \begin{array}{c} \chi^{(m)}_0(\tau)  \\ \chi^{(m)}_1(\tau)  \\ \sqrt{2}\chi^{(m)}_2(\tau)
    \end{array}\right) ,
\end{equation}
for $m$ odd and
\begin{equation}
\label{spin(m) Smatrix}
    \left( \begin{array}{c} \chi^{(m)}_0(-1/\tau)  \\ \chi^{(m)}_1(-1/\tau)  \\ \chi^{(m)}_2(-1/\tau) 
    \end{array}\right) 
    = \frac12\left( \begin{array}{ccc} 1 & 1 & 2  \\ 1 & 1 & -2  \\
    1 & -1 & 0  
    \end{array}\right)     
    \left( \begin{array}{c} \chi^{(m)}_0(\tau)  \\ \chi^{(m)}_1(\tau)  \\ \chi^{(m)}_2(\tau) 
    \end{array}\right) ,
\end{equation}
for $m$ even. From the above $S$-matrices, it is straightforward to read off the form of diagonal modular invariant partition functions,
\begin{align}
\begin{split}
\label{partition of so(m)}
    Z&=\big|\chi^{(m)}_0\big|^2+\big|\chi^{(m)}_1\big|^2+ \big|\sqrt{2}\chi^{(m)}_2\big|^2, 
\end{split}
\end{align}
for positive integer $m \ge 2$. When $m=1$, \eqref{partition of so(m)} reproduces partition function of the Ising model. We remark that the $\text{SO}(m)_1$ WZW models possess the Verlinde line $\mathcal{L}_{h=\frac{1}{2}}$ in common, which is associated with $\mathbb{Z}_2$ symmetry. Then, an application of the generalized Jordan-Wigner transformation yields a tensor product of $m$ copies of the Majorana-Weyl free fermions. This is the reason why characters of the $\text{SO}(m)_1$ WZW models \eqref{spin(m) Smatrix} can be decomposed by the holomorphic partition function of the Majorana-Weyl free fermions.

Now let us consider the triple product of the level-one WZW model for ${\text{SO}(m)}$. A diagonal partition function of triple product theory is readily followed from \eqref{partition of so(m)}. The triple product theory involves the Verlinde lines $\mathcal{L}_{h=1/2}$, $\mathcal{L}_{h=1}$ and $\mathcal{L}_{h=3/2}$ that are associated with the $\mathbb{Z}_2$ symmetry. The action of $\mathbb{Z}_2$ is easy to obtain from the $S$-matrix, see the formula \eqref{Velinde line}. 

Of particular interest here is an application of fermionization with the Verlinde line $\mathcal{L}_{h=3/2}$. After some computation, we find that the NS-sector partition function of the fermionized theory is given by
\begin{align}
Z^{\text{NS}} = \big|f_0^{\text{NS}}\big|^2 + 3 \big|f_1^{\text{NS}}\big|^2 + 3 \big|f_2^{\text{NS}}\big|^2\,,
\end{align}
where three functions $f_0^{\text{NS}}, f_1^{\text{NS}}, f_2^{\text{NS}}$ correspond to the characters of primaries with $h=0,\frac{1}{2},\frac{m}{8}$. In terms of the characters of $\text{SO}(m)_1$ WZW model, they are expressed as follows,
\begin{align}
\label{spin NS sector}
\begin{split}
f^{\text{NS}}_0 &= (\chi^{(m)}_0)^3 + (\chi^{(m)}_1)^3, \qquad f^{\text{NS}}_1 = \chi^{(m)}_0\chi^{(m)}_1(\chi^{(m)}_0+\chi^{(m)}_1), \\
f^{\text{NS}}_2 &= (\sqrt{2}\chi^{(m)}_2)^2(\chi^{(m)}_0+\chi^{(m)}_1),
\end{split}
\end{align}
for $m \ge 2$ and their $S$-transformation rule reads
\begin{equation}
\label{eq:Smatrixc32}
   \left(
   \begin{array}{c}
       f^{\text{NS}}_0(-1/\tau)   \\
       f^{\text{NS}}_1(-1/\tau)  \\
       f^{\text{NS}}_2(-1/\tau)  
    \end{array}
    \right)
     = \frac14\left( \begin{array}{ccc} 1 & 3 & 6 \\ 1 & 3 & -2 \\
    2 & -  2  & 0 
    \end{array}\right)    \left(
   \begin{array}{c}
       f^{\text{NS}}_0(\tau)   \\
       f^{\text{NS}}_1(\tau)  \\
       f^{\text{NS}}_2(\tau)  
    \end{array}
    \right).
\end{equation}

On the one hand, one can read off the partition functions of different spin structures using the generalized Jordan-Wigner transformation. The R-sector partition function is contributed by the following three characters,
\begin{align}
\label{spin R sector}
\begin{split}
f^{\text{R}}_0 &= (\chi^{(m)}_0)^2 \chi^{(m)}_2 + (\chi^{(m)}_1)^2 \chi^{(m)}_2, \quad f^{\text{R}}_1 =  \chi^{(m)}_0  \chi^{(m)}_1  \chi^{(m)}_2, \quad f^{\text{R}}_2 = (\chi^{(m)}_2)^3,
\end{split}
\end{align}
while the $\tilde{\text{R}}$-sector partition function is constant. Based on the fact that the characters \eqref{spin NS sector} and \eqref{spin R sector} agree with the class I solutions of the BPS second-order modular differential equations, we conclude that the class I solutions are identified with the fermionized $\left(\text{SO}(m)_1\right)^{3}$ theory.

Let us make side comments on the fusion rule algebra of the NS-sector. It turns out that the NS-sector of the fermionized $\left(\text{SO}(m)_1\right)^{3}$ theories has the consistent fusion rule algebra. To see this, we consider an extended $S$-matrix 
\begin{align}
\begin{split}
\widehat{S} = \frac{1}{4}
\left(
\begin{array}{ccccccc}
    1 & 1 & 1 & 1 & 2 & 2 & 2   \\
    1 & 1 & 1 & 1 & -2 & -2 & 2  \\
    1 & 1 & 1 & 1 & -2 & 2 & -2  \\
    1 & 1 & 1 & 1 & 2 & -2 & -2  \\
    2 & -2 & -2 & 2 & 0 & 0 & 0   \\
    2 & -2 & 2 & -2 & 0 & 0 & 0   \\
    2 & 2 & -2 & -2 & 0 & 0 & 0   
\end{array}
\right),
\end{split}
\end{align}
which acts on the vector-valued modular form
\begin{align}
(f_0^{\text{NS}}, f_1^{\text{NS}},  f_1^{\text{NS}},  f_1^{\text{NS}},  f_2^{\text{NS}},  f_2^{\text{NS}},  f_2^{\text{NS}}).
\end{align}
One can show that the above extended $S$-matrix provides a consistent fusion rule algebra.

As a second remark, we note that the fermionized $\left(\text{SO}(m)_1\right)^{3}$ WZW models with $c\le 24$ satisfies a special relations of the form
\begin{align}
\label{bilinear som13}
\begin{split}
   & f_0^{(m)}f_0^{(8-m)}+3f_1^{(m)}f_1^{(8-m)}+3 f_2^{(m)}f_2^{(8-m)}= K(\tau) ,   \\
  & f_0^{(m)}f_0^{(16-m)}+3f_1^{(m)}f_1^{(16-m)}+3 f_2^{(m)}f_2^{(16-m)}= K(\tau)^2-192.
   \end{split}
\end{align}  
The degenerated case with $m=8$ is the  $\left(\text{SO}(8)_1\right)^{3}$ theory of $c=12$. In this case, the bilinear relation \eqref{bilinear som13} simply reduced to $f_0^{(8)} + \frac{3}{2} f_2^{(8)} = K(\tau)$, which indicate that two characters of the fermionized $\left(\text{SO}(8)_1\right)^{3}$ WZW models are combined to produce $K(\tau)$. We observe that similar happens for $m=16$. More precisely, the linear combinations of two characters $f_0^{(16)}$ and $f_2^{(16)}$ yield $K(\tau)^2-192$. 

\paragraph{Comparison with simple current}

Let us now make some comments on the connection to the so-called simple current (see \cite{Blumenhagen:2009zz} for a comprehensive review). A simple current, by definition, means a primary field $J$ whose fusion with any other primaries $\phi_i$ takes a simple form:
\begin{equation}
[J] \times [\phi_i] = [\phi_{J(i)}]\,.
\end{equation}
On the right-hand side, there is only a single primary field denoted by $\phi_{J(i)}$. Therefore the fusion coefficient reads $N_{J,i}^{k} = \delta_{k, J(i)}$. For the purpose of this paper, we further assume that $[J]$ fusing with itself gives the identity. In other words, the fusion product with $[J]$ is a $\mathbb{Z}_2$ automorphism of all conformal primaries by virtue of the associativity of the OPE. $[J]$ organizes the primaries into orbits of length $l_i = 2/p$ where $p$ is 1 or 2,
\begin{equation}
\left(\phi_i,\, \phi_{J(i)},\, \cdots \phi_{J^{l_i-1}(i)}\right)\,.
\end{equation}
For simplicity, we abbreviate the fields $\phi_{J^{\alpha}(i)}$ in this orbit as $(\alpha, i)$. We can also define a monodromy charge associated with $\phi_i$ in terms of their weights,
\begin{equation}
Q(\phi_i) = h(J) + h(\phi_i) - h(\phi_{J(i)})\ \ \text{mod}\ 1 \,.
\end{equation}
It is not difficult to show that all the elements in a given orbit share the same monodromy charge, and as a corollary, the conformal weight of $J$ is either an integer or a half-integer. 

The existence of a simple current enables us to construct a non-diagonal partition function out of the diagonal one \cite{Blumenhagen:2009zz},
\begin{equation}\label{eq:simplecurrent}
Z = \sum_{\substack{\alpha,i\\Q(\phi_i) \in \mathbb{Z}}}\sum_{\beta = 0}^{1} \chi_{(\alpha+\beta,i)}(\tau) \overline{\chi}_{(\alpha,i)}(\overline{\tau}) = \sum_{\substack{ i (l_i = 2)\\Q(\phi_i) \in \mathbb{Z}}} |\chi_{(0,i)} + \chi_{(1,i)}|^2 + \sum_{\substack{ i (l_i = 1)\\Q(\phi_i) \in \mathbb{Z}}} 2|\chi_{(0,i)}|^2\,.
\end{equation}
It can be shown that the above partition function is always invariant under the modular $S$ transformation. Moreover, when $h(J) \in \mathbb{Z}$, since the monodromy charge $Q(\phi_i)$ is integral, the  difference between $h(\phi_i)$ and $h(\phi_{J(i)})$ is always an integer. So \eqref{eq:simplecurrent} is further invariant under $T$ transformation hence under the full $\text{SL}_2(\mathbb{Z})$. As an example, this corresponds to the orbifold construction using the $\mathbb{Z}_2$ symmetry generator with $h = 1$ in the $\left(\text{SO}(m)_1\right)^{3}$ theory.

On the other hand, if $h(J)$ is a half-integer which we may call a fermionic simple current, the difference between $h(\phi_i)$ and $h(\phi_{J(i)})$ is always half-integral. In this case \eqref{eq:simplecurrent} is only invariant under $T^2$ transformation, which altogether generates the symmetry group $\Gamma_\theta$. Namely, we actually construct the partition function in the NS sector. The other sectors can also be obtained after $S$ and $T$ transformations.

In section \ref{sec:fermionization}, we already introduce the Verlinde line operator $\mathcal{L}$ associated with a conformal primary. Using identities of the $S$-matrix, one can show that each simple current $[J]$ gives rise to an $\mathcal{L}_J$ that generates a $\mathbb{Z}_2$ symmetry of the underlying CFT.\footnote{We further conjecture that they are in one-to-one correspondence with each other.} As an example, the fermionization of the $\left(\text{SO}(m)_1\right)^{3}$ theory shown above can be partially understood from the simple current perspective.

\subsubsection{Class I$^{'}$ solutions: One-parameter family} Among the fermionized $\left(\text{SO}(m)_1\right)^{3}$ theory, We find that the one-parameter solutions arise for $c=12,18,24$ $(m=8, 12, 16)$. For these three cases, the NS sector involves a primary of $h=\frac{m}{8}$, which become half-integers and integers. Therefore, one can combine two independent solutions of  $h=\frac{1}{2}$ and $h=\frac{m}{8}$ to produce a vector-valued modular form of two components. 

Let us illustrate the above explicitly. For $c=12$, the analytic form of solution involves two free parameters $a_2$ and $b_1$, 
\begin{align}
\label{c=12 solution}
\begin{split}
f_0^{\rm{NS}}(\tau) &= (a_2-20) \frac{\lambda}{16(1-\lambda)} +  \frac{16}{\lambda} - 8\\
&=q^{-\frac{1}{2}} \left( 1 + a_2 q + (8a_2-160) q^{\frac{3}{2}} + (44a_2-942) q^2  + \cdots \right)\,, \\
f_1^{\rm{NS}}(\tau) &= 1 + b_1 \frac{\lambda}{16(1-\lambda)}\\
&= 1 + b_1 q^{1/2} + 8 b_1 q + 44 b_1 q^{3/2} + 192 b_1 q^2 + 718 b_1 q^{5/2} + \cdots\,, \\
f_2^{\rm{NS}}(\tau) &= \frac{\lambda}{16(1-\lambda)}\\
&= q^{1/2} + 8 q+ 44 q^{3/2} + 192 q^2 + 718 q^{5/2} + 2400 q^3 + 7352 q^{7/2} + \cdots\,,
\end{split}
\end{align}
where the $S$-transformation of the above characters are governed by following $S$-matrix,
\begin{align}
S = 
\left(
\begin{array}{ccc}
   \frac{1}{256}(a_2-20)  & \ \ \frac{69}{8} - \frac{a_2}{32} & \ \ -\frac{1}{256}(a_2 - 276)(a_2-8b_1+236) \\
    \frac{b_1}{256} & \ \ 1-\frac{b_1}{32} & \ \ \frac{1}{256} b_1(8b_1-a_2-236)\\
    \frac{1}{256} & \ \ -\frac{1}{32} & \ \ \frac{1}{256}(20-a_2+8b_1)
\end{array}
\right)\,.
\end{align}
To have a consistent fusion rule algebra and non-negative integer coefficients in $q$-series, two free parameters $a_2$ and $b_1$ ought to be restricted as follows,
\begin{align}
22 \le a_2 < 276, \quad 0<b_1<3\,.
\end{align}
When $a_1$ have the value of $36, 84, 132$ and $148$, the solutions \eqref{c=12 solution} are related to the WZW model for $(\text{SU}(2)_1)^{12}$, $(\text{SO}(8)_1)^3$, $(\text{SO}(12)_1)^2$ and $\text{SO}(16)_1 \times \text{SO}(8)_1$, by the fermionization \eqref{fermionization01}\footnote{We pause to remark that the vacuum character with $a_2=20$ reproduces the Mckay-Thompson series for class 4C. However, $a_2=20$ cannot provide the consistent fusion rule algebra. Furthermore, the vacuum character involves negative integer coefficients.}.

For $c=18$, the infinitely many solutions of the third-order MLDE can be formulated as follows,
\begin{align}
\label{bowwow}
\begin{split}
f_0^{\rm{NS}}(\tau) &= g_0(\tau) + (n-4096) g_1(\tau), \quad n \in \mathbbm{Z}_{\ge 0}, \\
f_1^{\rm{NS}}(\tau) &= g_1(\tau), \quad f_2^{\rm{NS}}(\tau) = g_2(\tau) + m  g_1(\tau),  \quad m \in \mathbbm{Z}_{\ge 0}\,.
\end{split}
\end{align}
Here the $q$-expansion of $g_0(\tau), g_1(\tau), g_2(\tau)$ are given by
\begin{align}
\begin{split}
g_0(\tau) &= q^{-\frac{3}{4}} \left( 1 + 198 q + 4800 q^{\frac32} + 51849 q^2 + 376704 q^{\frac52} + 2149302 q^3  + \cdots \right), \\
g_1(\tau) &= q^{\frac{3}{4}} \left( 1 + 12 q^{\frac{1}{2}} + 90 q + 520 q^{\frac{3}{2}} + 2523 q^2 + 10764 q^{\frac{5}{2}} + 
 41534 q^3 + \cdots \right), \\
g_2(\tau) &= q^{-\frac{1}{4}} \left( 1 + 12 q^{\frac{1}{2}} + 66 q + 232 q^{\frac{3}{2}} + 639 q^2 + 1596 q^{\frac{5}{2}} + 
 3774 q^3 + \cdots \right).
\end{split}
\end{align}
One can show that the $S$-matrix of the characters $f_i^{\rm{NS}}(\tau)$ is given by
\begin{align}
\left(
\begin{array}{c}
    f_0^{\rm{NS}}\left(-\frac{1}{\tau}\right)   \\
    f_1^{\rm{NS}}\left(-\frac{1}{\tau}\right)   \\
    f_2^{\rm{NS}}\left(-\frac{1}{\tau}\right)   
\end{array}
\right)
=
\left(
\begin{array}{ccc}
    \frac{n}{4096} & \ \ \frac{(n-4096)(12m-n-4096)}{4096} & \ \ 12-\frac{3n}{1024}\\
    \frac{1}{4096} & \ \ \frac{12m-n}{4096} & \ \ -\frac{3}{1024} \\
    \frac{m}{4096} & \ \ \frac{m(12m-n-4096)}{4096} & \ \ 1-\frac{3m}{1024}
\end{array}
\right)
\left(
\begin{array}{c}
    f_0^{\rm{NS}}\left(\tau\right)   \\
    f_1^{\rm{NS}}\left(\tau\right)   \\
    f_2^{\rm{NS}}\left(\tau\right)   
\end{array}
\right),
\end{align}
The parameters $m,n$ are constrained in order to have an extended $S$-matrix as follows:
\begin{align}
n < 4096, \quad 12m < n+4096.
\end{align}
Therefore among the infinitely many solutions of $c=18$, only a finite number of solutions with $n<4096$ and $m \le 682$ can potentially  have the consistent fusion rule algebra. For $n=1024$ and $m=24$, \eqref{bowwow} can be related to the NS-sector characters of the fermionized $SO(12)_1^{3}$ WZW model.

For $c=24$, the one-parameter solutions of the NS-sector third-order MLDE can be expressed in terms of the analytic solutions \eqref{three sol with hhalf}. The explicit expressions are given below.
\begin{align}
\begin{split}
g_0^{\rm{NS}}(\tau) &= f_0^{\text{NS}}(\tau) + (n-98580) f_1^{\text{NS}}(\tau), \quad n \ge 33044, \\
g_1^{\rm{NS}}(\tau) &= f_2^{\text{NS}}(\tau) + m f_1^{\text{NS}}(\tau),  \quad m \ge 0, \quad g_2^{\rm{NS}}(\tau) = f_1^{\text{NS}}(\tau).
\end{split}
\end{align}
Unless $n \ge 33044$, the vacuum solution $g_0^{\rm{NS}}(\tau)$ have the negative coefficients in higher order of $q$. For $n=49428$ and $m=1024$, the above solutions can be identified with the characters of the fermionized $SO(16)_1^{3}$ WZW model.

\subsubsection{Identification of Class II solutions}
\label{sec:Identification II}

In this subsection, we find the relations between class II solutions and the WZW models with help of the generalized Jordan-Wigner transformation. After all, we find that characters of the certain WZW models can be used to express the class II solutions. 

\paragraph{($\mathcal N=2$ minimal model)$^2$}

The NS-sector and R-sector characters of the first unitary $\mathcal N=2$ supersymmetric minimal model are known to solve the BPS  second-order MLDE \cite{Bae:2020xzl}. The NS-sector partition function of the $\mathcal N=2$ minimal model of our interest
\begin{align}
Z^{\text{NS}} = |f^{\text{NS}}_0(\tau)|^2 + 2 |f^{\text{NS}}_1(\tau)|^2,
\end{align}
consists of the two functions 
\begin{align}
    \label{characters of SM(6,4)}
    \begin{split}
    f^{\text{NS}}_0(\tau) &= q^{-\frac{1}{24}} \Big( 1 + q + 2 q^\frac{3}{2} + 2 q^2 + 2 q^\frac{5}{2} + \cdots \Big) , \\
    f^{\text{NS}}_1(\tau) &= q^{\frac{1}{6} - \frac{1}{24}} \Big( 1 + q^{\frac{1}{2}} + q + q^{\frac{3}{2}} + 2 q^2 + 3 q^{\frac{5}{2}} + \cdots \Big),
    \end{split}
\end{align}
where $f^{\text{NS}}_0(\tau)$ and $f^{\text{NS}}_1(\tau)$ correspond to the conformal characters of the vacuum and primary of $h=\frac{1}{6}$, respectively.

Let us take a tensor product of the above $\mathcal N=2$ supersymmetric minimal models. The central charge of product theory is two, and the NS-sector partition function involves three characters. Explicitly, the NS-sector partition function has a form of
\begin{align}
Z^{\text{NS}, \otimes 2} = |g^{\text{NS}}_0(\tau)|^2 + 4 |g^{\text{NS}}_1(\tau)|^2 + 4 |g^{\text{NS}}_2(\tau)|^2,
\end{align}
where individual characters $g_i(\tau)$ read
\begin{align}
\label{c=2 sol}
\begin{split}
g^{\text{NS}}_0(\tau) &= f^{\text{NS}}_0(\tau) f^{\text{NS}}_0(\tau) = q^{-\frac{1}{12}} + 2 q^{\frac{11}{12}} + 4 q^{\frac{17}{12}} + 5 q^{\frac{23}{12}} + 8 q^{\frac{29}{12}} + 
 14 q^{\frac{35}{12}} + \cdots , \\
g^{\text{NS}}_1(\tau) &= f^{\text{NS}}_0(\tau) f^{\text{NS}}_1(\tau) = q^{\frac{1}{12}} + q^{\frac{7}{12}} + 2 q^{\frac{13}{12}} + 4 q^{\frac{19}{12}} + 7 q^{\frac{25}{12}} + 
 10 q^{\frac{31}{12}} + \cdots ,\\
g^{\text{NS}}_2(\tau) &= f^{\text{NS}}_1(\tau) f^{\text{NS}}_1(\tau) = q^{\frac{1}{4}} + 2 q^{\frac{3}{4}} + 3 q^{\frac{5}{4}} + 4 q^{\frac{7}{4}} + 7 q^{\frac{9}{4}} + 12 q^{\frac{11}{4}} + \cdots.
\end{split}
\end{align}
The above three characters solve the NS-sector BPS third-order MLDE. It is easy to check that similar holds for the R-sector solutions. For this reason, we claim that the class II solution of $c=2$ can be identified to the tensor product of the first unitary $\mathcal N=2$ supersymmetric minimal model. 

\paragraph{Fermionization of (SU(2)$_3)^2$ WZW model}

Let us discuss the tensor product of two $\text{SU}(2)_3$ WZW models. This product theory is equivalent to the level-three WZW model for $\text{SO}(4)$. The central charge of product theory is  $c=18/5$ and it involves ten primaries of conformal weights
\begin{align}
h = \left\{ 0, \frac{3}{4},\frac{2}{5},\frac{3}{20},\frac{3}{2},\frac{23}{20},\frac{9}{10},\frac{4}{5},\frac{11}{20},\frac{3}{10} \right\}.
\end{align}
Since the center symmetry of SU$(2)$ is $\mathbb{Z}_2$, the global symmetry of product theory is given by $\mathbb{Z}_2 \times \mathbb{Z}_2$. We take the $\mathbb{Z}_2$ subgroup of it which is generated by the Verlinde line $\mathcal{L}_{h=3/2}$. An action of the Verlinde line $\mathcal{L}_{h=3/2}$ can be obtained from the $S$-matrix, as discussed in the previous section.

To analyze the partition function of fermionized theory, we apply the generalized Jordan-Wigner transformation using the Verlinde line $\mathcal{L}_{h=3/2}$. After some computation, we find that the NS-sector characters are given by
\begin{align}
\label{spin43 NS characters}
\begin{split}
    f^{\rm{NS}}_{0}(\tau) &= \chi^{\mathfrak{a}_{1,3}}_0(\tau) \chi^{\mathfrak{a}_{1,3}}_0(\tau) + \chi^{\mathfrak{a}_{1,3}}_{\frac{3}{4}}(\tau) \chi^{\mathfrak{a}_{1,3}}_{\frac{3}{4}}(\tau),
    \\ 
    f^{\rm{NS}}_{1}(\tau) & = \chi^{\mathfrak{a}_{1,3}}_{\frac{3}{20}}(\tau) \chi^{\mathfrak{a}_{1,3}}_{\frac{3}{20}}(\tau) + \chi^{\mathfrak{a}_{1,3}}_{\frac{2}{5}}(\tau) \chi^{\mathfrak{a}_{1,3}}_{\frac{2}{5}}(\tau),
    \\
    f^{\rm{NS}}_{2}(\tau) & = \chi^{\mathfrak{a}_{1,3}}_0(\tau) \chi^{\mathfrak{a}_{1,3}}_{\frac{2}{5}}(\tau) + \chi^{\mathfrak{a}_{1,3}}_{\frac{3}{4}}(\tau) \chi^{\mathfrak{a}_{1,3}}_{\frac{3}{20}}(\tau),
\end{split}
\end{align}
where $\chi^{\mathfrak{a}_{1,3}}_{h}(\tau)$ denote characters of the $\text{SU}(2)_3$ WZW model for the representation of weight $h$. We find that three characters \eqref{spin43 NS characters} agree with the BPS solution with $c=18/5$ in $q$-expansion, therefore we identify $c=18/5$ NS-sector solutions to the NS-sector characters of fermionized $\text{SU}(2)_3$ WZW model.

It is also straightforward to read off the R-sector characters with help of the generalized Jordan-Wigner transformation. We find
\begin{align}
\begin{split} 
    f^{\rm{R}}_{0}(\tau) & = \chi^{\mathfrak{a}_{1,3}}_{0}(\tau) \chi^{\mathfrak{a}_{1,3}}_{\frac{3}{20}}(\tau) + \chi^{\mathfrak{a}_{1,3}}_{\frac{3}{4}}(\tau)\chi^{\mathfrak{a}_{1,3}}_{\frac{2}{5}}(\tau),
    \\
    f^{\rm{R}}_{1}(\tau) & = 2\chi^{\mathfrak{a}_{1,3}}_{\frac{2}{5}}(\tau) \chi^{\mathfrak{a}_{1,3}}_{\frac{3}{20}}(\tau), \quad f^{\rm{R}}_{2}(\tau) = 2\chi^{\mathfrak{a}_{1,3}}_{0}(\tau) \chi^{\mathfrak{a}_{1,3}}_{\frac{3}{4}}(\tau),
\end{split}
\end{align}
and $q$-expansion of above characters agree with the R-sector solutions presented in table \ref{R-sector list}.

\paragraph{Fermionization of ($\text{SU}(2)_6)^2/{\mathbb{Z}_2}$}

Let us first consider the tensor product of two $\text{SU}(2)_6$ WZW models. It is RCFT with $c=9/2$ and exhibits non-anomalous $\mathbb{Z}_2$ symmetry that is generated by the Verlinde line $\mathcal{L}_{h=3}$. The representation of $h=3$ has the form of $[6;6] \otimes [6;6]$, which is presented in terms of the Dynkin labels of $\text{SU}(2)_6$.

As a first step, let us take an $\mathbb{Z}_2$ orbifold via the Verlinde line $\mathcal{L}_{h=3}$. The resulting orbifold partition function is contributed by 13 primaries and especially it involves the representation of conformal weight $h=3/2$. In terms of the Dynkin labels of $\text{SU}(2)_6$, the above representation with $h=3/2$ can be written as $[0;6] \otimes [6;0] + [6;0] \otimes [0;6]$. The next step is to apply the generalized Jordan-Wigner transformation to find the partition functions of each spin structure. After some computation, we find that the NS-sector characters are given by
\begin{align}
\label{su2 6 NS sector}
\begin{split}
f_0^{\text{NS}}(\tau)&= \left(\chi_0^{\mathfrak{a}_{1,6}}(\tau) + \chi_{\frac{3}{2}}^{\mathfrak{a}_{1,6}}(\tau)\right)^2, \\
f_1^{\text{NS}}(\tau)&= \left(\chi_0^{\mathfrak{a}_{1,6}}(\tau) + \chi_{\frac{3}{2}}^{\mathfrak{a}_{1,6}}(\tau)\right) \left(\chi_{\frac{1}{4}}^{\mathfrak{a}_{1,6}}(\tau) + \chi_{\frac{3}{4}}^{\mathfrak{a}_{1,6}}(\tau)\right), \\
f_2^{\text{NS}}(\tau)&= \left(\chi_{\frac{1}{4}}^{\mathfrak{a}_{1,6}}(\tau) + \chi_{\frac{3}{4}}^{\mathfrak{a}_{1,6}}(\tau)\right)^2,
\end{split}
\end{align}
where $\chi^{\mathfrak{a}_{1,6}}_{h}(\tau)$ denote characters of the $\text{SU}(2)_6$ WZW model for the representation of weight $h$. The characters \eqref{su2 6 NS sector} agree with the NS-sector solution of the BPS third-order fermionic MLDE with $c=9/2$, therefore we conclude that those solutions are related to the $(\text{SU}(2)_6)^2$ theory via fermionization. 

The Ramond sector characters can be obtained from the fermionization. In terms of the characters of the $\text{SU(2)}_6$ WZW model, it is given by
\begin{align}
\begin{split}
f_0^{\text{R}}(\tau) &= \left( \chi^{\mathfrak{a}_{1,6}}_{\frac{3}{32}}(\tau) + \chi^{\mathfrak{a}_{1,6}}_{\frac{35}{32}}(\tau) \right)^2, \quad f_1^{\text{R}}(\tau) = 2 \left( \chi^{\mathfrak{a}_{1,6}}_{\frac{3}{32}}(\tau) + \chi^{\mathfrak{a}_{1,6}}_{\frac{35}{32}}(\tau) \right) \chi^{\mathfrak{a}_{1,6}}_{\frac{15}{32}}(\tau), \\
f_2^{\text{R}}(\tau) &= 4 \left(\chi^{\mathfrak{a}_{1,6}}_{\frac{15}{32}}(\tau)\right)^2,
\end{split}
\end{align}
and they are matched with the R-sector solution listed in table \ref{R-sector list 1}. We further remark that the $\widetilde{\text{R}}$ sector partition function becomes a constant, therefore satisfies the SUSY criterion discussed in \cite{Bae:2021lvk}.

\paragraph{Fermionization of $\text{Sp}(4)_3$ WZW model}

Here we discuss a fermionization of the $\text{Sp}(4)_3$ WZW model. The central charge of this theory is five and there are ten primaries involving $[0;3,0]$ which is a primary of conformal weight $h=3/2$. To fermionize the theory, we introduce the Verlinde line $\mathcal{L}_{h=3/2}$ that is associated with a non-anomalous $\mathbb{Z}_2$ symmetry. Applying the generalized Jordan-Wigner transformation, we find that the NS-sector partition function consists of the following three characters,
\begin{align}
\label{sp43}
\begin{split}
    f^{\rm{NS}}_{0}(\tau) &= \chi^{{\mathfrak c}_{2,3}}_0(\tau) + \chi^{{\mathfrak c}_{2,3}}_{\frac{3}{2}}(\tau),
    \quad  f^{\rm{NS}}_{1}(\tau) = \chi^{{\mathfrak c}_{2,3}}_{\frac{1}{2}}(\tau) + \chi^{{\mathfrak c}_{2,3}}_{1}(\tau),
    \\
    f^{\rm{NS}}_{2}(\tau) & = \chi^{{\mathfrak c}_{2,3}}_{\frac{1}{3}}(\tau) + \chi^{{\mathfrak c}_{2,3}}_{\frac{5}{6}}(\tau),
\end{split}
\end{align}
and the above fermionic characters turn out to solve the NS-sector BPS third-order MLDE. Similarly, the R-sector characters read
\begin{align}
\begin{split} 
    f^{\rm{R}}_{0}(\tau) & = \sqrt{2}\chi^{{\mathfrak c}_{2,3}}_{\frac{5}{8}}(\tau), \quad
    f^{\rm{R}}_{1}(\tau)  = \sqrt{2}\chi^{{\mathfrak c}_{2,3}}_{\frac{7}{8}}(\tau), \quad f^{\rm{R}}_{2}(\tau) = \chi^{{\mathfrak c}_{2,3}}_{\frac{5}{24}}(\tau) + \chi^{{\mathfrak c}_{2,3}}_{\frac{29}{24}}(\tau).
\end{split}
\end{align}
and they are matched with the R-sector solutions presented in table \ref{R-sector list}.

\paragraph{Fermionization of $\text{Sp}(6)_2$ WZW model}

We repeatedly apply the fermionization to the $\text{Sp}(6)_2$ WZW model. It is an RCFT with $c=7$ and includes ten primaries. Especially, we focus on the primary $[0;0,0,2]$ whose conformal weight is $h=3/2$. The Verlinde line $\mathcal{L}_{h=3/2}$ has a role of the generator of non-anomalous $\mathbb{Z}_2$ symmetry and this line defect provides us the characters of fermionic WZW models. Utilizing the Jordan-Wigner transformation \eqref{fermionization01}, we obtain the following NS-sector characters
\begin{align}
\label{chofSP6}
\begin{split}
    f^{\rm{NS}}_{0}(\tau) &= \chi^{{\mathfrak c}_{3,2}}_0(\tau) + \chi^{{\mathfrak c}_{3,2}}_{\frac{3}{2}}(\tau), \quad f^{\rm{NS}}_{1}(\tau) = \chi^{{\mathfrak c}_{3,2}}_{\frac{1}{2}}(\tau) + \chi^{{\mathfrak c}_{3,2}}_{1}(\tau),
    \\
    f^{\rm{NS}}_{2}(\tau) & = \chi^{{\mathfrak c}_{3,2}}_{\frac{2}{3}}(\tau) + \chi^{{\mathfrak c}_{3,2}}_{\frac{7}{6}}(\tau),
\end{split}
\end{align}
and R-sector characters
\begin{align}
\begin{split} 
    f^{\rm{R}}_{0}(\tau) & = \sqrt{2}\chi^{{\mathfrak c}_{3,2}}_{\frac{5}{8}}(\tau), \quad f^{\rm{R}}_{1}(\tau)  = \sqrt{2}\chi^{{\mathfrak c}_{3,2}}_{\frac{7}{8}}(\tau), \quad f^{\rm{R}}_{2}(\tau) = \chi^{{\mathfrak c}_{3,2}}_{\frac{7}{24}}(\tau) + \chi^{{\mathfrak c}_{3,2}}_{\frac{31}{24}}(\tau).
\end{split}
\end{align}
The $q$-expansion of above NS-sector and R-sector characters can be identified with the solutions of the fermionic BPS third-order MLDE with $c=7$. Thus we claim the solutions with $c=7$ describe the fermionized $\text{Sp}(6)_2$ WZW model.

\paragraph{Fermionization of $\text{SU}(4)_4/\mathbb{Z}_2$}

Let us now discuss the identification of solutions with $c=15/2$. Our goal is to apply fermionization to an orbifold theory $\text{SU}(4)_4/\mathbb{Z}_2$ and show that its characters can be identified with the class II solutions with $c=15/2$.

The first step is to apply \eqref{orbifold} to the partition function of $\text{SU}(4)_4$ WZW model, which will be denoted as $\mathcal{B}$. We note that the WZW model of interest involves 35 primary and especially we will pay attention to a primary $[0;0,0,4]$ with conformal weight $h=2$. One can choose a $\mathbb{Z}_2$ subgroup of center symmetry $\mathbb{Z}_4$ which is generated by the Verlinde line associated with a representation $[0;0,0,4]$. With help of the Verlinde line $\mathcal{L}_{h=2}$, one can show that the partition function of an $\mathbb{Z}_2$ orbifold theory has the form of
\begin{align}
\label{partition function of A3k4Z2}
\begin{split}
Z_{\mathcal{B}/{\mathbb{Z}_2}}&= |\chi^{{\mathfrak a}_{3,4}}_0 + \chi^{{\mathfrak a}_{3,4}}_2|^2 + 2|\chi^{{\mathfrak a}_{3,4}}_{\frac{9}{16}} + \chi^{{\mathfrak a}_{3,4}}_{\frac{25}{16}}|^2 + |\chi^{{\mathfrak a}_{3,4}}_{\frac{5}{16}} + \chi^{{\mathfrak a}_{3,4}}_{\frac{21}{16}}|^2 + |\chi^{{\mathfrak a}_{3,4}}_{\frac{1}{2}} + \chi^{{\mathfrak a}_{3,4}}_{\frac{3}{2}}|^2 \\
& + 4 |\chi^{{\mathfrak a}_{3,4}}_{\frac{3}{2}}|^2  + 4 |\chi^{{\mathfrak a}_{3,4}}_{\frac{21}{16}}|^2  + 4 |\chi^{{\mathfrak a}_{3,4}}_{1}|^2 + 2 |\chi^{{\mathfrak a}_{3,4}}_{\frac{3}{4}}|^2 + 2 |\chi^{{\mathfrak a}_{3,4}}_{\frac{5}{4}}|^2 + 2 |\chi^{{\mathfrak a}_{3,4}}_{\frac{15}{16}}|^2.
\end{split}
\end{align}

An orbifold theory $\mathcal{B}/{\mathbb{Z}_2}$ possesses the Verlinde line $\mathcal{L}_{h=3/2}$ and it generates a non-anomalous $\mathbb{Z}_2$ symmetry. We use it to fermionize $\mathcal{B}/{\mathbb{Z}_2}$. By applying the Jordan-Wigner transformation to \eqref{partition function of A3k4Z2} with $\mathcal{L}_{h=3/2}$, we find that the characters of NS-sector are given by
\begin{align}
\label{A3/Z2 NS}
\begin{split}
    f^{\rm{NS}}_{0}(\tau) &= \chi^{{\mathfrak a}_{3,4}}_0(\tau) + 2\chi^{{\mathfrak a}_{3,4}}_{\frac{3}{2}}(\tau) + \chi^{{\mathfrak a}_{3,4}}_2(\tau), \quad f^{\rm{NS}}_{1}(\tau)  = \chi^{{\mathfrak a}_{3,4}}_{\frac{3}{4}}(\tau) + \chi^{{\mathfrak a}_{3,4}}_{\frac{5}{4}}(\tau),
    \\ 
    f^{\rm{NS}}_{2}(\tau) & = \chi^{{\mathfrak a}_{3,4}}_{\frac{1}{2}}(\tau) + 2\chi^{{\mathfrak a}_{3,4}}_{1}(\tau) + \chi^{{\mathfrak a}_{3,4}}_{\frac{3}{2}}(\tau),
\end{split}
\end{align}
and the R-sector characters read
\begin{align}
\label{A3/Z2 R}
\begin{split}
    f^{\rm{R}}_{0}(\tau) &= \chi^{{\mathfrak a}_{3,4}}_{\frac{5}{16}}(\tau) + 2\chi^{{\mathfrak a}_{3,4}}_{\frac{21}{16}}(\tau) + \chi^{{\mathfrak a}_{3,4}}_{\frac{21}{16}}(\tau), \\
    f^{\rm{R}}_{1}(\tau) & = 2\chi^{{\mathfrak a}_{3,4}}_{\frac{9}{16}}(\tau) + 2\chi^{{\mathfrak a}_{3,4}}_{\frac{25}{16}}(\tau), \quad
    f^{\rm{R}}_{2}(\tau)  = \chi^{{\mathfrak a}_{3,4}}_{\frac{15}{16}}(\tau).
\end{split}
\end{align}
By comparing \eqref{A3/Z2 NS} and \eqref{A3/Z2 R} with the solutions listed in table \ref{NS-sector list 2} and \ref{R-sector list 2}, we conclude that the solutions with $c=15/2$ can be understood as the characters of a fermionized $\text{SU}(4)_4/\mathbb{Z}_2$ theory.

\paragraph{Fermionization of $(\text{Sp}(6)_1)^2$}

We propose to interpret the solutions with $c=\frac{42}{5}$ as the characters of fermionized $(\text{Sp}(6)_1)^2$ WZW model. To this end, let us fermionize the tensor product of two $\text{Sp}(6)_1$ WZW models. The product theory possesses $\mathbb{Z}_2$ symmetry generated by the Verlinde line $\mathcal{L}_{h=\frac{3}{2}}$. We find that a fermionization yield the NS-sector characters of the form
\begin{align}
\label{aaaaaa}
\begin{split}
    f^{\rm{NS}}_{0}(\tau) &= \chi^{\mathfrak{c}_{3,1}}_0(\tau) \chi^{\mathfrak{c}_{3,1}}_0(\tau) + \chi^{\mathfrak{c}_{3,1}}_{\frac{3}{4}}(\tau) \chi^{\mathfrak{c}_{3,1}}_{\frac{3}{4}}(\tau),
    \\ 
    f^{\rm{NS}}_{1}(\tau) & = \chi^{\mathfrak{c}_{3,1}}_{\frac{3}{5}}(\tau) \chi^{\mathfrak{c}_{3,1}}_{\frac{3}{5}}(\tau) + \chi^{\mathfrak{c}_{3,1}}_{\frac{7}{20}}(\tau) \chi^{\mathfrak{c}_{3,1}}_{\frac{7}{20}}(\tau), \\
    f^{\rm{NS}}_{2}(\tau) & = \chi^{\mathfrak{c}_{3,1}}_{0}(\tau) \chi^{\mathfrak{c}_{3,1}}_{\frac{3}{5}}(\tau) + \chi^{\mathfrak{c}_{3,1}}_{\frac{3}{4}}(\tau) \chi^{\mathfrak{c}_{3,1}}_{\frac{7}{20}}(\tau),
\end{split}
\end{align}
and the above characters perfectly matched to the NS-sector BPS solution with $c=\frac{42}{5}$. In a similar way, one can show that the R-sector characters of $(\text{Sp}(6)_1)^2$ WZW model agree with the R-sector BPS solution with $c=\frac{42}{5}$. An explicit expression of the R-sector characters are given by
\begin{align}
\begin{split}
f^{\rm{R}}_{0}(\tau) &= 2\chi^{\mathfrak{c}_{3,1}}_{0}(\tau) \chi^{\mathfrak{c}_{3,1}}_{\frac{3}{4}}(\tau), \quad f^{\rm{R}}_{1}(\tau) = \chi^{\mathfrak{c}_{3,1}}_{0}(\tau) \chi^{\mathfrak{c}_{3,1}}_{\frac{7}{20}}(\tau) + \chi^{\mathfrak{c}_{3,1}}_{\frac{3}{4}}(\tau) \chi^{\mathfrak{c}_{3,1}}_{\frac{3}{5}}(\tau),\\
f^{\rm{R}}_{2}(\tau) &= 2\chi^{\mathfrak{c}_{3,1}}_{\frac{3}{5}}(\tau) \chi^{\mathfrak{c}_{3,1}}_{\frac{7}{20}}(\tau),
\end{split}
\end{align}
and one can show that the above characters solve the R-sector BPS solution with $c=\frac{42}{5}$.

\paragraph{Fermionization of $(\text{SU}(6)_1)^2$}

Our next goal is to show that the partition function of fermionized $(\text{SU}(6)_1)^2$ WZW model can be described by the BPS solution with $c=10$. We first note that the level-one WZW model for $\text{SU}(6)$ involves six primaries of weights $h = 0, \frac{5}{12}, \frac{2}{3}, \frac{3}{4}, \frac{2}{3}, \frac{5}{12}$. Fortunately, in this case, the characters of each representation can be distinguished by their conformal weights. Therefore we denote the characters of $\text{SU}(6)_1$ WZW model as $\chi_{h}^{\mathfrak{a}_{5,1}}(\tau)$ where $h$ denote the conformal weight of certain representation. 

The tensor product theory has 36 primaries and especially it involves a primary of $h=3/2$. Indeed, one can show that the Verlinde line for a primary of $h=3/2$ is related to the non-anomalous $\mathbb{Z}_2$ symmetry. It turns out that a fermionization with $\mathcal{L}_{h=3/2}$ provides the NS-sector partition function of the form
\begin{align}
\label{fermionic A5k1 square}
\begin{split}
Z^{\text{NS}} = |f^{\text{NS}}_0(\tau)|^2 + 4 |f^{\text{NS}}_1(\tau)|^2 + 4 |f^{\text{NS}}_2(\tau)|^2,
\end{split}
\end{align}
where the $q$-expansions of three characters appeared in the above partition function are given by
\begin{align}
\label{blablabla}
\begin{split}
f^{\text{NS}}_0(\tau) &= (\chi_0^{\mathfrak{a}_{5,1}})^2 +  (\chi_{\frac{3}{4}}^{\mathfrak{a}_{5,1}})^2  = q^{-5/12} + 70 q^{7/12}+400 q^{13/12}+1745 q^{19/12}+\cdots, \\
f^{\text{NS}}_1(\tau) &= (\chi_{\frac{5}{12}}^{\mathfrak{a}_{5,1}})^2 +  (\chi_{\frac{2}{3}}^{\mathfrak{a}_{5,1}})^2 = 36 q^{5/12}+225 q^{11/12}+1080 q^{17/12}+4230 q^{23/12} + \cdots, \\
f^{\text{NS}}_2(\tau) &= \chi_0^{\mathfrak{a}_{5,1}} \chi_{\frac{2}{3}}^{\mathfrak{a}_{5,1}} +  \chi_{\frac{5}{12}}^{\mathfrak{a}_{5,1}} \chi_{\frac{3}{4}}^{\mathfrak{a}_{5,1}} = 15 q^{1/4} + 120 q^{3/4}+666 q^{5/4}+2760 q^{7/4}+\cdots.
\end{split}
\end{align}
Based on the comparison between \eqref{blablabla} and table \ref{NS-sector list 2}, we claim the the class II solution with $c=10$ can be understood as characters of the fermionized $(\text{SU}(6)_1)^2$ theory.

On the one hand, it is straightforward to check that the generalized Jordan-Wigner transformation provides the following R-sector partition function.
\begin{align}
\begin{split}
Z^R = 4|f^{\text{R}}_0(\tau)|^2 + 4 |f^{\text{R}}_1(\tau)|^2 +  |f^{\text{R}}_2(\tau)|^2\,.
\end{split}
\end{align}
Here, each R-sector characters exhibit the following $q$-expansions.
\begin{align}
\begin{split}
f^{\text{R}}_0(\tau) &= \chi_0^{\mathfrak{a}_{5,1}} \chi_{\frac{5}{12}}^{\mathfrak{a}_{5,1}} +  \chi_{\frac{3}{4}}^{\mathfrak{a}_{5,1}} \chi_{\frac{2}{3}}^{\mathfrak{a}_{5,1}}  = 6 + 600 q + 10440 q^2 + 102120 q^3 +\cdots\,, \\
f^{\text{R}}_1(\tau) &= 2\chi_{\frac{5}{12}}^{\mathfrak{a}_{5,1}} \chi_{\frac{2}{3}}^{\mathfrak{a}_{5,1}} = 180 q^{2/3}+4392 q^{5/3}+49860 q^{8/3}+388080 q^{11/3}+\cdots\,, \\
f^{\text{R}}_2(\tau) &= 2\chi_{0}^{\mathfrak{a}_{5,1}} \chi_{\frac{3}{4}}^{\mathfrak{a}_{5,1}} = 40 q^{1/3} + 1720 q^{4/3}+23360 q^{7/3}+202064 q^{10/3}+\cdots\,.
\end{split}
\end{align}
The above R-sector characters perfectly match with the R-sector solution of the BPS MLDE, as presented in table \ref{R-sector list 2}.

\paragraph{Bilinear Pairs}

Note that the class II solutions involve four pairs whose sum of the central charge is 12. We remark that four bilinear relations of the form
\begin{align}
\begin{split}
f_0^{NS} \widetilde{f}_0^{NS} + 4 f_{1/6}^{NS} \widetilde{f}_{5/6}^{NS} + 4 f_{1/3}^{NS} \widetilde{f}_{2/3}^{NS} &= K(\tau), \quad \text{for $c=2$ and $\tilde{c}=10$}, \\
f_0^{NS} \widetilde{f}_0^{NS} + f_{3/10}^{NS} \widetilde{f}_{7/10}^{NS} + 2 f_{2/5}^{NS} \widetilde{f}_{3/5}^{NS} &= K(\tau), \quad \text{for $c=18/5$ and $\tilde{c}=42/5$}, \\
f_0^{NS} \widetilde{f}_0^{NS} + 2 f_{1/4}^{NS} \widetilde{f}_{3/4}^{NS} +  f_{1/2}^{NS} \widetilde{f}_{1/2}^{NS} &= K(\tau), \quad \text{for $c=9/2$ and  $\tilde{c}=15/2$}, \\
f_0^{NS} \widetilde{f}_0^{NS} +  f_{1/3}^{NS} \widetilde{f}_{2/3}^{NS} +  f_{1/2}^{NS} \widetilde{f}_{1/2}^{NS} &= K(\tau), \quad \text{for $c=5$ and  $\tilde{c}=7$} ,
\end{split}
\end{align}
are established among the class II solutions. Here we denote the characters and central charge of dual paired theory by $\widetilde{f}^{NS}_{h}$ and $\tilde{c} = 12-c$, respectively. We will make separate comments on the above bilinear relations in section \ref{deconstrution}.

\subsubsection{Comments on the Class III solutions}

The class III involves three solutions with $c=39/2$, $c=22$ and $c=66/5$. We remark that the first two solutions can be understood from solutions of the BPS second-order MLDE \cite{Bae:2020xzl}. The BPS second-order MLDE is known to have solutions with $(c=39/4, h=3/4)$ and $(c=11, h=5/6)$, it is very natural to expect that their tensor product theories appear as the solutions of the BPS third-order MLDE.

Let us discuss with more details. To analyze the solution with $c=39/2$, we start with the $(\text{Sp}(6)_1)^2$ WZW model. The product theory possesses non-anomalous $\mathbb{Z}_2$ symmetry that is inherited from the center symmetry of Sp$(6)$. Especially, we use the Verlinde line $\mathcal{L}_{h=3}$ to construct the $\mathbb{Z}_2$ orbifold theory. The partition function of orbifold theory consists of 13 primaries of conformal weights
\begin{align}
h = \left\{ 0,\frac{3}{2},\frac{5}{4},\frac{3}{4},\frac{45}{16},\frac{29}{16},\frac{39}{16},\frac{5}{4},\frac{11}{4},\frac{5}{2},2,\frac{39}{16},\frac{33}{16} \right\}.
\end{align}
We next consider the fermionization of an orbifold theory using the Verlinde line $\mathcal{L}_{h=3/2}$ associated with $\mathbb{Z}_2$ symmetry. The NS-sector partition function is now given by
\begin{align}
\begin{split}
Z^{NS} = |f_0^{NS}|^2 + 2|f_1^{NS}|^2 + |f_2^{NS}|^2,
\end{split}
\end{align}
where $f_i^{NS}$ are expressed in terms of characters of the $\text{Sp}(6)_1$ WZW model as follows.
\begin{align}
\label{characters of c61Z2}
\begin{split}
f_0^{NS} &= (\chi_0^{\frak{c}_{6,1}}+\chi_{3/2}^{\frak{c}_{6,1}})^2, \quad f_1^{NS} = (\chi_{0}^{\frak{c}_{6,1}}+\chi_{3/2}^{\frak{c}_{6,1}})(\chi_{3/4}^{\frak{c}_{6,1}}+\chi_{5/4}^{\frak{c}_{6,1}}), \\
f_2^{NS} &= (\chi_{3/4}^{\frak{c}_{6,1}}+\chi_{5/4}^{\frak{c}_{6,1}})^2.
\end{split}
\end{align}
By expanding \eqref{characters of c61Z2} in $q$ and comparing with the NS-sector solution in table \ref{NS-sector list}, the solutions with $c=39/2$ can be identified with the fermionized $(\text{Sp}(6)_1)^2/{\mathbb{Z}_2}$ theory.

We next focus on the solution with $c=22$. To identify it, we consider the tensor product of two $\text{SU}(12)_1$ WZW models. The product theory has the Verlinde line $\mathcal{L}_{h=3}$ that generates $\mathbb{Z}_2$ symmetry. With help of the formula \eqref{orbifold} and the above Verlinde line, it is straightforward to compute the orbifold partition function. We find that partition function of an orbifold theory $(\text{SU}(12)_1)^2/\mathbb{Z}_2$ is contributed by 36 primaries involving a primary of weight $h=3/2$. Indeed, we check that the Verlinde line $\mathcal{L}_{h=3/2}$ has a role of $\mathbb{Z}_2$ generator. After applying \eqref{fermionization01} to the partition function of an orbifold theory, one can show that the NS-sector partition function is given by
\begin{align}
\begin{split}
Z^{NS} = |f_0^{NS}|^2 + 4 |f_1^{NS}|^2 + 4 |f_2^{NS}|^2,
\end{split}
\end{align}
where the characters are expressed as follows.
\begin{align}
\begin{split}
f_0^{NS} &= (\chi_0^{\frak{a}_{11,1}}+\chi_{3/2}^{\frak{a}_{11,1}})^2, \quad f_1^{NS} = (\chi_0^{\frak{a}_{11,1}}+\chi_{3/2}^{\frak{a}_{11,1}})(\chi_{4/3}^{\frak{a}_{11,1}}+\chi_{5/6}^{\frak{a}_{11,1}}), \\
f_2^{NS} &= (\chi_{4/3}^{\frak{a}_{11,1}}+\chi_{5/6}^{\frak{a}_{11,1}})^2.
\end{split}
\end{align}
The above three characters $f_0^{NS},f_1^{NS},f_2^{NS}$ can be identified with the solution with $c=22$. 

We finally remark that an identification of the solution with $c=66/5$ is unclear. The vacuum character involves 198 spin-one conserved currents, none of the classical Lie groups can be used to explain the number 198. It is plausible to check that if this solution can be interpreted as the tensor product of WZW model or coset theory.

\section{Bilinear Relations of Fermionic RCFT} \label{deconstrution}

The main idea of Monster deconstruction is to decompose the stress-energy tensor $T(z)$ of Monster CFT into the sum of two stress-energy tensors $t_1(z)$ and $t_2(z)$. The Monster deconstruction produces two disjoint CFTs $\mathcal{M}_1$ and $\mathcal{M}_2$ that are associated with $t_1(z)$ and $t_2(z)$. By disjoint CFTs we mean every operator in $\mathcal{M}_1$ has regular OPE with any operator in $\mathcal{M}_2$. It has been known that the process of the above decomposition can be formulated by using the mathematical notion of the commutant subalgebra of Monster VOA. The explicit examples of Monster deconstruction have been discussed in
\cite{Bae:2018qfh,Bae:2020pvv} with choosing $t_1(z)$ as the stress-energy tensor of the minimal models or $\mathbbm{Z}_k$ parafermion model. As a consequence, the Monster deconstruction leads us to specific RCFTs which exhibit sporadic groups in their modular invariant partition function. 

In the previous section, we faced examples of fermionic RCFTs that establish a bilinear relation. Such bilinear relation implies that  deconstruction could happen even for the fermionic RCFT. To explore the bilinear relation more, we start from the  $\mathcal{N}=1$ SCFT with $c=12$. 
This theory has been constructed with eight bosons compactified on $E_8$ root lattice and eight free fermions \cite{Frenkel:1988xz}. The free-fermion currents are removed by considering a $\mathbb{Z}_2$ orbifold. From the lattice construction, one can show that the NS sector partition function is given by a modular form of $\Gamma_\theta$ which we denote by $K(\tau)$ in what follows,
\begin{align}
K(\tau) = q^{-\frac{1}{2}} + 276 q^{\frac{1}{2}} + 2048 q + 11202 q^{\frac{3}{2}} + 49152 q^2 + 184024 q^{\frac{5}{2}} + \cdots.
\end{align}
We explore the deconstruction of the above $\mathcal{N}=1$ SCFT using the level-three WZW models for $SO(m)$ as a basic building block. More precisely, we consider the fermionized $SO(m)_3$ WZW models that can be obtained by applying \eqref{fermionization01}. For notational convenience, we write the fermionized $SO(m)_3$ WZW model as $\mathcal{F}^{A}(m)$ where the superscript $A$ stand for the spin-structure. The next step is to find a dual theory whose partition function deconstructs $K(\tau)$ with the NS-sector partition function of $\mathcal{F}^{NS}(m)$. A necessary condition for having a dual pair is 
\begin{align}
\label{commutant pair central charge}
c + \tilde{c} = 12,
\end{align}
where $c$ and $\tilde{c}$ are the central charges of $\mathcal{F}^{NS}(m)$  and its dual pair $\widetilde{\mathcal{F}}^{NS}(m)$, respectively. The central charges of $\mathcal{F}^{NS}(m)$ for first few $m$ are presented in table \ref{suzuki and dual suzuki chain}.
Note that the central charge of $\mathcal{F}^{NS}(10)$ is larger than $12$ thus the dual pair cannot be the unitary CFT. In this section, we will focus on $\mathcal{F}^{NS}(m)$  and $\widetilde{\mathcal{F}}^{NS}(m)$ up to $m \le 7$. 

We constrain the dual characters $\tilde{f}^{NS}_i(q)$ of $\widetilde{\mathcal{F}}^{NS}(m)$ to satisfy a bilinear relation with the characters $f^{NS}_i(q)$ of $\mathcal{F}^{NS}(m)$. Explicitly, a bilinear relation of our interest is expected to take the form
\begin{align}
K(\tau) = \sum_i f^{NS}_i(q) \tilde{f}^{NS}_i(q).
\end{align}
This bilinear relation enables us to explore the $S$-matrix and characters of dual theory. The central charge of dual theory is obtained by a relation \eqref{commutant pair central charge} and the conformal weights of each primary of dual theory can be deduced from the above bilinear relation. Once the central charge, weights of primaries and $S$-matrix are known, there are several ways to find the dual characters $\tilde{f}^{NS}_i(q)$. For $m \le 5$, we use the MLDE to determine the dual characters. For $m \ge 6$, we apply the Rademacher expansion to obtain the first few coefficients of $\tilde{f}_i(q)$. We summarize the details of ${\mathcal{F}}^{NS}(m)$ and $\widetilde{\mathcal{F}}^{NS}(m)$ in table \ref{suzuki and dual suzuki chain}.

The characters of the bilinear pairs $\mathcal{F}^{A}(m)$ and $\widetilde{\mathcal{F}}^{A}(m)$ should be identified with the solutions of the fermionic MLDE. Specifically, the characters of pair theories $(c, \tilde{c})= (1,11)$ and $(9/4, 39/4)$ have been discussed in \cite{Bae:2020xzl} and they can be obtained from the BPS second-order MLDE. Furthermore, all the class II solutions of the BPS third-order MLDE can be understood with the theories listed in table  \ref{suzuki and dual suzuki chain}. It is natural to expect that the solutions for $m=6$ and $m=7$ are associated with the BPS fourth-order MLDE. 

As a side remark, we comment on the relation between $\mathcal{F}^{A}(m)$ and the $\mathcal{N}=1$ supersymmetric minimal models which describe the unitary supersymmetric RCFTs of $c \le \frac{3}{2}$. The central charge and conformal weights of the $\mathcal{N}=1$ supersymmetric minimal models $\mathcal{M}(k)$ are given by
\begin{align}
\begin{split}
    c&=\frac32 \Big[ 1- \frac{8}{k(k+2)} \Big], \quad  k=3,4,5, \cdots, \\
    h_{m,n} &= \frac{[(k+2)m-k n]^2-4}{8 k(k+2)}
  +\frac{1}{32}[1-(-1)^{m-n}],
\end{split}
\end{align}
where $m,n\in \mathbb{Z}$ and $1\le m <k, 1\le n<k+2 $. The NS and R sectors have even and odd values of $m-n$, respectively.  The character formulas for the NS and R algebra are given by \cite{Goddard:1986ee}
\begin{align}
\label{character}
\begin{split}
 \chi^{NS}_{m,n}(\tau) &= \zeta^{k}_{m,n}(q) \prod_{\ell=1}^{\infty}\left( \frac{1+q^{\ell-\frac{1}{2}}}{1-q^\ell} \right) \\
 \chi^{R}_{m,n}(\tau) &= \zeta^{k}_{m,n}(q) q^{\frac{1}{16}} \prod_{\ell=1}^{\infty}\left( \frac{1+q^{\ell-\frac{1}{2}}}{1-q^\ell} \right),
\end{split}
\end{align}
where
\begin{align}
\begin{split}
\zeta^k_{m,n}(q) &= \sum_{\alpha \in \mathbb{Z}} \left( q^{\gamma^k_{m,n}(\alpha)} - q^{\delta^k_{m,n}(\alpha)}\right), \\
\gamma^k_{m,n}(\alpha)&= \frac{\left(2k(k+1)\alpha - m(k+2) + pk\right)^2-4}{8k(k+2)}, \\
\delta^k_{m,n}(\alpha) &=  \frac{\left(2k(k+1)\alpha + m(k+2) + pk\right)^2-4}{8k(k+2)} .
\end{split}
\end{align}
One can use the characters of the product theory $\bigotimes_{i=1}^{m-1} \mathcal{M}(2i+2)$ to express the partition functions of the theories $\mathcal{F}^{A}(m)$. For instance, we need to consider the tensor product theory $\mathcal{M}(4) \otimes \mathcal{M}(6) \otimes \mathcal{M}(8)$ to describe the partition function of the fermionized SO$(4)_3$ WZW model. As a consistent check, the central charge of the above product theory $c=\frac{18}{5}$ agree with that of the SO$(4)_3$ WZW model.

\begin{table}[]
\def\arraystretch{1.4}
    \centering
\begin{tabular}{c|c|c|c||c|c|c}
 \multicolumn{4}{c||}{${\mathcal{F}}^{NS}(m)$} & \multicolumn{3}{c}{$\widetilde{\mathcal{F}}^{NS}(m)$}  \\ [1mm] 
 \hline
  $m$ & $c$ &$ h^{\rm NS}$ &   ${\mathcal{B}}$ & $\tilde c$ & $\tilde h^{\rm NS} $  & $\widetilde{\mathcal{B}}$   \\ \hline  
 & & & & 12 &  0  & ${\rm SO}(24)_1 $  \\ [1mm] 
 \hline
 2 & 1  & $\frac16$ & ${\rm SO}(2)_3$   & 11 & $\frac56$  & ${\rm SU}(12)_1 $ \\ [1mm]
 3 & $\frac{9}{4}$ & $\frac14$  &  ${\rm SO}(3)_3$ & $\frac{39}{4}$ &    $\frac34$ &  ${\rm Sp}(12)_1$   \\ [1mm] 
 4 &$\frac{18}{5}$   & $\frac{3}{10}, \frac25$  &   ${\rm SO}(4)_3  $  & $\frac{42}{5}$ &    $\frac{7}{10},\frac35$ &   ${\rm Sp}(6)_1^2$     \\ [1mm]
 5 &$5$ & $\frac13,\frac12$ &   ${\rm SO}(5)_3$ & 7 & $\frac12,\frac23$  & ${\rm Sp}(6)_2$   \\ [1mm]
 6 &$\frac{45}{7}$ & $\frac{5}{14}, \frac{9}{14}, \frac{4}{7}$  &  ${\rm SO}(6)_3$ & $\frac{39}{7}$ & $\frac{5}{14}, \frac{9}{14}, \frac{3}{7}$  &   \\[1mm]
 7 & $\frac{63}{8} $ & $\frac{3}{8}, \frac{5}{8}, \frac{3}{4}$   & ${\rm SO}(7)_3$ & $\frac{33}{8}$ & $\frac{5}{8}, \frac{3}{8}, \frac{1}{4}$   &    \\
 8 & $\frac{28}{3} $ & $\frac{7}{18}, \frac{8}{9}, \frac{5}{6}, \frac{2}{3}$ & ${\rm SO}(8)_3$ & $\frac{8}{3}$ & $\frac{1}{9}, \frac{1}{6}, \frac{1}{3}, \frac{11}{18} $ &   \\
\end{tabular}    
    \caption{The detailed data of ${\mathcal{F}}^{NS}(m)$ and $\widetilde{\mathcal{F}}^{NS}(m)$ for $2 \le m \le 8$. It is not clear if the bosonized theories of $\widetilde{\mathcal{F}}^{NS}(6)$, $\widetilde{\mathcal{F}}^{NS}(7)$ and $\widetilde{\mathcal{F}}^{NS}(8)$ can be understood as the WZW models or not.}
    \label{suzuki and dual suzuki chain}
\end{table}

\subsubsection*{$U(1)_3$ and $\widetilde{\mathcal{F}}(2)$}

The level-$k$ $U(1)$ theories are constructed by compactifying the free boson to the circle of radius $R=\sqrt{k}$. The partition function is known to be contributed by finitely many theta functions $\Theta_{m,k}(\tau)$ for even $k$. More precisely, the partition function has the form of 
\begin{align}
\begin{split}
Z = \frac{1}{|\eta(\tau)|^2} \sum_{m = -\frac{k}{2} + 1} ^{\frac{k}{2}} |\Theta_{m,k}(\tau)|^2\,,
\end{split}
\end{align}
where
\begin{align}
\begin{split}
\Theta_{m,k}(\tau) =  \sum_{n \in \mathbb{Z}} q^{\frac{k}{2} (n + \frac{m}{2k})^2}, \qquad \text{for} \quad  -\frac{k}{2} + 1 \le m \le \frac{k}{2}.
\end{split}
\end{align}
The partition function cannot be invariant under the $T$-transformation when $k$ is an odd integer. Instead, the partition function of $U(1)$ theory with odd level is invariant under $T^2 : \tau \rightarrow \tau + 2$ and $S : \tau \rightarrow -1/\tau$, which is the characteristic property of the NS-sector partition function.

The partition function of level-three $U(1)$ theory is described by the following functions,
\begin{align}
\label{c1 characters}
\begin{split}
f^{NS}_0(\tau) &= q^{-\frac{1}{24}} \left( 1 + q + 2 q^{\frac{3}{2}} + 2q^2 + 2q^{\frac{5}{2}} + 3q^3 + 4q^{\frac{7}{2}} + \cdots \right), \\
f^{NS}_1(\tau) &= q^{\frac{1}{8}} \left( 1 + q^{\frac{1}{2}} + q + q^{\frac{3}{2}} + 2 q^2 + 3 q^{\frac{5}{2}} + 3 q^3 + 4 q^{\frac{7}{2}} + 6 q^4 + \cdots \right),
\end{split}
\end{align}
where their ${S}$-matrix is given by
\begin{align}
\label{modular matrix L1}
S = 
\frac{1}{\sqrt{3}}
\left(
\begin{array}{cc}
  1   &  2 \\
  1   &  -1
\end{array}
\right).
\end{align}
We remark that two functions \eqref{c1 characters} can be expressed by the NS-sector characters of the $\mathcal{N}=1$ supersymmetric minimal model with $c=1$. We refer the reader to \cite{Bae:2020xzl} for the details.

Let us now consider the NS sector partition function of the dual theory $\widetilde{\mathcal{F}}^{NS}(2)$. This theory involves two primaries of conformal weights $h=0,\frac{5}{6}$ in the NS sector and their characters are given by \cite{Bae:2021lvk}
\begin{align}
\label{c11 characters}
\begin{split}
  \tilde{f}^{NS}_0(q) &= q^{-\frac{11}{24}} \left(1+143 q+924 q^{3/2}+4499 q^2+18084 q^{5/2} + \cdots \right),  \\
  \tilde{f}^{NS}_1(q) & =q^{\frac{3}{8}}  \left(66+495q^{1/2}+2718q+11649q^{3/2}+42174 q^2+ \cdots \right).
\end{split}
\end{align}
One can show that the characters $\tilde{f}^{NS}_0(q)$ and $\tilde{f}^{NS}_1(q)$ satisfy a duality relation of the form
\begin{align}
\begin{split}
  K(\tau) = f^{NS}_0(q) \tilde{f}^{NS}_0(q) + 2 f^{NS}_1(q) \tilde{f}^{NS}_1(q).
\end{split}
\end{align}
The above duality relation implies that the NS sector partition function is given by
\begin{align}
\widetilde{Z}^{NS}(\tau,\bar{\tau}) = |\tilde{f}^{NS}_0(\tau)|^2 +2 |\tilde{f}^{NS}_1(\tau)|^2.
\end{align}

We notice that the Fourier coefficients of the dual characters $\tilde{f}_0(q)$ and $\tilde{f}_1(q)$ can be decomposed into the dimension of irreducible representations of $3.Suz$. More precisely, the coefficients of the dual characters are decomposed as follows.
\begin{align}
\label{Suz Decomposition}
\begin{split}
924 &= {\bf{1}} + {\bf{143}} + {\bf{780}}, \quad 4499 =  {\bf{1}} + {\bf{143}} + {\bf{143}} + {\bf{780}} + {\bf{3432}} \\
495 &= {\bf{66}} + {\bf{429}}, \quad 2718 = {\bf{66}} + {\bf{78}} + {\bf{429}} + {\bf{2145}}, \quad \cdots .
\end{split}
\end{align}
Using the number decomposition in  \eqref{Suz Decomposition}, one can compute the twined character referring to the character table of $3.Suz$. For instance, the 2A twined characters read 
\begin{align}
\begin{split}
  \tilde{f}_0^{2A}(q) &= q^{-\frac{11}{24}} \left(1+15 q+28 q^{3/2}+19 q^2 + \cdots \right),  \\
  \tilde{f}_1^{2A}(q) & =q^{\frac{3}{8}}  \left(2-17q^{1/2}-34q + \cdots \right).
\end{split}
\end{align}
It turns out that the above 2A twined characters are combined with the characters of $U(1)_3$ to produce a duality relation 
\begin{align}
\label{2A twined suz}
f_0(q) \tilde{f}^{2A}_0(q) +2  f_1(q) \tilde{f}_1^{2A}(q)= \frac{1}{\sqrt{q}} + 20 \sqrt{q} - 62 q^{\frac{3}{2}} + \cdots.
\end{align}
We suggest that the right-hand side of duality relation \eqref{2A twined suz} can be identified with the modular form 
\begin{align}
T_{4C}(q) = 16 \left( \frac{\vartheta_3(q)}{\vartheta_2(q)}\right)^4 - 8 = \frac{1}{\sqrt{q}} + 20 \sqrt{q} - 62 q^{\frac{3}{2}} + \cdots,
\end{align}
and is known as the Mckay-Thompson series of class 4C.

Now we turn to the characters of  $\widetilde{\text{NS}}$ and R-sector. For the superconformal minimal model with $c=1$, the $\widetilde{\text{NS}}$ and R-sector characters are related to the NS-sector character as follow.
\begin{align}
f_i^{\widetilde{NS}}(\tau) = f_i^{NS}(\tau+1), \qquad f_i^{R}(\tau) = f_i^{NS}\left(-\frac{1}{\tau}+1\right).
\end{align}
To find the characters of $\widetilde{\text{NS}}$ and R-sector, one can utilize the analytic solutions of the second order MLDE \cite{Bae:2020xzl}. As a consequence, one obtains the following $\widetilde{\text{NS}}$-sector characters 
\begin{align}
\label{NStilde c11}
\begin{split}
f^{\widetilde{NS}}_0(\tau) &=  q^{-\frac{1}{24}} \left( 1 + q - 2 q^{\frac{3}{2}} + 2q^2 - 2q^{\frac{5}{2}} + 3q^3  + \cdots \right), \\
f^{\widetilde{NS}}_1(\tau) &=  q^{\frac{1}{8}} \left( 1 - q^{\frac{1}{2}} + q - q^{\frac{3}{2}} + 2 q^2 - 3 q^{\frac{5}{2}} + 3 q^3  + \cdots \right),
\end{split}
\end{align}
and the R-sector characters
\begin{align}
\label{Rsector c11}
\begin{split}
f^{R}_0(\tau) &=  1 + 2 q + 4 q^2 + 6 q^3 + 10 q^4 + 16 q^5 + 24 q^6 + 36 q^7  + \cdots, \\
f^{R}_1(\tau) &=  q^{\frac{1}{3}} \left( 1 + q + 2 q^2 + 4 q^3 + 6 q^4 + 9 q^5 + 14 q^6  + \cdots \right),
\end{split}
\end{align}
for ${\mathcal{F}}(2)$. The $\widetilde{\text{NS}}$ and R-sector characters of a dual theory $\widetilde{\mathcal{F}}(2)$ read \footnote{It has been indicated that the characters \eqref{c11 characters}, \eqref{NStilde c11} and \eqref{Rsector c11} can be obtained by applying the fermionization to the level-one $SU(12)$ WZW model \cite{Bae:2021lvk}. The $\tilde{R}$ sector partition function becomes constant and it can be considered as a signal of the supersymmetry.}
\begin{align}
\begin{split}
  \tilde{f}^{\widetilde{NS}}_0(q) &=q^{-\frac{11}{24}} \left(1+143 q-924 q^{3/2}+4499 q^2-18084 q^{5/2} + \cdots \right),  \\
  \tilde{f}^{\widetilde{NS}}_1(q) & =q^{\frac{3}{8}}  \left(66-495q^{1/2}+2718q-11649q^{3/2}+42174 q^2+ \cdots \right),
\end{split}
\end{align}
and
\begin{align}
\begin{split}
  \tilde{f}^{R}_0(q) &= 1 + 132 q + 2706 q^2 + 30008 q^3 + 237204 q^4 + 1498728 q^5 + \cdots,  \\
  \tilde{f}^{R}_1(q) & =\tilde{f}'^{R}_1(q) = q^{\frac{2}{3}} \left( 55 + 1628 q + 21131 q^2 + 183868 q^3 + 1241174 q^4 + \cdots \right).
\end{split}
\end{align}
Then, it is easy to see that they form bilinear relations of the form
\begin{align}
\begin{split}
 K(\tau + 1) &=f^{\widetilde{NS}}_0(q) \tilde{f}^{\widetilde{NS}}_0(q) +2 f^{\widetilde{NS}}_1(q) \tilde{f}^{\widetilde{NS}}_1(q), \\
 K\left(-\frac{1}{\tau}+1\right)  &=24 f^{R}_0(q) \tilde{f}^{R}_0(q) + 16 f^{R}_1(q) \tilde{f}^{R}_1(q) \\
 &= 24 + 4096 q + 98304 q^2 + 1228800 q^3 + 10747904 q^4 + \cdots.
\end{split}
\end{align}

\subsubsection*{SO$(3)_3$ WZW and $\widetilde{\mathcal{F}}(3)$}

Here we comment on the fermionic theory which is obtained from the SO$(3)_3$ WZW model, equivalently SU$(2)_6$ WZW model. An explicit illustration of the generalized Jordan-Wigner transformation applied to the SU$(2)_6$ WZW model is presented in \cite{Bae:2021lvk}. As a consequence of the fermionization, the NS-sector partition function is described by the following two functions
\begin{align}
\label{L2 character}
\begin{split}
f_0^{NS}(\tau) &= q^{-\frac{3}{32}} \left( 1 + 3q + 7 q^{\frac{3}{2}} + 9 q^2 + 12 q^{\frac{5}{2}} + 22 q^3 + 36 q^{\frac{7}{2}} + \cdots \right), \\
f_1^{NS}(\tau) &= q^{\frac{5}{32}} \left( 3 + 5 q^{\frac{1}{2}} + 9q + 15 q^{\frac{3}{2}} + 27 q^2 + 45 q^{\frac{5}{2}} + 66q^3 + \cdots \right).
\end{split}
\end{align}
It is known that the above two functions solve the BPS fermionic second-order MLDE with $c=9/4$ and $h=1/4$ \cite{Bae:2020xzl}. 

The dual theory $\widetilde{\mathcal{F}}(3)$ can be understood as the $\text{Sp}(12)_1$ WZW model. An application of the Jordan-Wigner transformation to the $\text{Sp}(12)_1$ WZW model with the Verlinde line $\mathcal{L}_{h=3/2}$ yields
\begin{align}
\begin{split}
\tilde{f}^{NS}_0(q) &= q^{-\frac{13}{32}}\left( 1 + 78 q + 429 q^{\frac{3}{2}} + 1794 q^2 + 6435 q^{\frac{5}{2}} + 20527 q^3 + \cdots \right),\\
\tilde{f}^{NS}_1(q) &= q^{\frac{11}{32}} \left( 65 + 429 q^{\frac{1}{2}} + 2145 q + 8437 q^{\frac{3}{2}} + 28236 q^2 + 84876 q^{\frac{5}{2}} + \cdots \right).
\end{split}
\end{align}
and they indeed satisfy the following duality relation.
\begin{align}
\tilde{f}^{NS}_0(q) {f}^{NS}_0(q) + \tilde{f}^{NS}_1(q) {f}^{NS}_1(q) = K(\tau)\,.
\end{align}

\subsubsection*{SO$(4)_3$ WZW and SO$(5)_3$ WZW}

The class II solutions with $c=18/5$ and $c=5$ are identified to the fermionized SO$(4)_3$ and SO$(5)_3$ WZW models, as discussed in section \ref{sec:Identification II}. Furthermore, their duality pairs can be understood as the $(\text{Sp}(6)_1)^2$ and $\text{Sp}(6)_2$ WZW models. Here, we demonstrate an alternative way of obtaining the dual theory of $\widetilde{\mathcal{F}}(5)$. The idea is to use the characters of the $\mathcal{N}=1$ supersymmetric minimal models. In terms of the characters of $\mathcal{M}(10)$ and ${\mathcal{F}}(4)$, one can find an alternative expression of the fermionized $\text{Sp}(4)_3$ WZW model, e.g., \eqref{sp43}. More precisely, the characters \eqref{sp43} can be recast as
\begin{align}
\label{L4 characters}
f^{\text{NS}}_0(q) &= g^{\text{NS}}_0(q) ( \chi^{k=10}_{1,1} + \chi^{k=10}_{1,11} )+ g^{\text{NS}}_1(q) ( \chi^{k=10}_{3,1} + \chi^{k=10}_{3,11})+ 2g^{\text{NS}}_2(q)  \chi^{k=10}_{5,1} , \nonumber \\
                &= q^{-\frac{5}{24}} \left( 1 + 10 q + 30 q^{\frac{3}{2}} + 65 q^2 + 146 q^{\frac{5}{2}} + 330 q^3 + \cdots  \right), \nonumber \\
f^{\text{NS}}_1(q) &= g^{\text{NS}}_0(q)  (\chi^{k=10}_{1,3} + \chi^{k=10}_{1,9})+ g^{\text{NS}}_1(q) (\chi^{k=10}_{3,3} + \chi^{k=10}_{3,9}) + 2f^{\text{NS}}_2(q)  \chi^{k=10}_{5,3}, \nonumber \\
                &= q^{\frac{1}{8}} \left( 5 + 14 q^{\frac{1}{2}} + 50 q + 140 q^{\frac{3}{2}} + 325 q^2 + 690 q^{\frac{5}{2}} + \cdots \right), \\
f^{\text{NS}}_2(q) &= g^{\text{NS}}_0(q)  (\chi^{k=10}_{1,5}+\chi^{k=10}_{1,7}) + g^{\text{NS}}_1(q) (\chi^{k=10}_{3,5} +\chi^{k=10}_{3,7}) + 2g^{\text{NS}}_2(q)  \chi^{k=10}_{5,5}, \nonumber \\
                &= q^{\frac{7}{24}} \left( 10 + 35 q^{\frac{1}{2}} + 100 q + 245 q^{\frac{3}{2}} + 566 q^2 + 1225 q^{\frac{5}{2}} + \cdots \right), \nonumber
\end{align}
where $g^{\text{NS}}_i(q)$ and $\chi^{k=10}_{m,n}$ denote the characters of the fermionized SO$(4)_3$ WZW model and $\mathcal{N}=1$ supersymmetric minimal model $\mathcal{M}(10)$, respectively. The $S$-matrix of the three characters \eqref{L4 characters} is given by
\begin{align}
S = \frac{1}{6}
\left(
\begin{array}{ccc}
   3-\sqrt{3}  & 2\sqrt{3} & 3+\sqrt{3} \\
    2\sqrt{3} & 2\sqrt{3} & -2\sqrt{3} \\
    3+\sqrt{3} & -2\sqrt{3} & 3-\sqrt{3}
\end{array}
\right)
\end{align}

The strategy of finding the characters of $\widetilde{\mathcal{F}}(5)$ is to use the bilinear relation of the form
\begin{align}
\label{Dual L4}
f^{\text{NS}}_0(q) \tilde{f}^{\text{NS}}_0(q) + f^{\text{NS}}_1(q) \tilde{f}^{\text{NS}}_1(q) + f^{\text{NS}}_2(q) \tilde{f}^{\text{NS}}_2(q) = K(\tau).
\end{align}
We also note that the characters of ${\mathcal{F}}(4)$ and $\widetilde{\mathcal{F}}(4)$ ought to satisfy a similar bilinear relation of the form
\begin{align}
\label{bbbbb}
\begin{split}
g^{\text{NS}}_0(q) \tilde{g}^{\text{NS}}_0(q) + g^{\text{NS}}_1(q) \tilde{g}^{\text{NS}}_1(q) + 2g^{\text{NS}}_2(q) \tilde{g}^{\text{NS}}_2(q) = K(\tau),
\end{split}
\end{align}
where now $g^{\text{NS}}_i(q)$ are  characters of the fermionized $(\text{Sp}(6)_1)^2$ WZW model, given in \eqref{aaaaaa}. By equalling \eqref{Dual L4} and \eqref{bbbbb}, one can show the following relations hold between $\tilde{f}^{\text{NS}}_i(q)$ and $\tilde{g}^{\text{NS}}_i(q)$.
\begin{align}
\begin{split}
\tilde{g}_0^{\text{NS}} &= (\chi^{k=10}_{1,1} + \chi^{k=10}_{1,11}) \tilde{f}_0^{\text{NS}} + (\chi^{k=10}_{1,3} + \chi^{k=10}_{1,9}) \tilde{f}_1^{\text{NS}} + (\chi^{k=10}_{1,5} + \chi^{k=10}_{1,7}) \tilde{f}_2^{\text{NS}}, \\
\tilde{g}_1^{\text{NS}} &= (\chi^{k=10}_{3,1} + \chi^{k=10}_{3,11}) \tilde{f}_0^{\text{NS}} + (\chi^{k=10}_{3,3} + \chi^{k=10}_{3,9}) \tilde{f}_1^{\text{NS}} + (\chi^{k=10}_{3,5} + \chi^{k=10}_{3,7}) \tilde{f}_2^{\text{NS}}. \\
\tilde{g}_2^{\text{NS}} &= 2\chi^{k=10}_{5,1} \tilde{f}_0^{\text{NS}} + 2\chi^{k=10}_{5,3} \tilde{f}_1^{\text{NS}} + 2\chi^{k=10}_{5,5}  \tilde{f}_2^{\text{NS}}.
\end{split}
\end{align}
From the above relations, it is straightforward to determine the dual characters $\tilde{f}^{\text{NS}}_i(q)$,
\begin{align}
\tilde{f}^{\text{NS}}_0(q) &= q^{-\frac{7}{24}} \left( 1 + 21 q + 84 q^{\frac{3}{2}} + 252 q^2 + 686 q^{\frac{5}{2}} + 1771 q^3 + \cdots  \right), \nonumber \\
\tilde{f}^{\text{NS}}_1(q) &= q^{\frac{3}{8}} \left( 21 + 90 q^{\frac{1}{2}} + 315 q + 966 q^{\frac{3}{2}} + 2646 q^2 + 6552 q^{\frac{5}{2}} + \cdots \right), \\
\tilde{f}^{\text{NS}}_2(q)  &= q^{\frac{5}{24}} \left( 14 + 70 q^{\frac{1}{2}} + 294 q + 945 q^{\frac{3}{2}} + 2604 q^2 + 6615 q^{\frac{5}{2}} + \cdots \right), \nonumber
\end{align}
which agree with the characters of the Sp$(6)_2$ WZW model, namely \eqref{chofSP6}.

\subsubsection*{SO$(6)_3$ WZW and $\widetilde{\mathcal{F}}(6)$}

So far, we discuss the fermionized level-three SO$(m)$ WZW models up to $m \le 5$ and their dual pairs. The characters of the above theories turn out to correspond to the solutions of the BPS second and third-order MLDE. From now on, we consider the level-three SO$(m)$ WZW model with a higher rank. We will show that the partition functions of each spin structure involve four characters for $m=6$ and $m=7$, and they solve the BPS fourth-order MLDE.

We first note that the SO$(6)_3$ WZW model is equivalent to the SU$(4)_3$ WZW model, a RCFT with $c=45/7$. It involves a primary of $h=3/2$ and a line defect $\mathcal{L}_{h=3/2}$ is associated with non-anomalous $\mathbb{Z}_2$ symmetry. Now the fermionization can be done via the generalized Jordan-Wigner transformation, \eqref{fermionization01}. We find that the NS-sector partition function consists of the following four functions,
\begin{align}
\label{NScharacterSO6}
\begin{split}
f_0^{NS}(q) &= \chi^{\mathfrak{a}_{3,3}}_{h=0} + \chi^{\mathfrak{a}_{3,3}}_{h=3/2}= q^{-\frac{15}{56}} \left( 1 + 15 q + 50 q^{\frac{3}{2}} + 135 q^2 + 366 q^{\frac{5}{2}} +  \cdots \right),\\
f_1^{NS}(q) &= \chi^{\mathfrak{a}_{3,3}}_{h=6/7} + \chi^{\mathfrak{a}_{3,3}}_{h=5/14} =  q^{\frac{5}{56}} \left( 6 + 20 q^{\frac{1}{2}} + 90 q + 300 q^{\frac{3}{2}} + 810 q^2+ \cdots \right),\\
f_2^{NS}(q) &= \chi^{\mathfrak{a}_{3,3}}_{h=4/7} + \chi^{\mathfrak{a}_{3,3}}_{h=15/14}= q^{\frac{17}{56}} \left( 15 + 64 q^{\frac{1}{2}} + 225 q + 660 q^{\frac{3}{2}} + 1725 q^2 + \cdots \right),\\
f_3^{NS}(q) &=\chi^{\mathfrak{a}_{3,3}}_{h=8/7} + \chi^{\mathfrak{a}_{3,3}}_{h=9/14} = q^{\frac{3}{8}} \left( 10 + 45 q^{\frac{1}{2}} + 150 q + 419 q^{\frac{3}{2}} + 1080 q^2 + \cdots \right),
\end{split}
\end{align}
where the above functions correspond to the NS-sector characters of the primaries of weight $h=0, 5/14, 4/7, 9/14$, respectively. The NS-sector characters \eqref{NScharacterSO6} form a $\Gamma_\vartheta$ invariant object of the form
\begin{align}
Z^{NS} = |f_0^{NS}(q)|^2 + |f_1^{NS}(q)|^2 +  |f_2^{NS}(q)|^2 + 2 |f_3^{NS}(q)|^2,
\end{align}
and $Z^{NS}$ can be considered as the NS-sector partition function. Here, the NS-primary of weight $h=9/14$ has degeneracy two, therefore $4 \times 4$ sized $S$-matrix of \eqref{NScharacterSO6} cannot be the symmetric one. Instead, one can find that the following extended $S$-matrix 
\begin{align}
\label{Smatrix SO6}
S = \frac{1}{\sqrt{7}}
\left(
\begin{array}{ccccc}
   2 \sin\left(\frac{\pi}{14}\right)  & 2 \sin\left(\frac{3\pi}{14}\right) & 2 \cos\left(\frac{\pi}{7}\right) & 1 & 1\\
   2 \sin\left(\frac{3\pi}{14}\right)  & 2 \cos\left(\frac{\pi}{7}\right) & -2\sin\left(\frac{\pi}{14}\right) & -1 & -1\\
   2 \cos\left(\frac{\pi}{7}\right)  & -2\sin\left(\frac{\pi}{14}\right) & -2\sin\left(\frac{3\pi}{14}\right) & 1 & 1 \\
   1  & -1 & 1 & -1-\sqrt{7} \beta & \sqrt{7} \beta \\
   1  & -1 & 1 & \sqrt{7} \beta & -1-\sqrt{7} \beta 
\end{array}
\right)
\end{align}
acting on the vector-valued modular form $(f_0^{NS}(q), f_1^{NS}(q), f_2^{NS}(q), f_3^{NS}(q), f_3^{NS}(q))$ with $\beta = -\sqrt{-\frac{3}{14}-\frac{i}{2\sqrt{7}}}$ is a symmetric matrix, thus can provide a consistent fusion rule algebra.

We further notice that the four characters \eqref{NScharacterSO6} can be identified with the solution of a below fourth-order MLDE,
\begin{equation}
\label{fourth order}
	\begin{split}
		&\Bigg[\mathcal{D}^4+\frac{1}{2}e_2\mathcal{D}^3+\left(\frac{127}{672}e_2^2-\frac{139}{504}E_4\right)\mathcal{D}^2+\left(\frac{10837}{1185408}e_2^3+\frac{1279}{98784}e_2 E_4\right)\mathcal{D}\\
		&+\frac{5}{4214784}e_2^4+\frac{475}{526848}e_2^2 E_4-\frac{955}{263424}E_4^2\Bigg]f^{NS}_i(\tau)=0\,.
	\end{split}
	\end{equation}
The above fourth-order MLDE can be considered as the BPS MLDE. To see this point, we note that the weight-eight coefficient function of generic fourth-order MLDE can be written as $\alpha e_2^4 + \beta e_2^2 E_4 + \gamma E_4^2$ with certain numerical coefficients $\alpha, \beta, \gamma$. Under the $TS$ transformation, one can show that
\begin{align}
\begin{split}
&\alpha e_2^4 + \beta e_2^2 E_4 + \gamma E_4^2  \\
&\longrightarrow \frac{1}{16} \left( 9 \gamma \vartheta_2^{16} + 6(2\beta+ \gamma) \vartheta_2^8 (\vartheta_3^4 + \vartheta_4^4 )^2 + (16 \alpha +4 \beta +\gamma) (\vartheta_3^4 + \vartheta_4^4 )^4 \right) \\
&\quad =16 \alpha +4 \beta +\gamma +q (1536 \alpha +1152 \beta +480 \gamma ) + \cdots.
\end{split}
\end{align}
Therefore, the condition of saturating R-sector unitary bound becomes $16 \alpha +4 \beta +\gamma =0$. Indeed, it is easy to see that  \eqref{fourth order} satisfy a constraint $16 \alpha +4 \beta +\gamma =0$.

It is straightforward to obtain an expression for the R-sector partition function with help of the \eqref{fermionization01}. Likewise the NS-sector, the R-sector partition function involves four primaries. Their characters take the form of
\begin{align}
\label{RcharacterSO6}
\begin{split}
f_0^{R}(q) &= \chi^{\mathfrak{a}_{3,3}}_{h=9/8} + \chi^{\mathfrak{a}_{3,3}}_{h=9/8}= q^{\frac{6}{7}} \left( 40 + 360 q + 2232 q^2 + 10640 q^3 +  \cdots \right),\\
f_1^{R}(q) &= \chi^{\mathfrak{a}_{3,3}}_{h=15/56} + \chi^{\mathfrak{a}_{3,3}}_{h=71/56} =   4 + 120 q + 1080 q^2 + 6360 q^3 + 29400 q^4+ \cdots ,\\
f_2^{R}(q) &= \chi^{\mathfrak{a}_{3,3}}_{h=55/56} + \chi^{\mathfrak{a}_{3,3}}_{h=55/56}= q^{\frac{5}{7}} \left( 72 + 760 q + 4920 q^2 + 24120 q^3 +\cdots \right),\\
f_3^{R}(q) &=\chi^{\mathfrak{a}_{3,3}}_{h=39/56} + \chi^{\mathfrak{a}_{3,3}}_{h=39/56} = q^{\frac{3}{7}} \left( 40 + 600 q + 4320 q^2 + 22600 q^3 + \cdots \right).
\end{split}
\end{align}
One can see that the conformal weight $h_1^R$ indeed saturates the unitarity bound, as desired.

Our next goal is to compute the characters of a dual theory $\widetilde{\mathcal{F}}^{NS}(6)$ that are combined with \eqref{NScharacterSO6} to produce $K(\tau)$. Therefore, the NS-sector of a dual theory is expected to be contributed from four primaries of weight $h^{NS}=\frac{5}{14},\frac{3}{7},\frac{9}{14}$. The strategy is to look for a bosonic RCFT $\widetilde{\mathcal{B}}$ that is related to $\widetilde{\mathcal{F}}^{NS}(6)$ via \eqref{fermionization01}. We assume that $\widetilde{\mathcal{B}}$ has central charge $c=12-39/7$ and involves 13 primaries of conformal weights
\begin{align}
h = \left\{0,\frac{3}{2},\frac{9}{14},\frac{8}{7},\frac{3}{7},\frac{13}{14},\frac{5}{14},\frac{6}{7}, \frac{3}{8},\frac{13}{56},\frac{45}{56},\frac{29}{56},\frac{69}{56} \right\}.
\end{align}
Furthermore, we demand that characters of the above 13 primaries share the same $S$-matrix with the $SU(4)_3$ WZW model. Now we demand the characters of dual theory paired with \eqref{NScharacterSO6} to form a bilinear relation
\begin{align}
    {f}^{NS}_0(q) \tilde{f}^{NS}_0(q) +{f}^{NS}_1(q) \tilde{f}^{NS}_1(q) + {f}^{NS}_2(q)  \tilde{f}^{NS}_2(q) + 2\tilde{f}^{NS}_3(q) \tilde{f}^{NS}_3(q)  = K(\tau)\,.
\end{align}
We combine the above bilinear relation with \eqref{Dual L4} to find the NS-sector characters of $\widetilde{\mathcal{F}}^{NS}(6)$. As a consequence, we obtain
\begin{align}
\begin{split}
\tilde{f}^{NS}_0(q) &= q^{-13/56} (1 + 9 q + 32 q^{\frac{3}{2}}+  84 q^2 + 192 q^{\frac{5}{2}} + 429 q^3 + 936 q^{\frac{7}{2}} +1899 q^4 + \cdots ),\\
\tilde{f}^{NS}_1(q) &= q^{23/56} (12 + 39 q^{\frac{1}{2}} + 108 q + 287 q^{\frac{3}{2}} + 702 q^2 + 1542 q^{\frac{5}{2}}+ 3177 q^3 + \cdots)\\
\tilde{f}^{NS}_2(q) &= q^{11/56} (8 + 30 q^{\frac{1}{2}} +  102 q + 286 q^{\frac{3}{2}} + 702 q^2 + 1584 q^{\frac{5}{2}} + 3385 q^3 + \cdots ),\\
\tilde{f}^{NS}_3(q) &= q^{1/8} (3 + 13 q^{\frac{1}{2}} + 51 q + 141 q^{\frac{3}{2}} + 338 q^2 + 780 q^{\frac{5}{2}} + 1701 q^3 + \cdots ).
\end{split}
\end{align}
In a similar way, the R-sector dual characters can be obtained. From the R-sector bilinear relation
\begin{align}
    {f}^{R}_0(q) \tilde{f}^{R}_0(q) +{f}^{R}_1(q) \tilde{f}^{R}_1(q) + {f}^{R}_2(q)  \tilde{f}^{R}_2(q) + 2\tilde{f}^{R}_3(q) \tilde{f}^{R}_3(q)  = K\bigg(-\frac{1}{\tau} + 1\bigg)\,,
\end{align}
we find that the R-sector dual characters are given by
\begin{align}
\begin{split}
\tilde{f}^{R}_0(q) &= q^{1/7} (2+48 q+312 q^2+1544 q^3 + \cdots ),\\
\tilde{f}^{R}_1(q) &= 3+74 q+552 q^2+2808 q^3 +11526 q^4 + \cdots ,\\
\tilde{f}^{R}_2(q) &= q^{2/7} (12+174 q+1126 q^2+5340 q^3 + \cdots ),\\
\tilde{f}^{R}_3(q) &= q^{4/7} (46+444 q+2604 q^2+11586 q^3 + \cdots ).\\
\end{split}
\end{align}

\subsubsection*{SO$(7)_3$ WZW and $\widetilde{\mathcal{F}}(7)$}

We now focus on the level-three SO$(7)$ WZW model with $c=63/8$ and its dual theory. SO$(7)_3$ WZW model has 13 primaries whose conformal weights are given by
\begin{align}
h = \left\{0,\frac{3}{2},\frac{21}{64},\frac{85}{64},\frac{3}{4},\frac{5}{4},\frac{81}{64},\frac{5}{8},\frac{9}{8},\frac{69}{64},\frac{3}{8},\frac{7}{8},\frac{49}{64} \right\}.
\end{align}
Specifically, the Verlinde line with primary of $h=3/2$ is associated with the $\mathbb{Z}_2$ symmetry. After applying the generalized Jordan-Wigner transformation, one can find that the NS-sector partition function consists of the below four characters,
\begin{align}
\begin{split}
f_0^{NS}(q)  &= \chi^{\mathfrak{b}_{3,3}}_{h=0} + \chi^{\mathfrak{b}_{3,3}}_{h=3/2} = q^{-\frac{21}{64}} \left( 1 + 21 q + 77 q^{\frac{3}{2}} + 252 q^2 + 798 q^{\frac{5}{2}} + \cdots \right),\\
f_1^{NS}(q) &= \chi^{\mathfrak{b}_{3,3}}_{h=3/4} + \chi^{\mathfrak{b}_{3,3}}_{h=5/4} = q^{\frac{3}{64}} \left( 7 + 27 q^{\frac{1}{2}} + 147 q + 567 q^{\frac{3}{2}} + 1764 q^2 + \cdots \right),\\
f_2^{NS}(q) &= \chi^{\mathfrak{b}_{3,3}}_{h=5/8} + \chi^{\mathfrak{b}_{3,3}}_{h=9/8} = q^{\frac{19}{64}} \left( 21 + 105 q^{\frac{1}{2}} + 441 q + 1512 q^{\frac{3}{2}} + 4467 q^2 + \cdots \right),\\
f_3^{NS}(q) &= \chi^{\mathfrak{b}_{3,3}}_{h=3/8} + \chi^{\mathfrak{b}_{3,3}}_{h=7/8} = q^{\frac{27}{64}} \left( 35 + 189 q^{\frac{1}{2}} + 735 q + 2352 q^{\frac{3}{2}} + 6741 q^2 + \cdots \right),
\end{split}
\end{align}
where their ${S}$-matrix is given by
\begin{align}
S = \frac{1}{\sqrt{2}} 
\left(
\begin{array}{cccc}
 \sin\left( \frac{\pi}{16} \right)    & \sin\left( \frac{3\pi}{16} \right) & \cos\left( \frac{3\pi}{16} \right) & \cos\left( \frac{\pi}{16} \right)\\
 \sin\left( \frac{3\pi}{16} \right)    & \cos\left( \frac{\pi}{16} \right) & \sin\left( \frac{\pi}{16} \right) & -\cos\left( \frac{3\pi}{16} \right)\\
 \cos\left( \frac{3\pi}{16} \right)    & \sin\left( \frac{\pi}{16} \right) & -\cos\left( \frac{\pi}{16} \right) & \sin\left( \frac{3\pi}{16} \right) \\
 \cos\left( \frac{\pi}{16} \right)   & -\cos\left( \frac{3\pi}{16} \right) & \sin\left( \frac{3\pi}{16} \right) & -\sin\left( \frac{\pi}{16} \right)
\end{array}
\right).
\end{align}
Note that the above ${S}$-matrix is a symmetric matrix and provides consistent fusion rule algebra coefficients. We further verify that the four characters $f^{NS}_i(q)$ satisfy a fourth-order MLDE of the form
	\begin{equation}
	\begin{split}
		&\Bigg[\mathcal{D}^4+\frac{9}{16}e_2\mathcal{D}^3+\left(\frac{471}{2048}e_2^2-\frac{1597}{4608}E_4\right)\mathcal{D}^2+\left(\frac{22127}{1769472}e_2^3+\frac{1603}{147456}e_2 E_4\right)\mathcal{D}\\
		&-\frac{903}{16777216}e_2^4+\frac{1911}{2097152}e_2^2 E_4-\frac{2919}{1048576}E_4^2\Bigg]f^{NS}_i(q)=0.
	\end{split}
	\end{equation}

On the one hand, the R-sector partition function of the fermionized SO$(7)_3$ WZW model reads
\begin{align}
\begin{split}
Z^{R} = \left|f_0^R(q)\right|^2 + \left|\sqrt{2}f_1^R(q)\right|^2 + \left|\sqrt{2}f_2^R(q)\right|^2 + \left|\sqrt{2}f_3^R(q)\right|^2,
\end{split}
\end{align}
with four characters having the following $q$-expansion.
\begin{align}
\begin{split}
f_0^{R}(q) &= \chi^{\mathfrak{b}_{3,3}}_{h=21/64} + \chi^{\mathfrak{b}_{3,3}}_{h=85/64} = 8 + 336 q + 4032 q^2 + 29568 q^3 + 164640 q^4+ \cdots \\
f_1^{R}(q) &= \chi^{\mathfrak{b}_{3,3}}_{h=81/64} = q^{\frac{15}{16}} \left( 112  + 1344 q + 10080 q^2 + 56944 q^{3}+266784 q^4 +  \cdots \right),\\
f_2^{R}(q) &= \chi^{\mathfrak{b}_{3,3}}_{h=69/64} = q^{\frac{19}{64}} \left( 112  + 1632 q + 13104 q^2 + 77280 q^{3}+373296 q^4 +  \cdots \right),\\
f_3^{R}(q) &= \chi^{\mathfrak{b}_{3,3}}_{h=49/64} = q^{\frac{27}{64}} \left( 48  + 1008 q + 9296 q^2 + 59472 q^{3}+303408 q^4 +  \cdots \right).
\end{split}
\end{align}

To discuss the dual theory $\widetilde{\mathcal{F}}(7)$, we first consider the dual bosonic theory of SO$(7)_3$ WZW model. By dual bosonic theory, we mean the bosonic RCFT whose central charge and conformal weights of primaries fulfill the conditions 
\begin{align}
\label{dual condition}
c + \tilde{c} = 12, \quad h_i + \tilde{h}_i \in \mathbb{N}/2,
\end{align}
where $\tilde{c}$ and $\tilde{h}_i$ denote the central charge and conformal weights of a putative dual theory. With the assumption that the characters of the dual bosonic theory are governed by the same $S$-matrix of the SO$(7)_3$ WZW model, one can exploit the Rademacher expansion. After some computation with the Rademacher expansion formula \eqref{eq:redemacherexp}, we find that the characters of the dual bosonic theory have the following $q$-expansions,
\begin{align}
\begin{split}
\tilde{f}_{0}(q) &= q^{-11/64} (1 + 3 q + 24 q^2 + 82 q^3 + 313 q^4 + \cdots ), \\
\tilde{f}_{\frac{3}{2}}(q) &= q^{85/64} (10 + 45 q + 165 q^2 + \cdots ), \\
\tilde{f}_{\frac{1}{4}}(q) &= q^{5/64} (3  + 33 q + 154 q^2 + 612 q^3 + \cdots), \\
\tilde{f}_{\frac{3}{4}}(q) &= q^{37/64} (10  + 75 q + 315 q^2 + 1114 q^3 + \cdots ), \\
\tilde{f}_{\frac{3}{8}}(q) &= q^{13/64} (5 + 30 q + 165 q^2 + 595 q^3 + \cdots ), \\
\tilde{f}_{\frac{7}{8}}(q) &= q^{45/64} (12 + 76 q + 318 q^2 + 1089 q^3 + \cdots ), \\
\tilde{f}_{\frac{5}{8}}(q) &= q^{29/64} (6  + 33 q + 154 q^2 + 536 q^3 + \cdots ), \\
\tilde{f}_{\frac{9}{8}}(q) &= q^{61/64} (15 + 72 q + 297 q^2 + 961 q^3 + \cdots ), \\
\tilde{f}_{\frac{11}{64}}(q) &= q^0(3 + 19 q+102 q^2+396 q^3+ \cdots), \\
\tilde{f}_{\frac{75}{64}}(q) &= q(19  + 102 q + 396 q^2 + 1298 q^3 + \cdots),\\
\tilde{f}_{\frac{15}{64}}(q) &= q^{1/16}(1 + 18 q + 84 q^2 + 327 q^3 + \cdots),\\
\tilde{f}_{\frac{27}{64}}(q) &= q^{1/4}(5 + 42 q + 207 q^2 + 766 q^3 + \cdots),\\
\tilde{f}_{\frac{47}{64}}(q) &= q^{9/16}(15 + 95 q + 414 q^2 + 1452 q^3 + \cdots).\\
\end{split}
\end{align}
Since the $S$-matrix of the above characters agrees with that of the SO$(7)_3$ WZW model, one can readily see that the $\mathbb{Z}_2$ symmetry is generated by $\mathcal{L}_{h=3/2}$. The four NS-sector characters can be computed by applying \eqref{fermionization01},
\begin{align}
\label{NS dual so7}
\begin{split}
\tilde{f}^{NS}_0(q) &= \tilde{f}_0(q) + \tilde{f}_{\frac{3}{2}}(q),  \quad \tilde{f}^{NS}_1(q)  = \tilde{f}_{\frac{1}{4}}(q) + \tilde{f}_{\frac{3}{4}}(q), \\
\tilde{f}^{NS}_2(q) &= \tilde{f}_{\frac{3}{8}}(q) + \tilde{f}_{\frac{7}{8}}(q), \quad \tilde{f}^{NS}_3(q) = \tilde{f}_{\frac{5}{8}}(q) + \tilde{f}_{\frac{9}{8}}(q).
\end{split}
\end{align}
In a similar way, we check that the R-sector partition function consists of the following four characters.
\begin{align}
\label{R dual so7}
\begin{split}
\tilde{f}^{R}_0(q) &= \tilde{f}_{\frac{11}{64}}(q) + \tilde{f}_{\frac{75}{64}}(q),  \quad \tilde{f}^{R}_1(q)  = \tilde{f}_{\frac{15}{64}}(q), \quad
\tilde{f}^{R}_2(q) = \tilde{f}_{\frac{27}{64}}(q), \quad \tilde{f}^{R}_3(q) = \tilde{f}_{\frac{47}{64}}(q).
\end{split}
\end{align}
A fermionic RCFT with $c=33/8$ indeed can be considered as the dual theory of the fermionized SO$(7)_3$ WZW model in that their characters satisfy following bilinear relations,
\begin{align}
\begin{split}
&K(\tau) = {f}^{NS}_0(q)  \tilde{f}^{NS}_0(q) + {f}^{NS}_1(q)  \tilde{f}^{NS}_1(q) + {f}^{NS}_2(q)  \tilde{f}^{NS}_2(q) + {f}^{NS}_3(q)  \tilde{f}^{NS}_3(q), \\
&K(-\frac{1}{\tau}+1) = {f}^{R}_0(q)  \tilde{f}^{R}_0(q) + 2{f}^{R}_1(q)  \tilde{f}^{R}_1(q) + 2{f}^{R}_2(q)  \tilde{f}^{R}_2(q) + 2{f}^{R}_3(q)  \tilde{f}^{R}_3(q).
\end{split}
\end{align}

We finally remark that the characters \eqref{NS dual so7} and \eqref{R dual so7} arose as the solutions of the fourth-order MLDE with $\tilde{\ell}=2$, due to the valence formula.

\section*{Acknowledgment} 
ZD would like to thank Kyung Hee University for hospitality during the completion of this manuscript. The work of J.B. is supported in part by the European Research Council(ERC) under the European Union's Horizon 2020 research and innovation programme (Grant No. 787185).
Z.D., K.L., S.L., and M.S. are supported in part by KIAS Individual Grant PG076901, PG006904, PG056502, and PG064201, respectively. K.L. is also supported in part by Korea National Research Foundation Individual Grant NRF-2017R1D1A1B06034369.

\appendix

\section{Generalized Hypergeometric Equation and Exact Solutions}\label{app:smatrix}

Here we derive the solution of the differential equation \eqref{eq2:hypergeomEq}. We show that the solution has the form of the generalized hypergeometric function of order three for $h^\text{NS}_1\neq \tfrac{1}{2}$ and $h^\text{NS}_2\neq \tfrac{1}{2}$. Otherwise, the solution is given by regular hypergeometric function when one of the primaries has conformal weight $\frac{1}{2}$. Once we have the solution in terms of the hypergeometric function, the structure of the $S$-matrix can be computed with help of the monodromy matrix.

\subsubsection*{Solving the hypergeometric equation}

\paragraph{Case 1: $h^\text{NS}_1\neq \tfrac{1}{2}$ and $h^\text{NS}_2\neq \tfrac{1}{2}$}

We start with the differential equation of the form
\begin{align}
	\left[\left(z \frac{d}{dz}\right)^3+\left(b_2+\frac{3}{2(z-1)}\right)\left(z \frac{d}{dz}\right)^2+\left(b_{11}+\frac{b_{12}}{z-1}\right)\left(z \frac{d}{dz}\right)+b_0\left(1+\frac{1}{z-1}\right)\right]\tilde{f}^\textsc{ns}_i=0\,,
\end{align}
where the expressions of coefficients are given in \eqref{coeff of ODE}. Let us multiply the factor $1-z$ to the above differential equation, to express it in the form of Beukers-Heckman \cite{Beukers1989}:
\begin{equation}
\label{BH equation}
	\left\{\left(\theta+\beta_0-1\right)\left(\theta+\beta_1-1\right)\left(\theta+\beta_2-1\right)-z\left(\theta+\alpha_0\right)\left(\theta+\alpha_1\right)\left(\theta+\alpha_2\right)\right\} \tilde{f}^\textsc{ns}_i=0\,,
\end{equation}
with $\theta = z\frac{d}{dz}$ and
\begin{equation}
\label{eq:HyperParametersGeneric}
\begin{split}
	&\alpha_0=-\frac{c}{12}\,,\quad \alpha_1=\frac{c}{24}+\frac{3}{4}-h^\textsc{ns}_1-h^\textsc{ns}_2-\frac{h^\textsc{r}_1-h^\textsc{r}_2}{2}\,, \quad \beta_0=1,\\
	&\alpha_2=\frac{c}{24}+\frac{3}{4}-h^\textsc{ns}_1-h^\textsc{ns}_2+\frac{h^\textsc{r}_1-h^\textsc{r}_2}{2}\,, \quad \beta_1=1-2h^\textsc{ns}_1\,, \quad \beta_2=1-2h^\textsc{ns}_2\,.
\end{split}
\end{equation}
Here we set $h^\textsc{ns}_0=0$ and $h^\textsc{r}_0=c/24$. Furthermore, one can plug (\ref{eq:3rdOrderRamondWeights}) into \eqref{eq:HyperParametersGeneric} so that the coefficients $\alpha_i$ and $\beta_i$ are written in terms of $c, h_1^{\text{NS}}, h_2^{\text{NS}}$ only.
The solution of \eqref{BH equation} is known to be the generalized hypergeometric function of order three, and an explicit expression is given below,
\begin{equation}
	\tilde{f}^\textsc{ns}_i=z^{1-\beta_i} \left._3F_2\right.(1+\alpha_0-\beta_i,1+\alpha_1-\beta_i,1+\alpha_2-\beta_i;1+\beta_0-\beta_i,\hat{\dots},1+\beta_2-\beta_i;z)\,.
\end{equation}
In the above expression, the hat means omission of the $1+\beta_i-\beta_i$ parameter.
Tracking back the various changes of variables, and fixing the normalization, we obtain
\begin{equation}
\begin{split}
	&f^\textsc{ns}_i=2^{\frac{c}{3} - 12 h^\textsc{ns}_i}\lambda^{-\frac{c}{12}}(1-\lambda)^{-\frac{c}{12}}\left(4\lambda(1-\lambda)\right)^{1-\beta_i} \\
	&\left._3F_2\right.(1+\alpha_0-\beta_i,1+\alpha_1-\beta_i,1+\alpha_2-\beta_i;1+\beta_0-\beta_i,\hat{\dots},1+\beta_2-\beta_i;4\lambda(1-\lambda))\,,
\end{split}
\end{equation}
as the exact expression for the NS-sector characters.

\paragraph{Case 2: $h^\text{NS}_1= \tfrac{1}{2}$}

In this singular case, the situation is slightly different. We see that the coefficient $b_0$ of \eqref{coeff of ODE} vanishes since $h^\text{NS}_1= \tfrac{1}{2}$. By introducing a new variable, 
\begin{equation}
	\xi_i:=\theta \tilde{\psi}^{\text{NS}}_{i},
\end{equation}
one can reduce the order of differential equation by one as follows,
\begin{equation}
\label{eq:hypergeomEq}
	\left[\theta^2+\left(b_2+\frac{3}{2(z-1)}\right)\theta+b_{11}+\frac{b_{12}}{z-1}\right]\xi_i=0\,.
\end{equation}
Now we multiply \eqref{eq:hypergeomEq} by $1-z$ to get
\begin{equation}
	\left\{\left(\theta+\beta_1-1\right)\left(\theta+\beta_2-1\right)-z\left(\theta+\alpha_1\right)\left(\theta+\alpha_2\right)\right\}\xi_i=0\,,
\end{equation}
with
\begin{equation}
\label{coeff a b}
	\begin{split}
		&\alpha_{1,2}= - h^\textsc{ns}_2   +\frac{1}{4}\\
		&\mp\frac{1}{12} \sqrt{c^2+27-(24 c +108)( \frac{1}{2} +h^\textsc{ns}_2)  +18 c+144 ( \frac{1}{4}+(h^\textsc{ns}_2) ^2)+36  h^\textsc{ns}_2}\,,\\
		&\beta_1=0, \quad \beta_2=1-2h^\textsc{ns}_2.
	\end{split}
\end{equation}
The three linearly independent solutions of \eqref{eq:hypergeomEq} are given by:
\begin{equation}
\label{int eqtn}
	\begin{split}
    	\tilde\psi^0&=\text{const.}\,,\\
		\tilde\psi^1&=\int^z \frac{\text{d}t}{t^{\beta_1}} \left._2F_1\right.(1+\alpha_1-\beta_1,1+\alpha_2-\beta_1;1+\beta_2-\beta_1;t)\,,\\
		\tilde\psi^2&=\int^z \frac{\text{d}t}{t^{\beta_2}} \left._2F_1\right.(1+\alpha_1-\beta_2,1+\alpha_2-\beta_2;1+\beta_1-\beta_2;t)\,.
	\end{split}
\end{equation}
By substituting \eqref{coeff a b} into \eqref{int eqtn}, one obtains
\begin{equation}
	\begin{split}
    	\tilde\psi^{\text{NS}}_0&=\text{const.}\,,\\
		\tilde\psi^{\text{NS}}_1&=\int^z \text{d}t \left._2F_1\right.\left(1-\frac{c}{12},\frac{3}{2}+\frac{c}{12}-2h;2-2h;t\right)\,, \\
		\tilde\psi^{\text{NS}}_2&=\int^z \frac{\text{d}t}{t^{1-2h}}\left._2F_1\right.\left(-\frac{c}{12}+2h,\frac{1}{2}+\frac{c}{12};2h;t\right)\,.
	\end{split}
\end{equation}
Then we arrive at the following three solutions
\begin{equation}
\label{eq:psiBasis}
	\begin{split}
    	\psi^{\text{NS}}_0&=2^\frac{c}{3}\lambda^{-\frac{c}{12}}(1-\lambda)^{-\frac{c}{12}}\,,\\
		\psi^{\text{NS}}_1&=2^{\frac{c}{3}-6}\frac{144 (2 h-1) }{c (c-24 h+6)}\,\lambda^{-\frac{c}{12}}(1-\lambda)^{-\frac{c}{12}}\left[\, _2F_1\left(-\frac{c}{12},\frac{c-24 h+6}{12} ;1-2 h;4\lambda(1-\lambda)\right)-1\right]\,,\\
		\psi^{\text{NS}}_2&=2^{\frac{c}{3}-8h}\,\lambda^{-\frac{c}{12}+2h}(1-\lambda)^{-\frac{c}{12}+2h}\, _2{F}_1\left(\frac{c+6}{12},2 h-\frac{c}{12};2 h+1;4\lambda(1-\lambda)\right)\,.
	\end{split}
\end{equation}
Let us perform the following linear combinations:
\begin{equation}
    \begin{split}
        &f^\textsc{ns}_0:=\psi^{\text{NS}}_0-2c\, \psi^{\text{NS}}_1\,,\\
        &f^\textsc{ns}_1:=\psi^{\text{NS}}_2\,,\\
        &f^\textsc{ns}_2:=\psi^{\text{NS}}_1\,,
    \end{split}
\end{equation}
Finally we obtain the closed-form expression of the characters in the case of $h^{\text{NS}}_1=1/2$. By denoting  $h^{\text{NS}}_2=h$, the analytic expressions are given by
\begin{equation}
	\begin{split}
    	f^\textsc{ns}_0&=2^\frac{c}{3}\lambda^{-\frac{c}{12}}(1-\lambda)^{-\frac{c}{12}}\left[1-\frac{9 (2 h-1) }{ 2(c-24 h+6)}\left[\, _2F_1\left(-\frac{c}{12},\frac{c-24 h+6}{12} ;1-2 h;4\lambda(1-\lambda)\right)-1\right]\right],\\
		f^\textsc{ns}_1&=2^{\frac{c}{3}-8h}\,\lambda^{-\frac{c}{12}+2h}(1-\lambda)^{-\frac{c}{12}+2h}\, _2{F}_1\left(\frac{c+6}{12},2 h-\frac{c}{12};2 h+1;4\lambda(1-\lambda)\right),\\
		f^\textsc{ns}_2&=\frac{9 (2 h-1)2^{\frac{c}{3}} }{4c (c-24 h+6)}\,\lambda^{-\frac{c}{12}}(1-\lambda)^{-\frac{c}{12}}\left[\, _2F_1\left(-\frac{c}{12},\frac{c-24 h+6}{12} ;1-2 h;4\lambda(1-\lambda)\right)-1\right].
	\end{split}
\end{equation}

\subsubsection*{Monodromy of the hypergeometric ODE and $S$-matrix}

We see that since each of the three characters depends solely on the parameter $$\lambda(1-\lambda)=\frac{16}{K+24}\,,$$
which is a function of the hauptmodul for $\Gamma_\vartheta$, hence is invariant under $S$ transformations. One might naively conclude that the modular ${S}$-matrix is simply an identity matrix. However this is surely wrong, and one should invoke monodromy arguments in order to show it. Once again, we split the derivation into two cases.

\paragraph{Case 1: $h^\text{NS}_1\neq \tfrac{1}{2}$ and $h^\text{NS}_2\neq \tfrac{1}{2}$}
	
Let us suppose that one is given a basis of local solutions in the solution space $V$ of the hypergeometric equation in a small neighborhood around a point $z_0\in\mathbb P^1\backslash\{0,1,\infty\}$. When analytically continuing the solution along a loop $\gamma_{z_\star}$ encircling one of the regular singularities $z_\star\in\{0,1,\infty\}$ in the anti-clockwise direction, the solution only comes back to itself up to a linear transformation $\text{M}(\gamma_{z_\star})\in\text{GL}(V)$:
	\begin{equation}
		X\mapsto \text{M}(\gamma_{z_\star})X\,.
	\end{equation}
	The fundamental group $\pi_1(\mathbb P^1\backslash\{0,1,\infty\})$ is generated by $\gamma_0$, $\gamma_1$ and $\gamma_\infty$, subject to the relation
	\begin{equation}
	\label{eq:homotopyRelation}
		\gamma_0\gamma_1\gamma_\infty=1\,.
	\end{equation} 
	We denoted by 
	\begin{equation}
		\text{M}:\pi_1(\mathbb P^1\backslash\{0,1,\infty\})\to\text{GL}(V)\,.
	\end{equation}
the monodromy representation defined by the hypergeometric equation. One can show\footnote{We refer the reader to some unpublished notes of Beukers on differential equations. Levelt in his Ph.D. thesis already provided this form of the matrices, but didn't construct explicitly the basis, hence providing the monodromy matrices only up to conjugation.} that in the so-called Mellin-Barnes basis of solutions to the hypergeometric equation, one has the following explicit form of the monodromy matrices:
	\begin{equation}
		\begin{split}
			\text{M}(\gamma_0)&=\begin{pmatrix}
				0&1&0\\0&0&1\\-B_3&-B_2&-B_1
			\end{pmatrix}\,,\\
			\text{M}(\gamma_\infty)^{-1}&=\begin{pmatrix}
				0&1&0\\0&0&1\\-A_3&-A_2&-A_1
			\end{pmatrix}\,.
		\end{split}
	\end{equation}
	These are the so-called companion matrices associated to the polynomials:
	\begin{equation}
		\begin{split}
			&\prod_{n=0}^2\left(t-e^{-2i\pi\beta_k}\right)=t^3+B_1t^2+B_2t+B_3\,,\\
			&\prod_{n=0}^2\left(t-e^{-2i\pi\alpha_k}\right)=t^3+A_1t^2+A_2t+A_3\,,\\
		\end{split}
	\end{equation}
where we recall that the hypergeometric parameters are fixed in terms of the RCFT data by (\ref{eq:HyperParametersGeneric}).
	From (\ref{eq:homotopyRelation}) we then deduce that:
	\begin{equation}
	\label{eq:monod1}
		\text{M}(\gamma_1)=\text{M}(\gamma_0)^{-1}\text{M}(\gamma_\infty)^{-1}\,.
	\end{equation}
Following the loop around $z=0$ counter-clockwise once corresponds to the transformation $\tau\to\tau+2$ since $z=q^{1/2}(1+\mathcal O(q^{1/2}))$, and therefore the monodromy matrix around $z=0$ corresponds to the representation of $T^2\in\Gamma_\vartheta$ on the solution space. We also know that $z=1\Leftrightarrow\lambda=\tfrac{1}{2}\Leftrightarrow\tau=i$, which is stabilized by $S$ transformations, and therefore the monodromy matrix around $z=1$ is conjugate to the representation of $S\in\Gamma_\vartheta$. Moreover, one can show from (\ref{eq:monod1}) that $\text{rank}(\text{M}(\gamma_1)-\text{id})=1$, therefore $\text{M}(\gamma_1)$ must have two eigenvalues equal to $1$, as a pseudo-reflection. In our CFT case we actually  know a little bit more from representation theory of $\text{SL}_2(\mathbb Z)$, namely that the $T$ and $S$ matrices are subject to the relations (\ref{eq:modularRelations}).
	
One should also recall that in order to obtain the hypergeometric equation, we performed a rescaling of the solutions by a factor proportional to $z^{-c/12}$, which contributes a phase $e^{-i\pi c/6}$ to the monodromy at $0$, and $e^{i\pi c/6}$ to the monodromy at $\infty$. These contributions compensate in the monodromy at $1$. 
	
Our basis of solutions is such that $T^2$ is diagonal, and is obtained from the Mellin-Barnes basis by diagonalizing the monodromy matrix around $z=0$:
	\begin{equation}
		\text{M}(\gamma_0)=P_1\begin{pmatrix}
			1&0&0\\0&e^{4i\pi h^\textsc{ns}_1}&0\\0&0&e^{4i\pi h^\textsc{ns}_2}
		\end{pmatrix}P_1^{-1}\,.
	\end{equation}
The matrix $P_1$ simply turns out to be a Vandermonde-like matrix:
	\begin{equation}
		P_1=\begin{pmatrix}
			1&1&1\\
			1&e^{4i\pi h^\textsc{ns}_1}&e^{4i\pi h^\textsc{ns}_2}\\
			1&e^{8i\pi h^\textsc{ns}_1}&e^{8i\pi h^\textsc{ns}_2}
		\end{pmatrix}\,.
	\end{equation}
Taking into account the extra phase factor, we find a diagonal $T^2$ matrix
	\begin{equation}
	\label{eq:Tmatrix}
		T^2=\begin{pmatrix}
			e^{-2i\pi \frac{c}{12}}&0&0\\0&e^{2i\pi\left(-\frac{c}{12}+2h^\textsc{ns}_1\right)}&0\\0&0&e^{2i\pi\left(-\frac{c}{12}+2h^\textsc{ns}_2\right)}
		\end{pmatrix}\,,
	\end{equation}
as expected. In the same basis, the monodromy matrix around $z=1$ reads:
	\begin{equation}
		P_1^{-1}\text{M}(\gamma_1)P_1\,,
	\end{equation}
which corresponds to the ${S}$-matrix up to conjugation by a matrix commuting with (\ref{eq:Tmatrix}). the extra conjugation matrix $P_2$ needed to reach our Frobenius basis of solutions nearby $z=0$ is simply diagonal, and is given by the proposition 2.7 of \cite{molag2017monodromy}. The $k^\text{th}$ diagonal entry reads:
	\begin{equation}
		(P_2)_{kk}=-\frac{(-1)^{3\beta_k}}{4}\frac{\Gamma(1+\alpha_0-\beta_k)\Gamma(1+\alpha_1-\beta_k)\Gamma(1+\alpha_2-\beta_k)}{\Gamma(1+\beta_0-\beta_k)\Gamma(1+\beta_1-\beta_k)\Gamma(1+\beta_2-\beta_k)}\prod_{\substack{i=0,\dots,2\\i\neq k}}\frac{1}{\sin(\pi(\beta_i-\beta_k))}\,.
	\end{equation}
One should also take into account the extra relative normalizing factors $2^{-12h^\textsc{ns}_i}$ in our solution basis by conjugating with
	\begin{equation}
	    P_3:=\begin{pmatrix}1&0&0\\0&2^{12h^\textsc{ns}_1}&0\\0&0&2^{12h^\textsc{ns}_2}\end{pmatrix}\,.
	\end{equation}
To summarize, the representation or the $S$-matrix on our basis of characters therefore reads:
	\begin{equation}
	\label{eq:nonDegSmatrix}
     S=P_3^{-1}P_2^{-1}P_1^{-1}\text{M}(\gamma_1)P_1P_2P_3\,,
	\end{equation}
	which more explicitly takes the form (the indices $m,n\in\{0,1,2\}$):
	\begin{equation}
	\begin{split}
	    &S_{mn}=2^{12(h^\textsc{ns}_n-h^\textsc{ns}_m)}\times\\
	    &\times\left\{\delta_{mn}-2ie^{-i\pi\sum_{k}(\alpha_k-\beta_k)}\prod_{k}\frac{\Gamma(1+\alpha_k-\beta_n)\Gamma(1+\beta_k-\beta_m)}{\Gamma(1+\beta_k-\beta_n)\Gamma(1+\alpha_k-\beta_m)}\frac{\prod_{k}\sin\left(\pi(\alpha_k-\beta_n)\right)}{\prod_{\substack{k(\neq n)}}\sin\left(\pi(\beta_k-\beta_n)\right)}\right\}\,,
	\end{split}
	\end{equation}
where we recall that the hypergeometric parameters are fixed by the CFT data in (\ref{eq:HyperParametersGeneric}).\footnote{We correct a typo of theorem 2.8 of \cite{molag2017monodromy}.}
	
This concludes the non-degenerate case, namely the case of RCFTs for which none of the NS conformal weight is equal to $\tfrac{1}{2}$. One can of course check that the modular relations (\ref{eq:modularRelations}) are well-satisfied.
	
\paragraph{Case 2: $h^\text{NS}= \tfrac{1}{2}$}
	
Let us now turn to the degenerate case in which one of the NS conformal weight is equal to $\tfrac{1}{2}$. We recall that $h^\textsc{ns}_0=0$, and that we denote by simply by $h$ the remaining NS conformal weight. We already saw that in this case, one can effectively reduce the MLDE down to a second order ODE, so one may expect that a more direct, computational way of obtaining the $S$-matrix may be available in that case, in the spirit what was done in the second order MLDE case.
	
After adding to the second order ODE solutions the constant solution and tracking back the various change of variables, we obtained the set of solutions (\ref{eq:psiBasis}). We perform on it the following change of basis:
	\begin{equation}
		\begin{pmatrix}
			\pi^0\\ \pi^1\\ \pi^2
		\end{pmatrix}
		=
		\begin{pmatrix}
			1&0&0\\ \frac{9}{4}\frac{2h-1}{c(c-24+6)}&1&0\\ 0&0&1
		\end{pmatrix}
		\begin{pmatrix}
			\psi^0\\ \psi^1\\ \psi^2
		\end{pmatrix}
		=:
		O_2
		\begin{pmatrix}
			\psi^0\\ \psi^1\\ \psi^2
		\end{pmatrix}\,.
	\end{equation}
But we know that Gauss hypergeometric function satisfies the quadratic change of variable identity:
	\begin{equation}
		\left._2F_1\right.\left(a,b;a+b+\frac{1}{2};4\lambda(1-\lambda)\right)=\left._2F_1\right.\left(2a,2b;a+b+\frac{1}{2};\lambda\right)\,,
	\end{equation}
and we recognize of course immediately our $z$-variable on the left-hand-side. This identity allows us to the rewrite the characters in a way that breaks the symmetry $\lambda\leftrightarrow 1-\lambda$, hence allowing us to read off the $S$-matrix by following the same direct method as we did for the second order MLDE, by using Gauss identity and Euler transformation formulae. We compute:
	\begin{equation}
		\begin{pmatrix}
			\pi^0(1-\lambda)\\ \pi^1(1-\lambda)\\ \pi^2(1-\lambda)
		\end{pmatrix}
		=
		O_1
		\begin{pmatrix}
			\pi^0(\lambda)\\ \pi^1(\lambda)\\ \pi^2(\lambda)
		\end{pmatrix}\,,
	\end{equation}
	with
	\begin{equation}
		O_1=\begin{pmatrix}
			1&0&0\\0&\frac{\Gamma\left(1-2h\right)\Gamma\left(2h\right)}{\Gamma\left(1-2h+\frac{c}{6}\right)\Gamma\left(2h-\frac{c}{6}\right)}&2^{8h}\frac{9}{4}\frac{2h-1}{c(c-24h+6)}\frac{\Gamma\left(1-2h\right)\Gamma\left(-2h\right)}{\Gamma\left(-\frac{c}{6}\right)\Gamma\left(\frac{c}{6}-4h+1\right)}\\0&2^{-8h}\frac{4}{9}\frac{c(c-24h+6)}{2h-1}\frac{\Gamma\left(2h+1\right)\Gamma\left(2h\right)}{\Gamma\left(\frac{c}{6}+1\right)\Gamma\left(4h-\frac{c}{6}\right)}&\frac{\Gamma\left(2h+1\right)\Gamma\left(-2h\right)}{\Gamma\left(-\frac{c}{6}+2h\right)\Gamma\left(\frac{c}{6}-2h+1\right)}
		\end{pmatrix}\,.
	\end{equation}
We finally recall that our Frobenius basis of solutions was obtained by:
	\begin{equation}
		\begin{pmatrix}
			f_\textsc{ns}^0\\ f_\textsc{ns}^1\\ f_\textsc{ns}^2
		\end{pmatrix}
		=
		\begin{pmatrix}
			1&-2c&0\\ 0&0&1\\ 0&1&0
		\end{pmatrix}
		\begin{pmatrix}
			\psi^0\\ \psi^1\\ \psi^2
		\end{pmatrix}
		=:
		O_3
		\begin{pmatrix}
			\psi^0\\ \psi^1\\ \psi^2
		\end{pmatrix}\,.
	\end{equation}
We therefore obtain as a final result for the $S$-matrix in the degenerate case:
	\begin{equation}
		 S=O_3O_2^{-1}O_1O_2O_3^{-1}\,,
	\end{equation}
	or in full glory
	\begin{equation}
	\label{Smatrixhhalf}
\scalemath{0.8}{
 S=
\begin{pmatrix}
 \frac{9 (2 h-1)  }{2 (c-24 h+6)}\left[\frac{\sin \left(\frac{1}{6} \pi  (c-12 h)\right)}{\sin (2 \pi  h)}+\frac{2 c-30 h+3}{9 (2 h-1)}\right] &  2^{8 h-2}\frac{ 3\Gamma (2-2 h) \Gamma (-2 h)}{\Gamma \left(-\frac{c}{6}\right) \Gamma \left(\frac{c}{6}-4 h+2\right)} & \frac{c (2 c-30 h+3) }{c-24 h+6}\left[\frac{\sin \left(\frac{1}{6} \pi  (c-12 h)\right)}{\sin (2 \pi  h)}+1\right] \\
 2^{-8 h}\frac{ \Gamma (2 h) \Gamma (2 h+1)}{\Gamma \left(\frac{c}{6}+1\right) \Gamma \left(4 h-\frac{c}{6}\right)} &  \frac{\sin \left(\frac{1}{6} \pi  (c-12 h)\right)}{\sin (2 \pi  h)} & 2^{-8 h+2}(2 c-30 h+3)\frac{  \Gamma (2 h-1) \Gamma (2 h+1)}{ 3\Gamma \left(\frac{c}{6}\right) \Gamma \left(4 h-\frac{c}{6}\right)} \\
 -\frac{9 (2 h-1) }{4 c (c-24 h+6)}\left[\frac{\sin \left(\frac{1}{6} \pi  (c-12 h)\right)}{\sin (2 \pi  h)}+1\right] & 2^{8 h-4}\frac{ \Gamma (2-2 h) \Gamma (-2 h)}{\Gamma \left(1-\frac{c}{6}\right) \Gamma \left(\frac{c}{6}-4 h+2\right)} & \frac{(-2 c+30 h-3) }{2 (c-24 h+6)}\left[\frac{ \sin \left(\frac{1}{6} \pi  (c-12 h)\right)}{\sin (2 \pi  h)}+\frac{18 h-9}{2 c-30 h+3}\right]\,. \\
\end{pmatrix}
}
\end{equation}

\paragraph{Examples and comments:} The $S$-matrices that we derived above correspond to our Frobenius basis of solutions with the three characters being normalized so that their q-expansion starts with coefficient 1. Accommodating for integer degeneracies $m_1$ and $m_2$ in the non-vacuum characters $f^\textsc{ns}_1$ and $f^\textsc{ns}_2$ simply corresponds to performing an extra conjugation to the $S$-matrix by the matrix $\text{diag}(1,m_1,m_2)$.

Let us now consider for illustration a non-degenerate theory with both NS conformal weights different from $\tfrac{1}{2}$. We take for instance the theory $(c,h^\textsc{ns}_1,h^\textsc{ns}_2)=(18/5,3/10,2/5)$, identified in section \ref{sec:Identification II} as a tensor product of $\mathcal N=1$ minimal models, with degeneracies $(m_1,m_2)=(4,3)$. Our generic expression (\ref{eq:nonDegSmatrix}) gives the following result:
\begin{equation}
    S=\frac{1}{10}\begin{pmatrix}
 5-\sqrt{5} & \sqrt{5}+5 & 4 \sqrt{5} \\
 \sqrt{5}+5 & 5-\sqrt{5} & -4 \sqrt{5} \\
 2 \sqrt{5} & -2 \sqrt{5} & 2 \sqrt{5} \\
    \end{pmatrix}\,,
\end{equation}
which is indeed the correct result.

As an example with degenerate $h^{\text{NS}}$, let us consider the theories $(c,h^\textsc{ns}_1,h^\textsc{ns}_1)=(3m/2,m/8,1/2)$ corresponding to $\left(\text{SO}(m)_1\right)^{3}$, namely tensor products of the Ising model, discussed in section \ref{sec:Identification I}. With degeneracies $(m_1,m_2)=( 2^{m-1},m)$, we obtain the $S$-matrix:
\begin{equation}
     S=\frac{1}{4}\begin{pmatrix}
    1&6&3\\2&0&-2\\1&-2&3
    \end{pmatrix}
\end{equation}
which very nicely agrees with (\ref{eq:Smatrixc32}) upon exchanging $f^\textsc{ns}_1$ and $f^\textsc{ns}_2$, as expected.

We finally comment on the some solutions in  section \ref{tab:thirdOrderClassification} which cannot have the consistent fusion rules. As an illustrative example, let us consider the $S$-matrix of solution with $c=1$ and $h^{\rm{NS}}=\frac{1}{6}, \frac{1}{2}$ using \eqref{Smatrixhhalf}. It is given by 
\begin{equation}
\label{3rdSmatrixc1}
    S=\frac{1}{\sqrt{3}}\begin{pmatrix}
 1 & 2 & 0 \\
 1 & -1 & 0 \\
 \frac{-1+\sqrt{3}}{2} & -1 &  \sqrt{3}
    \end{pmatrix}
\end{equation}
and satisfies a relation 
\begin{align}
S^T \cdot \rm{diag}(1,2,0) \cdot S = \rm{diag}(1,2,0).
\end{align}
Therefore, the solution with $h^{\rm NS}=\frac{1}{2}$ does not contribute to the NS-sector partition function. One cannot construct the NS sector partition function contributed by three characters $f_0(\tau), f_{\frac{1}{6}}(\tau), f_{\frac{1}{2}}(\tau)$ altogether, since an extended matrix of \eqref{3rdSmatrixc1} cannot exist. Therefore, no consistent fusion rule can be obtained for hypothetical theory with $(c,h^{\rm{NS}}_1, h^{\rm{NS}}_2) = (1,\frac{1}{6}, \frac{1}{2})$. 

For the same reason, we cannot have consistent fusion rule algebra for the following five solutions of class IV:
\begin{align}
(c,h^{\rm{NS}}_1,h^{\rm{NS}}_2) = \left(\frac{3}{2}, \frac{1}{2}, \frac{1}{4}\right), \ \ \left(3, \frac{1}{2}, \frac{1}{3}\right), \ \ \left(9, \frac{1}{2}, \frac{2}{3}\right), \ \ \left(\frac{21}{2}, \frac{1}{2}, \frac{3}{4}\right), \ \ \left(11, \frac{1}{2}, \frac{5}{6}\right), \nonumber
\end{align}
thus they cannot be understood as the consistent two-dimensional RCFT.

\section{Classification of the third-order bosonic MLDE}
\label{app:bosonic_3MLDE_appendix}

\begin{table}[t]
    \centering
\scalebox{0.85}{
\begin{tabular}{c|c|c|c|c|c|c|c}
  No. & $(c,h_1,h_2,a_1)$ & No. & $(c,h_1,h_2,a_1)$ & No. & $(c,h_1,h_2,a_1)$ & No. & $(c,h_1,h_2,a_1)$ \\ \hline  
 1 & $(\frac{m}{2},\frac{1}{2},\frac{m}{16},\frac{m(m-1)}{2})$  & 25 & $(20,\frac{7}{5},\frac{8}{5},120)$ & 49 & $(14,\frac{3}{2},\frac{3}{4},266)$  & 73 & $(\frac{108}{5},\frac{4}{5},\frac{12}{5},1404)$  \\ [1mm]  
 2 & $(8,\frac{1}{2},1,\alpha)$  & 26 & $(\frac{41}{2},\frac{3}{2},\frac{25}{16},123)$ & 50 & $(10,\frac{3}{2},\frac{1}{4},270)$  & 74 & $(\frac{132}{5},\frac{3}{5},\frac{16}{5},1536)$  \\ [1mm]
 3 & $(16,\frac{3}{2},1,\alpha)$  & 27 & $(\frac{68}{5},\frac{4}{5},\frac{7}{5},136)$ & 51 & $(\frac{27}{2},\frac{3}{2},\frac{11}{16},270)$  & 75 & $(\frac{45}{2},\frac{5}{2},\frac{13}{16},1640)$  \\ [1mm]
 4 & $(\frac{47}{2},\frac{3}{2},\frac{31}{16},0)$  & 28 & $(\frac{36}{5},\frac{3}{5},\frac{4}{5},144)$ & 52 & $(\frac{21}{2},\frac{3}{2},\frac{5}{16},273)$  & 76 & $(\frac{116}{5},\frac{4}{5},\frac{13}{5},1711)$  \\ [1mm]
 5 & $(\frac{164}{5},\frac{11}{5},\frac{12}{5},0)$  & 29 & $(\frac{52}{7},\frac{4}{7},\frac{6}{7},156)$ & 53 & $(13,\frac{3}{2},\frac{5}{8},273)$  & 77 & $(23,\frac{5}{2},\frac{7}{8},2323)$  \\ [1mm]
 6 & $(\frac{236}{7},\frac{16}{7},\frac{17}{7},0)$  & 30 & $(\frac{39}{2},\frac{3}{2},\frac{23}{16},156)$ & 54 & $(11,\frac{3}{2},\frac{3}{8},275)$  & 78 & $(30,\frac{7}{2},\frac{3}{4},2778)$  \\ [1mm]
 7 & $(\frac{4}{7},\frac{1}{7},\frac{3}{7},1)$  & 31 & $(19,\frac{3}{2},\frac{11}{8},171)$ & 55 & $(\frac{25}{2},\frac{3}{2},\frac{9}{16},275)$  & 79 & $(\frac{61}{2},\frac{7}{2},\frac{13}{16},3599)$  \\ [1mm]
 8 & $(\frac{4}{5},\frac{1}{5},\frac{2}{5},2)$  & 32 & $(\frac{37}{2},\frac{3}{2},\frac{21}{16},185)$ & 56 & $(\frac{23}{2},\frac{3}{2},\frac{7}{16},276)$  & 80 & $(\frac{156}{5},\frac{4}{5},\frac{18}{5},3612)$  \\ [1mm]
 9 & $(\frac{12}{5},\frac{1}{5},\frac{3}{5},3)$  & 33 & $(18,\frac{3}{2},\frac{5}{4},198)$ & 57 & $(\frac{100}{7},\frac{5}{7},\frac{11}{7},325)$  & 81 & $(\frac{47}{2},\frac{5}{2},\frac{15}{16},4371)$  \\ [1mm]
 10 & $(\frac{12}{7},\frac{2}{7},\frac{3}{7},6)$  & 34 & $(\frac{60}{7},\frac{3}{7},\frac{8}{7},210)$ & 58 & $(\frac{84}{5},\frac{6}{5},\frac{7}{5},336)$  & 82 & $(31,\frac{7}{2},\frac{7}{8},5239)$  \\ [1mm]
 11 & $(23,\frac{3}{2},\frac{15}{8},23)$  & 35 & $(\frac{35}{2},\frac{3}{2},\frac{19}{16},210)$ & 59 & $(\frac{116}{7},\frac{8}{7},\frac{10}{7},348)$  & 83 & $(\frac{276}{5},\frac{4}{5},\frac{33}{5},13110)$  \\ [1mm]
 12 & $(4,\frac{2}{5},\frac{3}{5},24)$  & 36 & $(\frac{44}{5},\frac{2}{5},\frac{6}{5},220)$ & 60 & $(\frac{68}{5},\frac{2}{5},\frac{9}{5},374)$  & 84 & $(4,\frac{1}{3},\frac{2}{3},16)$  \\ [1mm]
 13 & $(\frac{108}{5},\frac{7}{5},\frac{9}{5},27)$  & 37 & $(\frac{68}{7},\frac{3}{7},\frac{9}{7},221)$ & 61 & $(\frac{108}{7},\frac{6}{7},\frac{11}{7},378)$  & 85 & $(20,\frac{4}{3},\frac{5}{3},80)$  \\ [1mm]
 14 & $(\frac{28}{5},\frac{2}{5},\frac{4}{5},28)$  & 38 & $(17,\frac{3}{2},\frac{9}{8},221)$ & 62 & $(\frac{100}{7},\frac{4}{7},\frac{12}{7},380)$  & 86 & $(20,\frac{3}{2},\frac{3}{2},140)$  \\ [1mm]
 15 & $(\frac{164}{7},\frac{11}{7},\frac{13}{7},41)$  & 39 & $(12,\frac{3}{5},\frac{7}{5},222)$ & 63 & $(\frac{76}{5},\frac{4}{5},\frac{8}{5},380)$  & 87 & $(12,\frac{2}{3},\frac{4}{3},156)$  \\ [1mm]
 16 & $(\frac{45}{2},\frac{3}{2},\frac{29}{16},45)$  & 40 & $(\frac{33}{2},\frac{3}{2},\frac{17}{16},231)$ & 64 & $(\frac{76}{5},\frac{3}{5},\frac{9}{5},437)$  & 88 & $(12,\frac{1}{3},\frac{5}{3},318)$  \\ [1mm]
 17 & $(\frac{116}{5},\frac{8}{5},\frac{9}{5},58)$  & 41 & $(\frac{31}{2},\frac{3}{2},\frac{15}{16},248)$ & 65 & $(\frac{108}{7},\frac{4}{7},\frac{13}{7},456)$  & 89 & $(20,\frac{1}{3},\frac{8}{3},728)$  \\ [1mm]
 18 & $(22,\frac{3}{2},\frac{7}{4},66)$  & 42 & $(\frac{124}{7},\frac{9}{7},\frac{10}{7},248)$ & 66 & $(\frac{84}{5},\frac{1}{5},\frac{12}{5},534)$  & 90 &  $(20,\frac{2}{3},\frac{7}{3},890)$  \\ [1mm]
 19 & $(\frac{156}{7},\frac{11}{7},\frac{12}{7},78)$  & 43 & $(\frac{44}{5},\frac{1}{5},\frac{7}{5},253)$ & 67 & $(18,\frac{5}{2},\frac{1}{4},598)$  & 91 & $(20,\frac{3}{4},\frac{9}{4},1004)$  \\ [1mm]
 20 & $(\frac{43}{2},\frac{3}{2},\frac{27}{16},86)$  & 44 & $(\frac{17}{2},\frac{3}{2},\frac{1}{16},255)$ & 68 & $(\frac{92}{5},\frac{3}{5},\frac{11}{5},690)$  & 92 & $(28,\frac{2}{3},\frac{10}{3},1948)$  \\ [1mm]
 21 & $(\frac{44}{7},\frac{4}{7},\frac{5}{7},88)$  & 45 & $(15,\frac{3}{2},\frac{7}{8},255)$ & 69 & $(\frac{108}{5},\frac{2}{5},\frac{14}{5},860)$  & 93 & $(44,\frac{1}{3},\frac{17}{3},3146)$  \\ [1mm]
 22 & $(\frac{92}{5},\frac{6}{5},\frac{8}{5},92)$  & 46 & $(9,\frac{3}{2},\frac{1}{8},261)$ & 70 & $(26,\frac{7}{2},\frac{1}{4},1118)$  & 94 & $(36,\frac{2}{3},\frac{13}{3},3384)$  \\ [1mm]
 23 & $(\frac{52}{5},\frac{3}{5},\frac{6}{5},104)$  & 47 & $(\frac{29}{2},\frac{3}{2},\frac{13}{16},261)$ & 71 & $(\frac{156}{7},\frac{5}{7},\frac{18}{7},1248)$  &  &    \\ [1mm]
 24 & $(21,\frac{3}{2},\frac{13}{8},105)$  & 48 & $(\frac{19}{2},\frac{3}{2},\frac{3}{16},266)$ & 72 & $(22,\frac{5}{2},\frac{3}{4},1298)$  &  &   \\ [1mm]  
\end{tabular}   
}
    \caption{Classification of the three-character RCFT with $c>0$.}
    \label{bosonic third order classification}
\end{table}

In this appendix, we provide the classification of three-character bosonic RCFTs. As discussed in the main text, we search for the values of $(c,h_1,h_2,a_1)$ such that allow \eqref{vacuum character of 3rd} to have non-negative integer Fourier coefficients. The central charge $c = \frac{m_1}{m_2}$ is restricted to be positive rational numbers, since we are interested in the unitary RCFTs. We scan the solutions by varying the positive integers $m_1, m_2, a_1$ in the range of \eqref{parameter range}. The resulting list of unitary solutions are presented in table \ref{bosonic third order classification}. We note that the solutions number 1 to 83 are reported in \cite{Hampapura:2016mmz,Gaberdiel:2016zke,Franc_2020} and number 84 to 94 are new solutions.

Let us make a few comments on the solutions. The solution number 1 can be identified with characters of the $\text{SO}(m)_1$ WZW models for any integer $m \ge 2$. For $m=1$, it correspond to the three characters of the Ising model. The solutions number 2 and 3 involves one free parameter $\alpha$. For any non-negative integer $\alpha$, they solve the third-order MLDE. If one of the conformal weight takes the value of $3/2$, that solution have a relation with a single-character solution \cite{Bae:2020xzl} via the bosonization. Except the solutions number 1 to 3, we find that there are finitely many solutions of the third-order MLDE.

Throughout table \ref{new bosonic solutions}, we provide $q$-expansion of solutions number 84 to 94. It would be tempting to investigate how many solutions of table \ref{bosonic third order classification} possess a consistent fusion rule algebra.

\begin{table}[h]
\centering
\scalebox{0.80}{
\begin{tabular}{c | c | L | L}
\hline
\text{No.} & c  & h & \text{$q$-expansion of the solution}  \\[0.3em]
\hline
\hline 
 \multirow{3}{*}{84} & \multirow{3}{*}{$4$} & 0 & q^{-\frac{1}{6}} \left(1 + 16 q + 98 q^2 + 364 q^3 + 1221 q^4 + 3528 q^5  +\cdots \right) \\[0.3em]
 & & \frac{1}{3} & q^{\frac{1}{6}} \left(1 + 11 q + 50 q^2 + 188 q^3 + 583 q^4 + 1646 q^5 + 4238 q^6 +\cdots \right) \\[0.3em]
& & \frac{2}{3} & q^{\frac{1}{2}} \left(1 + 6 q + 27 q^2 + 92 q^3 + 279 q^4 + 756 q^5 + 1913 q^6 + \cdots \right) \\[0.3em]
\hline 
\multirow{3}{*}{85} & \multirow{3}{*}{$20$} & 0 & q^{-\frac{5}{6}} \left(1 + 80 q + 46790 q^2 + 2654800 q^3 + 68308625 q^4 + \cdots \right) \\[0.3em]
& & \frac{4}{3} & q^{\frac{1}{2}} \left(5 + 840 q + 34398 q^2 + 744040 q^3 + 10930320 q^4 + \cdots \right) \\[0.3em]
& & \frac{5}{3} & q^{\frac{5}{6}} \left(4 + 355 q + 11240 q^2 + 209810 q^3 + 2791180 q^4 + \cdots \right) \\[0.3em]
\hline
\multirow{3}{*}{86} & \multirow{3}{*}{$20$} & 0 & q^{-\frac{5}{6}} \left(1 + 140 q + 69950 q^2 + 3983800 q^3 + 102455165 q^4 + \cdots \right) \\[0.3em]
& & \frac{3}{2} & q^{\frac{2}{3}} \left(5 + 592 q + 21160 q^2 + 424000 q^3 + 5918900 q^4 +  \cdots \right) \\[0.3em]
& & \frac{3}{2} & q^{\frac{2}{3}} \left(5 + 592 q + 21160 q^2 + 424000 q^3 + 5918900 q^4 + \cdots \right) \\[0.3em]
\hline
\multirow{3}{*}{87} & \multirow{3}{*}{$12$} & 0 & q^{-\frac{1}{2}} \left(1 + 156 q + 7542 q^2 + 122488 q^3 + 1254867 q^4 + \cdots \right) \\[0.3em]
& & \frac{2}{3} & q^{\frac{1}{6}} \left(1 + 92 q + 1913 q^2 + 22220 q^3 + 185749 q^4 +  \cdots \right) \\[0.3em]
& & \frac{4}{3} & q^{\frac{5}{6}} \left(1 + 28 q + 380 q^2 + 3488 q^3 + 24928 q^4 + \cdots \right) \\[0.3em]
\hline
\multirow{3}{*}{\text{88}} & \multirow{3}{*}{$12$} & 0 & q^{-\frac{1}{2}} \left(1 + 318 q + 8514 q^2 + 126862 q^3 + 1269771 q^4 +\cdots \right) \\[0.3em]
& & \frac{1}{3} & q^{-\frac{1}{6}} \left(1 + 178 q + 4634 q^2 + 61924 q^3 + 566277 q^4 +\cdots \right) \\[0.3em]
& & \frac{5}{3} & q^{\frac{7}{6}} \left(1 + 23 q + 272 q^2 + 2286 q^3 + 15318 q^4 + 87091 q^5 + \cdots \right) \\[0.3em]
\hline 
\multirow{3}{*}{89} & \multirow{3}{*}{$20$} & 0 & q^{-\frac{5}{6}} \left(1 + 728 q + 106406 q^2 + 3894424 q^3 + 82707185 q^4 + \cdots \right) \\[0.3em]
& & \frac{1}{3} & q^{-\frac{1}{2}} \left(2 + 717 q + 83124 q^2 + 3031214 q^3 + 62776974 q^4 + \cdots \right)  \\[0.3em]
& & \frac{8}{3} & q^{\frac{11}{6}} \left(1 + 40 q + 806 q^2 + 11076 q^3 + 117599 q^4 + \cdots \right) \\[0.3em]
\hline 
\multirow{3}{*}{90} &\multirow{3}{*}{$20$} & 0 & q^{-\frac{5}{6}} \left(1 + 890 q + 55700 q^2 + 2695300 q^3 + 68460905 q^4 + \cdots \right) \\[0.3em]
& & \frac{2}{3} & q^{-\frac{1}{6}} \left(5 + 728 q + 58000 q^2 + 1822700 q^3 + 33995325 q^4 + \cdots \right) \\[0.3em]
& & \frac{7}{3} & q^{\frac{3}{2}} \left(10 + 423 q + 9180 q^2 + 134925 q^3 + 1516500 q^4 + \cdots \right) \\[0.3em]
\hline 
\multirow{3}{*}{91} &\multirow{3}{*}{$20$} & 0 & q^{-\frac{5}{6}} \left(1 + 1004 q + 4286 q^2 + 9080 q^3 + 29821 q^4 + \cdots \right) \\[0.3em]
& & \frac{3}{4} & q^{-\frac{1}{12}} \left(1 + 86 q + 6257 q^2 + 186934 q^3 + 3373364 q^4 + \cdots \right) \\[0.3em]
& & \frac{9}{4} & q^{\frac{17}{12}} \left(13 + 578 q + 13031 q^2 + 196614 q^3 + 2252637 q^4 + \cdots \right) \\[0.3em]
\hline 
\multirow{3}{*}{92} & \multirow{3}{*}{$28$} & 0 & q^{-\frac{7}{6}} \left(1 + 1948 q + 424724 q^2 + 23967552 q^3 +  \cdots \right) \\[0.3em]
& & \frac{2}{3} & q^{-\frac{1}{2}} \left(25 + 9084 q + 862632 q^2 + 59416816 q^3 +  \cdots \right) \\[0.3em]
& & \frac{10}{3} & q^{\frac{13}{6}} \left(11 + 628 q + 18439 q^2 + 368372 q^3 + 5618761 q^4 + \cdots \right) \\[0.3em]
\hline 
\multirow{3}{*}{93} & \multirow{3}{*}{$44$} & 0 & q^{-\frac{11}{6}} \left(1 + 3146 q + 1906130 q^2 + 467324061 q^3 + + \cdots \right) \\[0.3em]
& & \frac{1}{3} & q^{-\frac{3}{2}} \left(13 + 20178 q + 8638218 q^2 + 1682129028 q^3 + \cdots \right) \\[0.3em]
& & \frac{17}{3} & q^{\frac{23}{6}} \left(19 + 1702 q + 76646 q^2 + 2317292 q^3 + 52998283 q^4 + \cdots \right) \\[0.3em]
\hline
\multirow{3}{*}{94} & \multirow{3}{*}{$36$} & 0 & q^{-\frac{3}{2}} \left(1 + 3384 q + 1337850 q^2 + 186743064 q^3 + \cdots \right) \\[0.3em]
& & \frac{2}{3} & q^{-\frac{5}{6}} \left(2 + 1294 q + 243835 q^2 + 19695524 q^3 + \cdots \right) \\[0.3em]
& & \frac{13}{3} & q^{\frac{17}{6}} \left(4 + 289 q + 10688 q^2 + 268714 q^3 + 5154084 q^4 + \cdots \right) \\[0.3em]
\hline
\end{tabular}
}
\caption{\label{new bosonic solutions} New solutions of the third-order MLDE. The overall normalization chosen to a minimal number which provide the positive integers up to higher-order of $q$.}
\end{table}

\newpage
\section{Solutions of the Fermionic BPS third-order MLDE}
\label{app:3MLDE_appendix}

\subsection*{NS sector characters}

\begin{table}[H]
\centering
\scalebox{0.9}{
\begin{tabular}{L | L | L}
\hline
 c  & h^{\textsc{ns}} & \text{NS-sector characters multiplied by $q^{\frac{c}{24}-h^{\textsc{ns}}}$}  \\[0.3em]
\hline
\hline \multirow{3}{*}{$1$} & 0 & 1+q+2 q^{3/2}+2 q^2+2 q^{5/2}+3 q^3+\cdots \\[0.3em]
& \frac{1}{6} & 1+\sqrt{q}+q+q^{3/2}+2 q^2+3 q^{5/2}+3 q^3+\cdots \\[0.3em]
& \frac{1}{2} & 1+q^{3/2}+q^2+q^{5/2}+q^3+\cdots \\[0.3em]
\hline \multirow{3}{*}{$\frac{3}{2}$} & 0 & 1+q^{3/2}+3 q^2+3 q^{5/2}+3 q^3+\cdots \\[0.3em]
& \frac{1}{8} & 1+\sqrt{q}+2 q+3 q^{3/2}+4 q^2+6 q^{5/2}+9 q^3+\cdots \\[0.3em]
& \frac{1}{2} & 1+\sqrt{q}+q+2 q^{3/2}+3 q^2+4 q^{5/2}+5 q^3+\cdots \\[0.3em]
\hline \multirow{3}{*}{$\frac{3}{2}$} & 0 & 1+3 q+4 q^{3/2}+3 q^2+6 q^{5/2}+9 q^3+\cdots \\[0.3em]
& \frac{1}{4} & 2+2\sqrt{q}+2q+4 q^{3/2}+8 q^2+10 q^{5/2}+ 12 q^3+\cdots \\[0.3em]
& \frac{1}{2} & 1+2 q^{3/2}+2 q^2+2 q^{5/2}+3 q^3+\cdots \\[0.3em]
\hline \multirow{3}{*}{$2$} & 0 & 1+2 q+4 q^{3/2}+5 q^2+8 q^{5/2}+14 q^3+\cdots \\[0.3em]
& \frac{1}{6} & 1+\sqrt{q}+2 q+4 q^{3/2}+7 q^2+10 q^{5/2}+14 q^3+\cdots \\[0.3em]
& \frac{1}{3} & 1+2 \sqrt{q}+3 q+4 q^{3/2}+7 q^2+12 q^{5/2}+17 q^3+\cdots \\[0.3em]
\hline \multirow{3}{*}{$3$} & 0 & 1+3 q+8 q^{3/2}+15 q^2+24 q^{5/2}+40 q^3+\cdots \\[0.3em]
& \frac{1}{4} & 2+4 \sqrt{q}+10 q+20 q^{3/2}+36 q^2+64 q^{5/2}+110 q^3+\cdots \\[0.3em]
& \frac{1}{2} & 2+4 \sqrt{q}+6 q+12 q^{3/2}+26 q^2+44 q^{5/2}+68 q^3+\cdots \\[0.3em]
\hline \multirow{3}{*}{$3$} & 0 & 1+6 q+14 q^{3/2}+18 q^2+30 q^{5/2}+64 q^3+\cdots \\[0.3em]
& \frac{1}{3} & 3+6 \sqrt{q}+ 12 q+ 24 q^{3/2}+ 48 q^2+ 84 q^{5/2}+ 135 q^3+\cdots \\[0.3em]
& \frac{1}{2} & 2+ 3 \sqrt{q} + 4 q+ 11 q^{3/2} +24 q^2+ 36 q^{5/2}+56 q^3+\cdots \\[0.3em]
\hline \multirow{3}{*}{$\frac{18}{5}$} & 0 & 1+6 q+16 q^{3/2}+27 q^2+48 q^{5/2}+98 q^3+\cdots \\[0.3em]
& \frac{3}{10} & 4+9 \sqrt{q}+24 q+54 q^{3/2}+108 q^2+201 q^{5/2}+360 q^3+\cdots \\[0.3em]
& \frac{2}{5} & 3+8 \sqrt{q}+18 q+36 q^{3/2}+74 q^2+144 q^{5/2}+252 q^3+\cdots \\[0.3em]
\hline \multirow{3}{*}{$\frac{9}{2}$} & 0 & 1+9 q+27 q^{3/2}+54 q^2+108 q^{5/2}+234 q^3+\cdots \\[0.3em]
& \frac{3}{8} & 4+12 \sqrt{q}+36 q+88 q^{3/2}+192 q^2+396 q^{5/2}+776 q^3+\cdots \\[0.3em]
& \frac{1}{2} & 3+9 \sqrt{q}+22 q+51 q^{3/2}+117 q^2+241 q^{5/2}+453 q^3+\cdots \\[0.3em]
\hline \multirow{3}{*}{$\frac{9}{2}$} & 0 & 1+6 q+14 q^{3/2}+27 q^2+66 q^{5/2}+147 q^3+\cdots \\[0.3em]
& \frac{1}{4} & 3+5 \sqrt{q}+18 q+51 q^{3/2}+116 q^2+234 q^{5/2}+459 q^3+\cdots \\[0.3em]
& \frac{1}{2} & 9+30 \sqrt{q}+79 q+180 q^{3/2}+393 q^2+810 q^{5/2}+1557 q^3+\cdots \\[0.3em]
\hline \multirow{3}{*}{$5$} & 0 & 1+10 q+30 q^{3/2}+65 q^2+146 q^{5/2}+330 q^3+\cdots \\[0.3em]
& \frac{1}{3} & 5+14 \sqrt{q}+50 q+140 q^{3/2}+325 q^2+690 q^{5/2}+1419 q^3+\cdots \\[0.3em]
& \frac{1}{2} & 10+35 \sqrt{q}+100 q+245 q^{3/2}+566 q^2+1225 q^{5/2}+2460 q^3+\cdots \\[0.3em]
\hline
\end{tabular}
}
\caption{\label{NS-sector list 1} The solutions of the NS-sector BPS third-order MLDE are listed in $q$-expansion.}
\end{table}

\begin{table}[H]
\centering
\scalebox{0.9}{
\begin{tabular}{L | L | L}
\hline
 c  & h^{\textsc{ns}} & \text{NS-sector characters multiplied by $q^{\frac{c}{24}-h^{\textsc{ns}}}$}  \\[0.3em]
\hline
\hline \multirow{3}{*}{$6$} & 0 & 1+18 q+64 q^{3/2}+159 q^2+384 q^{5/2}+942 q^3+\cdots \\[0.3em]
& \frac{1}{2} & 8+32 \sqrt{q}+112 q+320 q^{3/2}+808 q^2+1888 q^{5/2}+4144 q^3 +\cdots \\[0.3em]
& \frac{1}{2} & 4+16 \sqrt{q}+56 q+160 q^{3/2}+404 q^2+944 q^{5/2}+2072 q^3+\cdots \\[0.3em]
\hline \multirow{3}{*}{$7$} & 0 & 1+21 q+84 q^{3/2}+252 q^2+686 q^{5/2}+1771 q^3+\cdots \\[0.3em]
& \frac{1}{2} & 14+70 \sqrt{q}+294 q+945 q^{3/2}+2604 q^2+6615 q^{5/2}+15758 q^3+\cdots \\[0.3em]
& \frac{2}{3} & 21+90 \sqrt{q}+315 q+966 q^{3/2}+2646 q^2+6552 q^{5/2}+15141 q^3+\cdots \\[0.3em]
\hline \multirow{3}{*}{$\frac{15}{2}$} & 0 & 1+30 q+125 q^{3/2}+390 q^2+1125 q^{5/2}+3055 q^3+\cdots \\[0.3em]
& \frac{1}{2} & 5+25 \sqrt{q}+115 q+400 q^{3/2}+1156 q^2+3035 q^{5/2}+7500 q^3+\cdots \\[0.3em]
& \frac{5}{8} & 16+80 \sqrt{q}+320 q+1040 q^{3/2}+2960 q^2+7696 q^{5/2}+18640 q^3+\cdots \\[0.3em]
\hline \multirow{3}{*}{$\frac{15}{2}$} & 0 & 1+15 q+70 q^{3/2}+240 q^2+672 q^{5/2}+1760 q^3+\cdots \\[0.3em]
& \frac{1}{2} & 15+90 \sqrt{q}+400 q+1350 q^{3/2}+3921 q^2+10400 q^{5/2}+25665 q^3+\cdots \\[0.3em]
& \frac{3}{4} & 20+84 \sqrt{q}+300 q+960 q^{3/2}+2700 q^2+6840 q^{5/2}+16244 q^3+\cdots \\[0.3em]
\hline \multirow{3}{*}{$\frac{42}{5}$} & 0 & 1+42 q+196 q^{3/2}+693 q^2+2184 q^{5/2}+6314 q^3+\cdots \\[0.3em]
& \frac{3}{5} & 14+84 \sqrt{q}+399 q+1448 q^{3/2}+4452 q^2+12432 q^{5/2}+32291 q^3+\cdots \\[0.3em]
& \frac{7}{10} & 36+196 \sqrt{q}+840 q+2940 q^{3/2}+8932 q^2+24549 q^{5/2}+62664 q^3+\cdots \\[0.3em]
\hline \multirow{3}{*}{$9$} & 0 & 1+45 q+216 q^{3/2}+828 q^2+2808 q^{5/2}+8481 q^3+\cdots \\[0.3em]
& \frac{1}{2} & 6+36 \sqrt{q}+206 q+852 q^{3/2}+2844 q^2+8416 q^{5/2}+23082 q^3+\cdots \\[0.3em]
& \frac{3}{4} & 32+192 \sqrt{q}+864 q+3136 q^{3/2}+9888 q^2+28224 q^{5/2}+74560 q^3+\cdots \\[0.3em]
\hline \multirow{3}{*}{$9$} & 0 & 1+63 q+324 q^{3/2}+1233 q^2+4086 q^{5/2}+12336 q^3+\cdots \\[0.3em]
& \frac{1}{2} & 6+ 30 \sqrt{q}+170 q + 717 q^{3/2} + 2418 q^2 + 7131 q^{5/2} + 19476 q^3+\cdots \\[0.3em]
& \frac{2}{3} & 27+162 q^{\frac{1}{2}}+783 q+2970 q^{\frac{3}{2}}+9531 q^2 + 27486 q^{5/2}+73494 q^3+\cdots \\[0.3em]
\hline \multirow{3}{*}{$10$} & 0 & 1+70 q+400 q^{3/2}+1745 q^2+6400 q^{5/2}+20710 q^3+\cdots \\[0.3em]
& \frac{2}{3} & 15+120 \sqrt{q}+666 q+2760 q^{3/2}+9600 q^2+29880 q^{5/2}+85440 q^3+\cdots \\[0.3em]
& \frac{5}{6} & 36+225 \sqrt{q}+1080 q+4230 q^{3/2}+14220 q^2+42831 q^{5/2}+119160 q^3+\cdots \\[0.3em]
\hline \multirow{3}{*}{$\frac{21}{2}$} & 0 & 1+63 q+343 q^{3/2}+1575 q^2+6174 q^{5/2}+20811 q^3+\cdots \\[0.3em]
& \frac{1}{2} & 7+49 \sqrt{q}+336 q+1617 q^{3/2}+6202 q^2+20678 q^{5/2}+62679 q^3+\cdots \\[0.3em]
& \frac{7}{8} & 64+448 \sqrt{q}+2240 q+8960 q^{3/2}+30912 q^2+95872 q^{5/2}+273728 q^3+\cdots \\[0.3em]
\hline \multirow{3}{*}{$\frac{21}{2}$} & 0 & 1+126 q+784 q^{3/2}+3612 q^2+13818 q^{5/2}+46389 q^3+\cdots \\[0.3em]
& \frac{1}{2} & 7+28 \sqrt{q}+189 q+938 q^{3/2}+3654 q^2+12152 q^{5/2}+36660 q^3+\cdots \\[0.3em]
& \frac{3}{4} & 56+392 q^{\frac{1}{2}}+2144 q+9128 q^{\frac{3}{2}}+32536 q^2 +102984 q^{5/2}+ 299544 q^3+\cdots \\[0.3em]
\hline \multirow{3}{*}{$11$} & 0 & 1+143 q+924 q^{3/2}+4499 q^2+18084 q^{5/2}+63052 q^3+\cdots \\[0.3em]
& \frac{1}{2} & 1+4 \sqrt{q}+29 q+150 q^{3/2}+607 q^2+2098 q^{5/2}+6554 q^3+\cdots \\[0.3em]
& \frac{5}{6} & 22+ 165 \sqrt{q} + 906 q +3883 q^{3/2} + 14058 q^2+ 90750 q^{5/2} + 134464 q^3+\cdots \\[0.3em]
\hline
\end{tabular}
}
\caption{\label{NS-sector list 2}  The solutions of the NS-sector BPS third-order MLDE are listed in $q$-expansion.}
\end{table}

\begin{table}[H]
\centering
\scalebox{0.9}{
\begin{tabular}{L | L | L}
\hline
 c  & h^{\textsc{ns}} & \text{NS-sector characters multiplied by $q^{\frac{c}{24}-h^{\textsc{ns}}}$}  \\[0.3em]
\hline
\hline \multirow{3}{*}{$12$} & 0 & 1+84 q+512 q^{3/2}+2754 q^2+12288 q^{5/2}+46168 q^3+\cdots \\[0.3em]
& \frac{1}{2} & 8+64 \sqrt{q}+512 q+2816 q^{3/2}+12288 q^2+45952 q^{5/2}+153600 q^3+\cdots \\[0.3em]
& 1 & 128+1024 \sqrt{q}+5632 q+24576 q^{3/2}+91904 q^2+307200
q^{5/2}+\cdots \\[0.3em]
\hline \multirow{3}{*}{$\frac{66}{5}$} & 0 & 1+198 q+1936 q^{3/2}+12045 q^2+58080 q^{5/2}+237006 q^3+\cdots \\[0.3em]
& \frac{4}{5} & 33+528 \sqrt{q}+3718 q + 19536 q^{3/2}+ 84909 q^2+330464 q^{5/2} +\cdots \\[0.3em]
& \frac{11}{10} & 144 +1089 \sqrt{q} +6336 q+29766 q^{3/2} +118272 q^2+ 416845 q^{5/2}  +\cdots \\[0.3em]
\hline \multirow{3}{*}{$\frac{27}{2}$} & 0 & 1+108 q+729 q^{3/2}+4509 q^2+22599 q^{5/2}+94239 q^3+\cdots \\[0.3em]
& \frac{1}{2} & 9+81 \sqrt{q}+741 q+4590 q^{3/2}+22545 q^2+93942 q^{5/2}+345267 q^3+\cdots \\[0.3em]
& \frac{9}{8} & 256+2304 \sqrt{q}+13824 q+65280 q^{3/2}+262656 q^2+940032
q^{5/2}+\cdots \\[0.3em]
\hline \multirow{3}{*}{$15$} & 0 & 1+135 q+1000 q^{3/2}+7005 q^2+39000 q^{5/2}+179490 q^3+\cdots \\[0.3em]
& \frac{1}{2} & 10+100 \sqrt{q}+1030 q+7100 q^{3/2}+38862 q^2+179140 q^{5/2}+\cdots \\[0.3em]
& \frac{5}{4} & 512+5120 \sqrt{q}+33280 q+168960 q^{3/2}+727040 q^2+2770944
q^{5/2}+\cdots \\[0.3em]
\hline \multirow{3}{*}{$\frac{33}{2}$} & 0 & 1+165 q+1331 q^{3/2}+10428 q^2+63888 q^{5/2}+322564 q^3+\cdots \\[0.3em]
& \frac{1}{2} & 11+121 \sqrt{q}+1386 q+10527 q^{3/2}+63635 q^2+322223 q^{5/2}+\cdots \\[0.3em]
& \frac{11}{8} & 1024+11264 \sqrt{q}+78848 q+428032 q^{3/2}+1959936 q^2+\cdots \\[0.3em]
\hline \multirow{3}{*}{$18$} & 0 & 1+198 q+1728 q^{3/2}+14985 q^2+100224 q^{5/2}+551862 q^3+\cdots \\[0.3em]
& \frac{1}{2} & 12+144 \sqrt{q}+1816 q+15072 q^{3/2}+99828 q^2+551632 q^{5/2}+\cdots \\[0.3em]
& \frac{3}{2} &2048+24576 \sqrt{q}+184320 q+1064960 q^{3/2}+5167104 q^2+\cdots \\[0.3em]
\hline \multirow{3}{*}{$\frac{39}{2}$} & 0 & 1+234 q+2197 q^{3/2}+20904 q^2+151593 q^{5/2}+905307 q^3+\cdots \\[0.3em]
& \frac{1}{2} & 13+169 \sqrt{q}+2327 q+20956 q^{3/2}+151034 q^2+905333
q^{5/2}+\cdots \\[0.3em]
& \frac{13}{8} & 4096+53248 \sqrt{q}+425984 q+2609152 q^{3/2}+13365248 q^2+\cdots \\[0.3em]
\hline \multirow{3}{*}{$\frac{39}{2}$} & 0 & 1 + 156 q + 858 q^{3/2} + 9672 q^2 + 79794 q^{5/2} + 504959 q^3+\cdots \\[0.3em]
& \frac{3}{4} & 65 + 429 q^{1/2} + 7215 q + 69784 q^{3/2} + 496197 q^2 + 
 2851068 q^{5/2} + \cdots \\[0.3em]
& \frac{3}{2} & 4225 + 55770 q^{1/2} + 462891 q + 2937220 q^{3/2} + 15510651 q^2 +\cdots \\[0.3em]
\hline \multirow{3}{*}{$21$} & 0 & 1+273 q+2744 q^{3/2}+28434 q^2+222264 q^{5/2}+1432291 q^3+\cdots \\[0.3em]
& \frac{1}{2} & 14+196 \sqrt{q}+2926 q+28420 q^{3/2}+221536 q^2+1432760
q^{5/2}+\cdots \\[0.3em]
& \frac{7}{4} & 8192+114688 \sqrt{q}+974848 q+6307840 q^{3/2}+34004992 q^2+\cdots \\[0.3em]
\hline \multirow{3}{*}{$22$} & 0 & 1+286 q+1848 q^{3/2}+29447 q^2+300432 q^{5/2}+2266594 q^3+\cdots \\[0.3em]
& \frac{5}{6} & 66+495 \sqrt{q}+12156 q+143418 q^{3/2}+1185162 q^2+7733913 q^{5/2}+\cdots \\[0.3em]
& \frac{5}{3} & 4356+65340 \sqrt{q}+603801 q+4228488 q^{3/2}+24487002 q^2+\cdots \\[0.3em]
\hline \multirow{3}{*}{$\frac{45}{2}$} & 0 & 1+315 q+3375 q^{3/2}+37845 q^2+317250 q^{5/2}+2195805 q^3+\cdots \\[0.3em]
& \frac{1}{2} & 15+225 \sqrt{q}+3620 q+37725 q^{3/2}+316368 q^2+2196940
q^{5/2}+\cdots \\[0.3em]
& \frac{15}{8} & 16384+245760 \sqrt{q}+2211840 q+15073280 q^{3/2}+85278720q^2+\cdots \\[0.3em]
\hline \multirow{3}{*}{$24$} & 0 & 1+360 q+4096 q^{3/2}+49428 q^2+442368 q^{5/2}+3274752 q^3+\cdots \\[0.3em]
& \frac{1}{2} & 16+256 \sqrt{q}+4416 q+49152 q^{3/2}+441376 q^2+3276800
q^{5/2}+\cdots \\[0.3em]
& 2 & 32768+524288 \sqrt{q}+4980736 q+35651584 q^{3/2}+211156992
q^2+\cdots \\[0.3em]
\hline
\end{tabular}}
\caption{\label{NS-sector list} The solutions of the NS-sector BPS third-order MLDE are listed in $q$-expansion.}
\end{table}

\subsection*{R sector characters multiplied by $q^{\frac{c}{24}-h^{\textsc{R}}}$}

\begin{table}[H]
\centering
\scalebox{0.9}{
\begin{tabular}{L | L | L}
\hline
 c  & h^{\textsc{r}} & \text{R-sector characters multiplied by $q^{\frac{c}{24}-h^{\textsc{r}}}$}  \\[0.3em]
\hline
\hline \multirow{3}{*}{$1$} & \frac{1}{24} & 1+2 q+4 q^2+6 q^3+10 q^4+\cdots \\[0.3em]
& \frac{1}{8} & 1+2 q+3 q^2+6 q^3+9 q^4+\cdots \\[0.3em]
& \frac{3}{8} & 1+q+2 q^2+4 q^3+6 q^4+\cdots \\[0.3em]
\hline \multirow{3}{*}{$\frac{3}{2}$} & \frac{1}{16} & \frac{1}{\sqrt{2}}+\sqrt{2} q+3 \sqrt{2} q^2+6 \sqrt{2} q^3+11 \
\sqrt{2} q^4 +\cdots \\[0.3em]
& \frac{3}{16} & \frac{1}{2 \sqrt{2}}+\frac{3 q}{2 \sqrt{2}}+\frac{3 \
q^2}{\sqrt{2}}+\frac{13 q^3}{2 \sqrt{2}}+6 \sqrt{2} q^4+\cdots \\[0.3em]
& \frac{9}{16} & \frac{1}{\sqrt{2}}+\sqrt{2} q+2 \sqrt{2} q^2+4 \sqrt{2} q^3+\frac{15 \
q^4}{\sqrt{2}}+\cdots \\[0.3em]
\hline \multirow{3}{*}{$\frac{3}{2}$} & \frac{1}{16} & 1+4 q+8 q^2+16 q^3+32 q^4+\cdots \\[0.3em]
& \frac{3}{16} & 1+3 q+6 q^2+13 q^3+24 q^4+\cdots \\[0.3em]
& \frac{5}{16} & 1+2 q+5 q^2+10 q^3+18 q^4+\cdots \\[0.3em]
\hline \multirow{3}{*}{$2$} & \frac{1}{12} & 1+4 q+12 q^2+28 q^3+60 q^4+\cdots \\[0.3em]
& \frac{5}{12} & 1+3 q+8 q^2+18 q^3+38 q^4+\cdots \\[0.3em]
& \frac{3}{4} & 1+2 q+5 q^2+12 q^3+24 q^4+\cdots \\[0.3em]
\hline \multirow{3}{*}{$3$} & \frac{1}{8} & 1+8 q+32 q^2+96 q^3+256 q^4+\cdots \\[0.3em]
& \frac{3}{8} & 1+6 q+21 q^2+62 q^3+162 q^4+\cdots \\[0.3em]
& \frac{5}{8} & 2+8 q+28 q^2+80 q^3+202 q^4+\cdots \\[0.3em]
\hline \multirow{3}{*}{$3$} & \frac{1}{8} & 1+10 q+36 q^2+110 q^3+298 q^4+\cdots \\[0.3em]
& \frac{3}{8} & 1+6 q+21 q^2+62 q^3+162 q^4+\cdots \\[0.3em]
& \frac{11}{24} & 1+5 q+18 q^2+52 q^3+134 q^4+\cdots \\[0.3em]
\hline \multirow{3}{*}{$\frac{18}{5}$} & \frac{3}{20} & 2+24 q+108 q^2+376 q^3+1128 q^4+\cdots \\[0.3em]
& \frac{11}{20} & 12+72 q+296 q^2+960 q^3+2736 q^4 +\cdots \\[0.3em]
& \frac{3}{4} & 8+36 q+144 q^2+460 q^3+1272 q^4 +\cdots \\[0.3em]
\hline \multirow{3}{*}{$\frac{9}{2}$} & \frac{3}{16} & \sqrt{2}+18 \sqrt{2} q+102 \sqrt{2} q^2+428 \sqrt{2} q^3+1494 \
\sqrt{2} q^4+\cdots \\[0.3em]
& \frac{9}{16} & 2 \sqrt{2}+18 \sqrt{2} q+90 \sqrt{2} q^2+348 \sqrt{2} q^3+1152 \
\sqrt{2} q^4+\cdots \\[0.3em]
& \frac{11}{16} & 3 \sqrt{2}+22 \sqrt{2} q+108 \sqrt{2} q^2+408 \sqrt{2} q^3+1325 \
\sqrt{2} q^4+\cdots \\[0.3em]
\hline \multirow{3}{*}{$\frac{9}{2}$} & \frac{3}{16} & 4+48 q+288 q^2+1216 q^3+4224 q^4 +\cdots \\[0.3em]
& \frac{9}{16} & 16+144 q+ 720 q^2+ 2784 q^3+ 9216 q^4 +\cdots \\[0.3em]
& \frac{15}{16} & 32+192 q+864 q^2+3136 q^3+9888 q^4+\cdots \\[0.3em]
\hline \multirow{3}{*}{$5$} & \frac{5}{24} & 4+80 q+520 q^2+2400 q^3+9040 q^4 +\cdots \\[0.3em]
& \frac{5}{8} & 16 \sqrt{2}+160 \sqrt{2} q+880 \sqrt{2} q^2+3680 \sqrt{2} q^3+13040 \sqrt{2} q^4 +\cdots \\[0.3em]
& \frac{7}{8} & 20 \sqrt{2}+144 \sqrt{2} q+740 \sqrt{2} q^2+2960 \sqrt{2} q^3+10120 \sqrt{2} q^4 +\cdots \\[0.3em]
\hline
\end{tabular}
}
\caption{\label{R-sector list 1} The solutions of the R-sector BPS third-order MLDE are listed in $q$-expansion.}
\end{table}

\begin{table}[H]
\centering
\scalebox{0.9}{
\begin{tabular}{L | L | L}
\hline
 c  & h^{\textsc{r}} & \text{R-sector characters multiplied by $q^{\frac{c}{24}-h^{\textsc{r}}}$}  \\[0.3em]
\hline
\hline \multirow{3}{*}{$6$} & \frac{1}{4} & 2+64 q+512 q^2+2816 q^3+12288 q^4 + \cdots \\[0.3em]
& \frac{3}{4} & 8+96 q+624 q^2+3008 q^3+12072 q^4+\cdots \\[0.3em]
& \frac{3}{4} & 8+96 q+624 q^2+3008 q^3+12072 q^4+\cdots \\[0.3em]
\hline \multirow{3}{*}{$7$} & \frac{7}{24} & 6+252 q+2520 q^2+16212 q^3+80892 q^4 +\cdots \\[0.3em]
& \frac{5}{8} & 14 \sqrt{2}+294 \sqrt{2} q+2428 \sqrt{2} q^2+14168 \sqrt{2} q^3+66388 \sqrt{2} q^4 +\cdots \\[0.3em]
& \frac{7}{8} & 64 \sqrt{2}+896 \sqrt{2} q+6720 \sqrt{2} q^2+36736 \sqrt{2} q^3+164864 \sqrt{2} q^4 +\cdots \\[0.3em]
\hline \multirow{3}{*}{$\frac{15}{2}$} & \frac{5}{16} & 2 \sqrt{2}+100 \sqrt{2} q+1100 \sqrt{2} q^2+7640 \sqrt{2} q^3+40620 \
\sqrt{2} q^4+\cdots \\[0.3em]
& \frac{13}{16} & 10 \sqrt{2}+180 \sqrt{2} q+1512 \sqrt{2} q^2+9040 \sqrt{2} q^3+43670 \
\sqrt{2} q^4+\cdots \\[0.3em]
& \frac{15}{16} & 16 \sqrt{2}+240 \sqrt{2} q+1920 \sqrt{2} q^2+11120 \sqrt{2} q^3+52560 \
\sqrt{2} q^4+\cdots \\[0.3em]
\hline \multirow{3}{*}{$\frac{15}{2}$} & \frac{5}{16} & 6+280 q+3120 q^2+21600 q^3+114880 q^4 +\cdots \\[0.3em]
& \frac{9}{16} & 20+552 q+5220 q^2+33480 q^3+169320 q^4 +\cdots \\[0.3em]
& \frac{15}{16} & 64+960 q+7680 q^2+44480 q^3+210240 q^4 +\cdots \\[0.3em]
\hline \multirow{3}{*}{$\frac{42}{5}$} & \frac{7}{20} & 6+392 q+5124 q^2+40488 q^3+240184 q^4 +\cdots \\[0.3em]
& \frac{3}{4} & 28+744 q+7560 q^2+ 52416 q^3+ 285936 q^4 +\cdots \\[0.3em]
& \frac{19}{20} & 168 + 3220 q+ 29904 q^2+195804 q^3+1028440 q^4 +\cdots \\[0.3em]
\hline \multirow{3}{*}{$9$} & \frac{3}{8} & 4+288 q+4224 q^2+36224 q^3+230400 q^4+\cdots \\[0.3em]
& \frac{7}{8} & 24+608 q+6480 q^2+47040 q^3+268664 q^4+\cdots \\[0.3em]
& \frac{9}{8} & 1+18 q+171 q^2+1158 q^3+6309 q^4+\cdots \\[0.3em]
\hline \multirow{3}{*}{$9$} & \frac{3}{8} & 1+84 q+1218 q^2+10456 q^3+66516 q^4+\cdots \\[0.3em]
& \frac{25}{24} & 1+20 q+197 q^2+1364 q^3+7546 q^4+\cdots \\[0.3em]
& \frac{9}{8} & 64+1152 q+10944 q^2+74112 q^3+403776 q^4 +\cdots \\[0.3em]
\hline \multirow{3}{*}{$10$} & \frac{5}{12} & 6+600 q+10440 q^2+102120 q^3+726120 q^4 +\cdots \\[0.3em]
& \frac{3}{4} & 40+1720 q+23360 q^2+202064 q^3+1327600 q^4 +\cdots \\[0.3em]
& \frac{13}{12} & 180+ 4392 q+ 49860 q^2+ 388080 q^3+ 2377440 q^4 +\cdots \\[0.3em]
\hline \multirow{3}{*}{$\frac{21}{2}$} & \frac{7}{16} & 4 \sqrt{2}+392 \sqrt{2} q+7448 \sqrt{2} q^2+77616 \sqrt{2} q^3+582232 \
\sqrt{2} q^4+\cdots \\[0.3em]
& \frac{15}{16} & 28 \sqrt{2}+952 \sqrt{2} q+12656 \sqrt{2} q^2+109792 \sqrt{2} \
q^3+731556 \sqrt{2} q^4+\cdots \\[0.3em]
& \frac{21}{16} & 128 \sqrt{2}+2688 \sqrt{2} q+29568 \sqrt{2} q^2+229376 \sqrt{2} \
q^3+1416576 \sqrt{2} q^4+\cdots \\[0.3em]
\hline \multirow{3}{*}{$\frac{21}{2}$} & \frac{7}{16} & 1+140 q+2632 q^2+27440 q^3+205856 q^4+\cdots \\[0.3em]
& \frac{19}{16} & 7+ 170 q +1981 q^2+15918 q^3+100723 q^4+\cdots \\[0.3em]
& \frac{21}{16} & 1+21 q+231 q^2+1792 q^3+11067 q^4+\cdots \\[0.3em]
\hline \multirow{3}{*}{$11$} & \frac{11}{24} & 1+132 q+2706 q^2+30008 q^3+237204 q^4+\cdots \\[0.3em]
& \frac{9}{8} & 55+ 1628 q + 21131 q^2+ 183868 q^3+1241174 q^4 +\cdots \\[0.3em]
& \frac{11}{8} & 1+22 q+253 q^2+2046 q^3+13134 q^4+\cdots \\[0.3em]
\hline
\end{tabular}
}
\caption{\label{R-sector list 2} The solutions of the R-sector BPS third-order MLDE are listed in $q$-expansion.}
\end{table}

\begin{table}[H]
\centering
\scalebox{0.9}{
\begin{tabular}{L | L | L}
\hline
 c  & h^{\textsc{r}} & \text{R-sector characters multiplied by $q^{\frac{c}{24}-h^{\textsc{r}}}$}  \\[0.3em]
\hline
\hline \multirow{3}{*}{$12$} & \frac{1}{2} & 8+1024 q+24576 q^2+307200 q^3+2686976 q^4 +\cdots \\[0.3em]
& 1 & 64+2816 q+45952 q^2+470528 q^3+3619136 q^4+\cdots \\[0.3em]
& \frac{3}{2} & 512+12288 q+153600 q^2+1343488 q^3+9280512 q^4+\cdots \\[0.3em]
\hline \multirow{3}{*}{$\frac{66}{5}$} & \frac{11}{20} & 3+726 q+20812 q^2+298584 q^3+2932842 q^4 +\cdots \\[0.3em]
& \frac{3}{4} & 11+1254 q+29868 q^2+391061 q^3+3626436 q^4+\cdots \\[0.3em]
& \frac{27}{20} & 9+ 321 q + 5106 q^2 + 53380 q^3 +426486 q^4 +\cdots \\[0.3em]
\hline \multirow{3}{*}{$\frac{27}{2}$} & \frac{9}{16} & 8 \sqrt{2}+1296 \sqrt{2} q+38448 \sqrt{2} q^2+569952 \sqrt{2}
q^3+5758128 \sqrt{2} q^4+\cdots \\[0.3em]
& \frac{17}{16} & 72 \sqrt{2}+3984 \sqrt{2} q+78624 \sqrt{2} q^2+940608 \sqrt{2}
q^3+8279096 \sqrt{2} q^4 +\cdots \\[0.3em]
& \frac{27}{16} & 1024 \sqrt{2}+27648 \sqrt{2} q+387072 \sqrt{2} q^2+3769344 \sqrt{2}
q^3+28809216 \sqrt{2} q^4 +\cdots \\[0.3em]
\hline \multirow{3}{*}{$15$} & \frac{5}{8} & 16+3200 q+115200 q^2+2004480 q^3+23203840 q^4+\cdots \\[0.3em]
& \frac{9}{8} & 160+10880 q+256192 q^2+3549440 q^3+35487520 q^4 +\cdots \\[0.3em]
& \frac{15}{8} & 4096+122880 q+1904640 q^2+20439040 q^3+171294720 q^4+\cdots \\[0.3em]
\hline \multirow{3}{*}{$\frac{33}{2}$} & \frac{11}{16} & 16 \sqrt{2}+3872 \sqrt{2} q+166496 \sqrt{2} q^2+3368640 \sqrt{2}
q^3+44369248 \sqrt{2} q^4+\cdots \\[0.3em]
& \frac{19}{16} & 176 \sqrt{2}+14432 \sqrt{2} q+400576 \sqrt{2} q^2+6376832 \sqrt{2}
q^3+71957072 \sqrt{2} q^4+\cdots \\[0.3em]
& \frac{33}{16} & 8192 \sqrt{2}+270336 \sqrt{2} q+4595712 \sqrt{2} q^2+53886976
\sqrt{2} q^3+\cdots \\[0.3em]
\hline \multirow{3}{*}{$18$} & \frac{3}{4} & 32+9216 q+466944 q^2+10891264 q^3+162201600 q^4+\cdots \\[0.3em]
& \frac{5}{4} & 384+37376 q+1209600 q^2+21967872 q^3+278189952 q^4+\cdots \\[0.3em]
& \frac{9}{4} & 32768+1179648 q+21823488 q^2+277610496 q^3+\cdots \\[0.3em]
\hline \multirow{3}{*}{$\frac{39}{2}$} & \frac{13}{16} & 32 \sqrt{2}+10816 \sqrt{2} q+638144 \sqrt{2} q^2+17024384 \sqrt{2} 
q^3+\cdots \\[0.3em]
& \frac{21}{16} & 416 \sqrt{2}+47424 \sqrt{2} q+1772160 \sqrt{2} q^2+36478208 \sqrt{2} 
q^3+\cdots \\[0.3em]
& \frac{39}{16} & 65536 \sqrt{2}+2555904 \sqrt{2} q+51118080 \sqrt{2} q^2+701169664 
\sqrt{2} q^3+\cdots \\[0.3em]
\hline \multirow{3}{*}{$21$} & \frac{7}{8} & 64+25088 q+1705984 q^2+51681280 q^3+967852032 q^4+\cdots \\[0.3em]
& \frac{11}{8} & 262144+11010048 q+236716032 q^2+3482845184 q^3+\cdots \\[0.3em]
& \frac{21}{8} & 896+118272 q+5058816 q^2+117312512 q^3+1843356032 q^4 +\cdots \\[0.3em]
\hline \multirow{3}{*}{$22$} & \frac{11}{12} & 144+38016 q+3288384 q^2+111513600 q^3+2263529664 q^4+\cdots \\[0.3em]
& \frac{19}{12} & 2640+426624 q+18473136 q^2+433390464 q^3+6940391040 q^4 +\cdots \\[0.3em]
& \frac{9}{4} & 48400+2865280 q+79596704 q^2+1424448256 q^3+18907560144 q^4 +\cdots \\[0.3em]
\hline \multirow{3}{*}{$\frac{45}{2}$} & \frac{15}{16} & 64 \sqrt{2}+28800 \sqrt{2} q+2236800 \sqrt{2} q^2+76440320 \sqrt{2} 
q^3+\cdots \\[0.3em]
& \frac{23}{16} & 524288 \sqrt{2}+23592960 \sqrt{2} q+542638080 \sqrt{2} q^2+8524922880 
\sqrt{2} q^3+\cdots \\[0.3em]
& \frac{45}{16} & 960 \sqrt{2}+145280 \sqrt{2} q+7057152 \sqrt{2} q^2+183344640
\sqrt{2} q^3+\cdots \\[0.3em]
\hline \multirow{3}{*}{$24$} & 1 & 128+65536 q+5767168 q^2+220987392 q^3+5104467968 q^4+\cdots \\[0.3em]
& \frac{3}{2} & 2097152+100663296 q+2466250752 q^2+41204842496 q^3  +\cdots \\[0.3em]
& 3 &2048+352256 q+19296256 q^2+558743552 q^3+10708527104 q^4 +\cdots \\[0.3em]
\hline
\end{tabular}
}
\caption{\label{R-sector list}The solutions of the R-sector BPS third-order MLDE are listed in $q$-expansion.}
\end{table}

\bibliographystyle{IEEEtran}
\bibliography{main}

\end{document}